\documentclass[prd,twocolumn,nofootinbib]{revtex4}
\usepackage{epsfig,psfrag,graphics,color,verbatim,amsmath}
\newcommand{\params}{\Theta}
\newcommand{\data}{d}
\newcommand{\like}{\mathcal{L}}
\newcommand{\be}{\begin{equation}}
\newcommand{\ee}{\end{equation}}
\newcommand{\bone}[1]{$\gamma_1$}

\newcommand{\msbar}{\overline{MS}} 

\newcommand{\mtpole}{M_t}
\newcommand{\alphas}{\alpha_s(M_Z)^{\overline{MS}}}
\newcommand{\alphaemmz}{\alpha_{\text{em}}(M_Z)^{\overline{MS}}}
\newcommand{\mbmbmsbar}{m_b(m_b)^{\msbar} }

\newcommand{\met}{\not \!\! E_T}

\newcommand\etal{{\it {et al.}}}
\newcommand{\Oh}{\Omega_\text{DM} h^2}
\newcommand{\OhKK}{\Omega_\text{KK} h^2}
\newcommand{\muW}{\mu_\text{WMAP}}
\newcommand{\siW}{\sigma_{\text{WMAP}}}
\newcommand{\siSI}{\sigma_\text{p}^\text{SI}}
\newcommand{\siSD}{\sigma_\text{p}^\text{SD}}
\newcommand\lsim{\mathrel{\rlap{\lower4pt\hbox{\hskip1pt$\sim$}}
    \raise1pt\hbox{$<$}}}
\newcommand\gsim{\mathrel{\rlap{\lower4pt\hbox{\hskip1pt$\sim$}}
    \raise1pt\hbox{$>$}}}
\newcommand{\superbayes}[1]{\texttt{SuperBayeS}}
\begin{document}

\title{Global fits of the Minimal Universal Extra Dimensions scenario} 
\author{Gianfranco Bertone\,$^{1,2}$,
  Kyoungchul Kong\,$^{3,4}$,
   Roberto Ruiz de Austri\,$^{5}$ and 
  Roberto Trotta\,$^{6}$}
\affiliation{${^1}$ Institut for Theoretical Physics, Univ. of
  Z\"urich, Winterthurerst. 190, 8057 Z\"urich CH} 
\affiliation{${^2}$ 
  Institut d'Astrophysique de Paris, UMR 7095-CNRS, Univ. P. et
  M. Curie, 98bis Bd Arago, 75014 Paris, France} 
\affiliation{${^3}$Theoretical Physics Department, SLAC, Menlo Park, CA 94025, USA}
\affiliation{${^4}$Department of Physics and Astronomy, University of Kansas, Lawrence, KS 66045, USA}
\affiliation{${^5}$ Instituto de F\'isica Corpuscular, IFIC-UV/CSIC,
  Valencia, Spain} 
\affiliation{${^6}$ Astrophysics Group, Imperial College London,
  Blackett Laboratory, Prince Consort Road, London SW7 2AZ, UK} 

\begin{abstract}
In theories with Universal Extra-Dimensions (UED), the $\gamma_1$ particle, first excited state of the hypercharge gauge boson, provides an excellent Dark Matter (DM) candidate.
Here we use a modified version of the \superbayes{} code to perform a Bayesian analysis of the minimal UED scenario, in order to assess its detectability at accelerators and with DM experiments. We derive in particular the most probable range of mass and scattering cross sections off nucleons, keeping into account cosmological and electroweak precision constraints. The consequences for the detectability of the $\gamma_1$ with direct and indirect experiments are dramatic. The spin-independent cross section probability distribution peaks at $\sim 10^{-11}$ pb, i.e. below the sensitivity of ton-scale experiments. The spin-dependent cross-section drives the predicted neutrino flux from the center of the Sun below the reach of present and upcoming experiments. The only strategy that remains open appears to be direct detection with ton-scale experiments sensitive to spin-dependent cross-sections. On the other hand, the LHC with 1 fb$^{-1}$ of data should be able to probe the current best-fit UED parameters. 
\end{abstract}

\maketitle

%%%%%%%%%%%%%%%%%%%%%%%%%%%%%%%%%%%%%%%%%%%%%%%%%%%%%%%%%%%%%%%%%%%%%
\section{Introduction.} \label{intro}
%%%%%%%%%%%%%%%%%%%%%%%%%%%%%%%%%%%%%%%%%%%%%%%%%%%%%%%%%%%%%%%%%%%%%

%{\bf (Gf: Maybe KC can start drafting this, especially with key-references on UED models. Then I can add references to DM searches, SuperBayes etc.)}

Dark matter (DM) studies are often carried out in the framework of Supersymmetric (SUSY) theories, but there are many alternative extensions of the Standard model of particle physics that lead to viable DM candidates. Among them, models with universal extra dimensions (UED), in which all Standard Model (SM) particles can propagate 
in the bulk of one or more compactified flat extra dimensions \cite{Appelquist:2000nn}, have received considerable attention, and they have been studied in relation to
collider phenomenology, 
%\cite{Rizzo:2001sd,Macesanu:2002db,Cheng:2002ab,Carone:2003ms,
%Bhattacharyya:2005vm,Battaglia:2005zf,Smillie:2005ar,Battaglia:2005ma,Datta:2005zs,
%Datta:2005vx,Kong:2006pi,Cembranos:2006gt,Bhattacherjee:2007wy,Bhattacherjee:2008ik,
%Konar:2009ae,Matsumoto:2009tb,Bhattacharyya:2009br,Bandyopadhyay:2009gd}, 
indirect low-energy constraints, 
%\cite{Agashe:2001ra,Agashe:2001xt,Appelquist:2001jz,%
%Petriello:2002uu,Appelquist:2002wb,Chakraverty:2002qk,Buras:2002ej,%
%Oliver:2002up,Buras:2003mk,Iltan:2003tn,Khalil:2004qk,Bashiry:2008en,Gogoladze:2006br,Haisch:2007vb}
cosmology, 
%\cite{Matsumoto:2006bf,Shah:2006gs,Li:2005aia,Feng:2003nr,Mazumdar:2003vg,Bringmann:2003sz,Kolb:2003mm}.
and dark matter 
% \cite{Servant:2002aq,Cheng:2002ej,Servant:2002hb,Majumdar:2002mw,Burnell:2005hm,Kong:2005hn,Arrenberg:2008wy,
%Majumdar:2003dj,Kakizaki:2005en,Kakizaki:2005uy,Bertone:2002ms,Bergstrom:2004cy,
%Baltz:2004ie,Bergstrom:2004nr,Bringmann:2005pp,Barrau:2005au,Birkedal:2005ep,Flacke:2008ne,Belanger:2008gy,
%Blennow:2009ag,Matsumoto:2007dp,Kakizaki:2006dz,Matsumoto:2005uh,Bertone:2009cb,Bertone:2010fn}.
  (see Refs. \cite{Hooper:2007qk,Datta:2010us} for recent reviews).
In the simplest and most popular version, there is a single extra dimension 
compactified on an interval, $S_1/Z_2$. Each SM particle has a whole tower of Kaluza-Klein
(KK) modes, labelled by an integer $n$, called KK number,
which is nothing but the number of quantum units of momentum which the SM particle
carries along the extra dimension. 

One of the peculiar features of UED theories is the conservation of the
KK number at tree level, which is a simple consequence 
of momentum conservation along the extra dimension. 
This implies that the lightest KK parity odd particle (LKP) is stable over cosmological timescales, and being cold and neutral, it provides a suitable WIMP
candidate \cite{Cheng:2002iz,Servant:2002aq,Cheng:2002ej,Servant:2002hb,Arrenberg:2008wy}.

In this paper we shall concentrate on the Minimal Universal Extra Dimensions (MUED) 
discussed in Ref. \cite{Cheng:2002ab}, where there are only two parameters 
in addition to the Higgs mass ($m_h$), namely the size of extra dimension ($R$) and the cutoff scale of the theory, $\Lambda$ (see Refs. \cite{Dobrescu:2004zi,Burdman:2006gy,Dobrescu:2007xf,Dobrescu:2007ec,Freitas:2007rh,Ghosh:2008ix,Ghosh:2008ji,Bertone:2009cb} for the case of two universal extra dimensions).
Often $\Lambda R$ is used instead of the cutoff scale itself, and we follow the same convention.

The LKP in MUED turns out to be the level-1 KK partner ($\gamma_1$) of the SM photon, which must therefore be stable by virtue of the conservation of KK parity. 
In the limit where $R^{-1} > v$, where 
$v$ is the vacuum expectation value of the Higgs, 
the Weinberg angle for KK gauge boson is negligible, and the KK gauge bosons are in fact almost weak 
eigenstate. Therefore the KK photon is almost the hypercharge KK gauge boson ($\gamma_n \approx B_n$) and 
the KK Z is almost the neutral weak eigenstate of $SU(2)_W$ ($Z_n \approx W_n^3$). 

%The fixed points in orbifold compactifications break translation invariance along the extra dimension, and the KK number is broken by
%bulk and brane radiative effects~\cite{Cheng:2002iz} down to a discrete conserved quantity, the so called {\it KK parity} $(-1)^n$. The geometrical origin of KK parity in the simplest ($S_1/Z_2$) case is the invariance under reflections with respect to the center of the interval. 

%KK parity also ensures that the KK-parity odd KK partners (e.g. those at level one)
%are always pair-produced in collider experiments, and the
%virtual effects from UED only appear at the loop level and are loop suppressed. 

While we do not know much about the DM particle, there are many direct, indirect and accelerator searches current undergoing, with the aim of identifying them \cite{Bergstrom:2000pn,Bertone:2004pz,book}. KK DM provides a valid alternative to the widely discussed Supersymmetric DM \cite{Jungman:1995df}, and it is often adopted as a case study scenario when trying to assess the capability of experimental strategies to discriminate among various DM candidates (e.g. \cite{Bertone:2007xj}). 

In this paper we perform a Bayesian analysis of the MUED scenario, in order to assess its detectability at accelerators and with DM experiments. We derive in particular the most probable range of mass and scattering cross sections off nucleons, keeping into account cosmological and electroweak precision constraints. As we shall see, this has dramatic implications for the detectability of KK DM.

The paper is organized as follows: in section \ref{sec:theory} we discuss the theoretical framework of the MUED scenario. In section \ref{sec:bayes} we provide some details on our statistical tools, including a discussion of the priors adopted in the Bayesian analysis of the MUED parameter space. In Sec. \ref{sec:results} we present the results and in \ref{sec:conc} we discuss their consequences and conclude.

\section{Theoretical framework}
\label{sec:theory}

%{\bf (Gf: KC please describe the theoretical framework. I guess this is also the place where we can describe the relic density calculations, including co-annihilations.)}

% {\bf (KC:  
% the model  \cite{Appelquist:2000nn},
% mass spectrum \cite{Cheng:2002iz},
% Relic density \cite{Servant:2002aq,Burnell:2005hm,Kong:2005hn},
% EW constraints \cite{Gogoladze:2006br}, 
% DD \cite{Cheng:2002ej,Servant:2002hb,Arrenberg:2008wy}, 
% ID \cite{Cheng:2002ej,Bertone:2002ms}, 
% Tevatron analysis \cite{Lin:2005ix}, 
% LHC reach (level-1 and level-2) \cite{Cheng:2002ab,Datta:2005zs}.)}

In the MUED, the vanishing boundary conditions are assumed for all KK particles at the cutoff scale 
(i.e., all KK particles at level-n are degenerate ($m_n = n/R$) at the cutoff scale), and therefore 
the mass spectrum at electroweak (EW) scale ($R^{-1}$) is completely determined 
by RG evolution between $R^{-1}$ and $\Lambda$ \cite{Cheng:2002iz} 
(there is also a contribution from EW symmetry breaking, which is small except for top quark).
Since the estimated cutoff scale is not too far away from $R^{-1}$, 
the resulting mass spectrum is somewhat degenerate due to short RG running.
As expected from RG running, the masses of the KK particles depend on how strongly they interact, therefore 
strongly interacting KK particles get larger corrections than weakly interacting particles.
In MUED, the KK gluon is the heaviest particle, followed by KK quarks, KK Z/W and KK leptons.

Due to KK-parity, contributions to electroweak observables do not appear at tree-level and 
this allows KK particles to be light enough so that they can be produced at current collider experiments.
This has been studied in Refs. \cite{Appelquist:2000nn,Appelquist:2002wb} and 
revisited more recently in Ref. \cite{Gogoladze:2006br} including subleading contributions 
as well as two loop corrections to the SM $\rho$ parameter.
 A lower bound on $R^{-1}$ from those oblique corrections is $\sim$600 GeV at 90\% C.L. with a Higgs mass of 115 GeV, which is the LEP limit.
However this constraint is significantly relaxed with increasing Higgs mass, 
allowing for a compactification scale as low as 300 GeV 
(other indirect low-energy constraints 
%\cite{Agashe:2001ra,Agashe:2001xt,Petriello:2002uu,Chakraverty:2002qk,Buras:2002ej,
%Oliver:2002up,Buras:2003mk,Iltan:2003tn,Khalil:2004qk,Bashiry:2008en,Haisch:2007vb}
 are comparable or weaker. See \cite{Hooper:2007qk,Datta:2010us} and references therein.).

In the rest of this section, we briefly review the calculation of the relic density.
Since KK particles in MUED are somewhat degenerate, it is important to include coannihilation effects.
The generalization of the relic density calculation including 
coannihilations is straightforward~\cite{Griest:1990kh,Servant:2002aq}.
Assume that the particles $\chi_i$ are labelled according to their masses, 
so that $m_i < m_j$ when $i < j$. 
The number densities $n_i$ of the various species $\chi_i$ obey
a set of Boltzmann equations. It can be shown that
under reasonable assumptions \cite{Griest:1990kh},
the ultimate relic density $n_\chi$ of the lightest species $\chi_1$ 
(after all heavier particles $\chi_i$ have decayed into it)
obeys the following simple Boltzmann equation

\begin{equation}
\frac{d n_\chi}{ d t} = -3 Hn_\chi - \langle \sigma_\text{eff} v \rangle ( n_\chi^2 - n^2_{eq})\ ,
\end{equation}
where $H$ is the Hubble parameter, $v$ is the relative velocity between the two incoming particles, 
$n_{eq}$ is the equilibrium number density and 
\begin{eqnarray}
\sigma_\text{eff}(x) &=& \sum_{ij}^N \sigma_{ij} \frac{g_i g_j}{g_\text{eff}^2} 
                 (1 + \Delta_i)^{3/2} (1 + \Delta_j)^{3/2}  \nonumber \\
               && ~~~ \otimes  \exp(-x ( \Delta_i + \Delta_j ))\ ,
\label{sigmaeff}\\
g_\text{eff}(x)   &=& \sum_{i=1}^N g_i (1+\Delta_i)^{3/2} \exp(-x \Delta_i)\ , 
\label{geff}\\
\Delta_i     &=& \frac{m_i - m_1}{m_1}\ , \text{~~~~} x = \frac{m_1}{T}.
\end{eqnarray}
Here $\sigma_{ij}\equiv \sigma(\chi_i\chi_j\to SM)$ are the various pair 
annihilation cross sections into final states with SM particles,
$g_i$ is the number of internal degrees of freedom of particle $\chi_i$ and 
$n_\chi \equiv \sum_{i=1}^N n_i$ is the density of $\chi_1$ we want to calculate.

By solving the Boltzmann equation analytically with appropriate approximations
~\cite{Griest:1990kh, Servant:2002aq}, the abundance of the lightest species
$\chi_1$ is given by 
\begin{equation}
\Omega_\chi h^2 \approx \frac{1.04 \times 10^9\ {\rm GeV}^{-1}}{M_{Pl}}
\frac{x_F}{\sqrt{g_\ast(x_F)}} \frac{1}{I_a+3 I_b/x_F }\ ,
\label{oh2coann}
\end{equation}
where the Planck mass scale is $M_{Pl} = 1.22\times 10^{19}$\,GeV and
$g_\ast$ is the total number of effectively massless degrees of freedom at temperature $T$:
\begin{equation}
g_\ast(T) = \sum_{i=\text{bosons}} g_i + \frac{7}{8} \sum_{i=\text{fermions}}g_i\ .
\label{gstar}
\end{equation}
The functions $I_a$ and $I_b$ are defined as
\begin{eqnarray}
I_a &=& x_F \int_{x_F}^\infty a_\text{eff}(x) x^{-2} d x \ ,
\label{I_a} \\
I_b &=& 2 x_F^2 \int_{x_F}^\infty b_\text{eff}(x) x^{-3} d x\ .
\label{I_b}
\end{eqnarray}
The freeze-out temperature, $x_F$, is found iteratively from
\begin{eqnarray}
x_F &=& \ln \Big [ c(c+2) \sqrt{\frac{45}{8}} \frac{g_\text{eff}(x_F)}{2\pi^3} 
                 \frac{m_1 M_{Pl}}{\sqrt{g_\ast(x_F) x_F}}  \nonumber  \\
    &&~~~~~~~~ \otimes \Big ( a_\text{eff}(x_F)+ 6 \frac{b_\text{eff}(x_F)}{x_F} \Big ) \Big ],
\label{xfcoann}
\end{eqnarray}
where the constant $c$ is determined empirically by comparing to numerical solutions of 
the Boltzmann equation and here we take $c=\frac{1}{2}$ as usual. 
$a_\text{eff}$ and $b_\text{eff}$ are the first two terms 
in the velocity expansion of $\sigma_\text{eff}$
\begin{equation}
\sigma_\text{eff}(x)\,v = a_\text{eff}(x) + b_\text{eff}(x)\, v^2 + {\cal O}(v^4)\ .
\label{sigmaeffv}
\end{equation}
Comparing Eqns.~(\ref{sigmaeff}) and (\ref{sigmaeffv}), one gets
\begin{eqnarray}
a_\text{eff}(x) &=& \sum_{ij}^N a_{ij} \frac{g_i g_j}{g_\text{eff}^2} 
                 (1 + \Delta_i)^{3/2} (1 + \Delta_j)^{3/2} \nonumber \\
          && ~~~ \otimes \exp(-x ( \Delta_i + \Delta_j ))\ , 
\label{aeff}\\
b_\text{eff}(x) &=& \sum_{ij}^N b_{ij} \frac{g_i g_j}{g_\text{eff}^2} 
                 (1 + \Delta_i)^{3/2} (1 + \Delta_j)^{3/2}\nonumber \\
           && ~~~ \otimes \exp(-x ( \Delta_i + \Delta_j ))\ ,
\label{beff}
\end{eqnarray}
where $a_{ij}$ and $b_{ij}$ are obtained from $\sigma_{ij}v = a_{ij} + b_{ij} v^2 + {\cal O}(v^4)$ and 
$v$ is the relative velocity between the two annihilating particles in the initial state.
Considering relativistic corrections %\cite{Srednicki:1988ce}
to the above treatment results in an additional subleading 
term which can be accounted for by the simple replacement 
\begin{equation}
b\to b-\frac{1}{4}a \, ,
\label{bcorr}
\end{equation}
in the above formulas.
For our calculation of the relic density, 
we use the cross sections given in Refs.~\cite{Servant:2002aq,Burnell:2005hm,Kong:2005hn}.
For resonance effect, which we do not include, see Refs. \cite{Kakizaki:2005en,Kakizaki:2005uy,Kakizaki:2006dz}.
In the MUED, coannihilation with $SU(2)_W$-singlet KK leptons ($e_1$) is important since 
it is the next-to-lightest KK particle, and $\frac{m_{e_1}-m_{\gamma_1}}{m_{e_1}} \sim 0.01$.
%{\bf (need to shorten explanation on relic density.)}

%%%%%%%%%%%%%%%%%%%%%%%%%%%%%%%%%%%%%%%%%%%%%%%%%%%%%%%%%%%%%%%%%%%%%
\section{Statistical analysis} \label{sec:bayes}
%%%%%%%%%%%%%%%%%%%%%%%%%%%%%%%%%%%%%%%%%%%%%%%%%%%%%%%%%%%%%%%%%%%%%

The free parameters of the model are the SM Higgs mass, $m_h$, the inverse radius of the UED, $R^{-1}$, and the cutoff scale $\Lambda$. For numerical reasons, we work with the following MUED parameters:
\be \label{eq:UEDpars}
\Psi = \left\{ m_h, R^{-1}, \Lambda R \right\}.
\ee
In particular, we adopt as a free parameter the number of KK levels $\Lambda R$ rather than $\Lambda$ itself. In our scan, $\Lambda R$ is considered as a real-valued variable, but we the round it to the nearest integer value when computing the observable quantities. We also include in our scan as nuisance parameters the relevant SM parameter set  
\be \label{nuipars:eq} \Phi = \left\{\mtpole,\ \mbmbmsbar,\ \alphaemmz,\ \alphas \right\}, 
\ee
where $\mtpole$ is the pole top quark mass, while the other three parameters (the bottom mass, the electromagnetic and the strong coupling constants) are all evaluated in the $\msbar$ scheme at the indicated scales. 

We denote by $\params = \left\{ \Psi, \Phi \right\}$ the vector of parameters entering the analysis, and by $\data$ the available data (described below). Bayes theorem reads
\be \label{eq:bayes}
P(\params|\data) = \frac{P(\data | \params)P(\params)}{P(\data)},
\ee 
where $P(\params|\data)$ is the posterior distribution on the parameters (after the observations have been taken into account), $P(\data | \params) = \like(\params)$ is the likelihood function (when considered as a function of $\params$ for fixed data $\data$) and $P(\params)$ is the prior distribution, which encompasses our state of knowledge about the value of the parameters before we have seen the data. Finally, the quantity in the denominator of Eq.~\eqref{eq:bayes} is the Bayesian evidence (or model likelihood), a normalizing constant which does not depend on $\params$ and which can be neglected when one is interested in parameter inference. 
Together with the model we must specify the priors for the parameters, which enter Bayes' theorem, Eq.~\eqref{eq:bayes}. 
As in any good Bayesian analysis it is important to asses the relevance of prior choices, we perform our scan using two different priors: 
\begin{itemize}
\item Flat prior: a uniform prior over the ranges $10 \text{ GeV } \leq m_h \leq 3$ TeV, $280  \text{ GeV } \leq R^{-1} \leq 3$ TeV and $1 \leq \Lambda R \leq 100$.
\item Log prior: a uniform prior over the ranges $1 \leq \log(m_h/\text{GeV}) \leq 3.5$, $2.4 \leq \log(R^{-1}/\text{GeV}) \leq 3.5$ and $1 \leq \Lambda R \leq 100$.
\end{itemize}
The lower bound on $R^{-1}$ comes from considering current collider limits from trilepton search at the Tevatron, giving $R^{-1} > 280$ GeV at 95\% C.L. with 100 pb $^{-1}$ of data \cite{Lin:2005ix} (while this limit corresponds to a value of $\Lambda R=20$, it does not depend strongly on $\Lambda R$). Notice that we keep a uniform prior over $\Lambda R$ for both choices of priors of the other two variables. We take a flat prior over the SM nuisance parameters, whose value is however directly constrained by the likelihood -- hence the choice of prior for those variables is unproblematic and it does not affect our results.

\begin{figure*}
\includegraphics[width=.48 \linewidth]{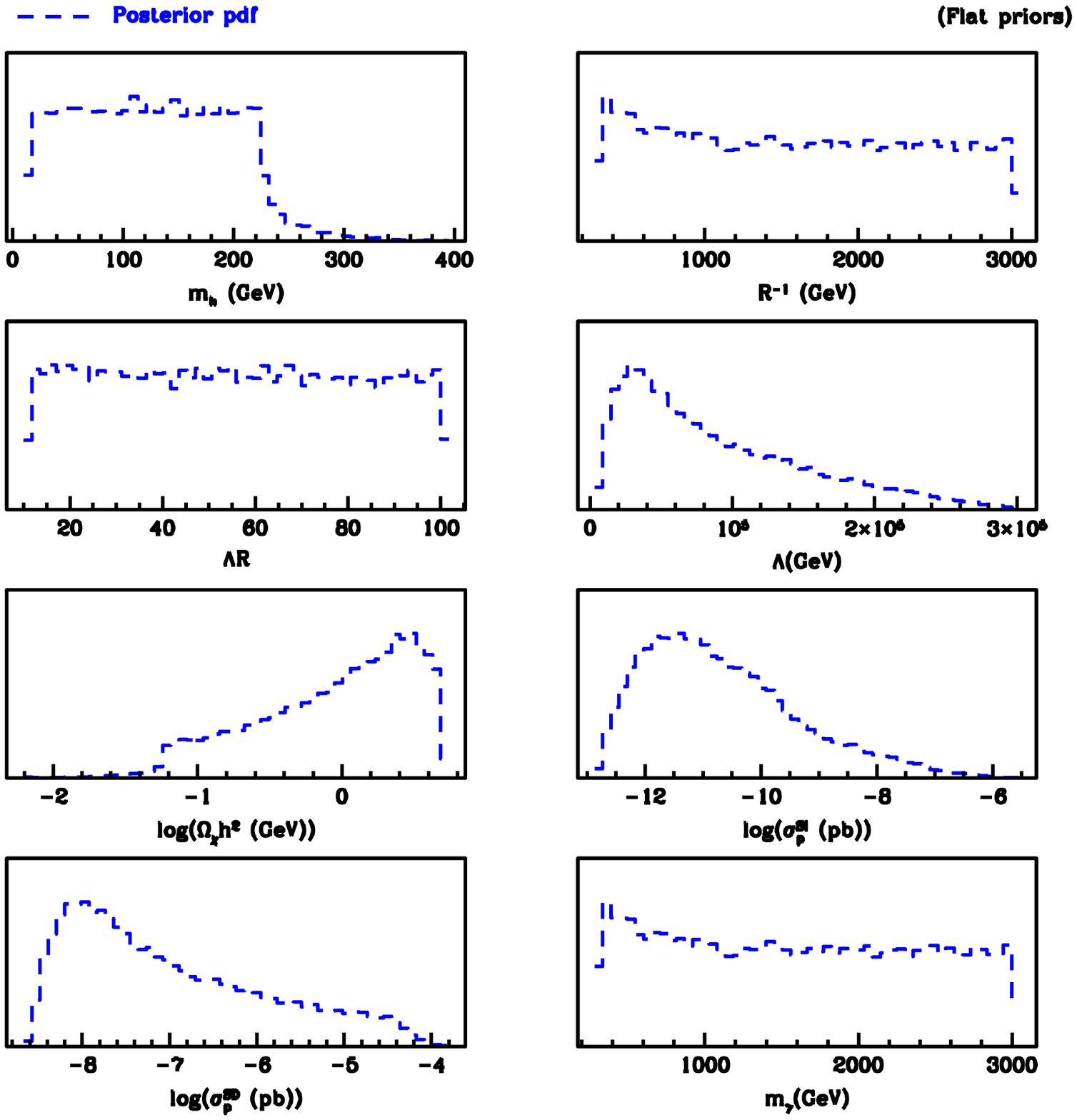} 
\includegraphics[width=.48 \linewidth]{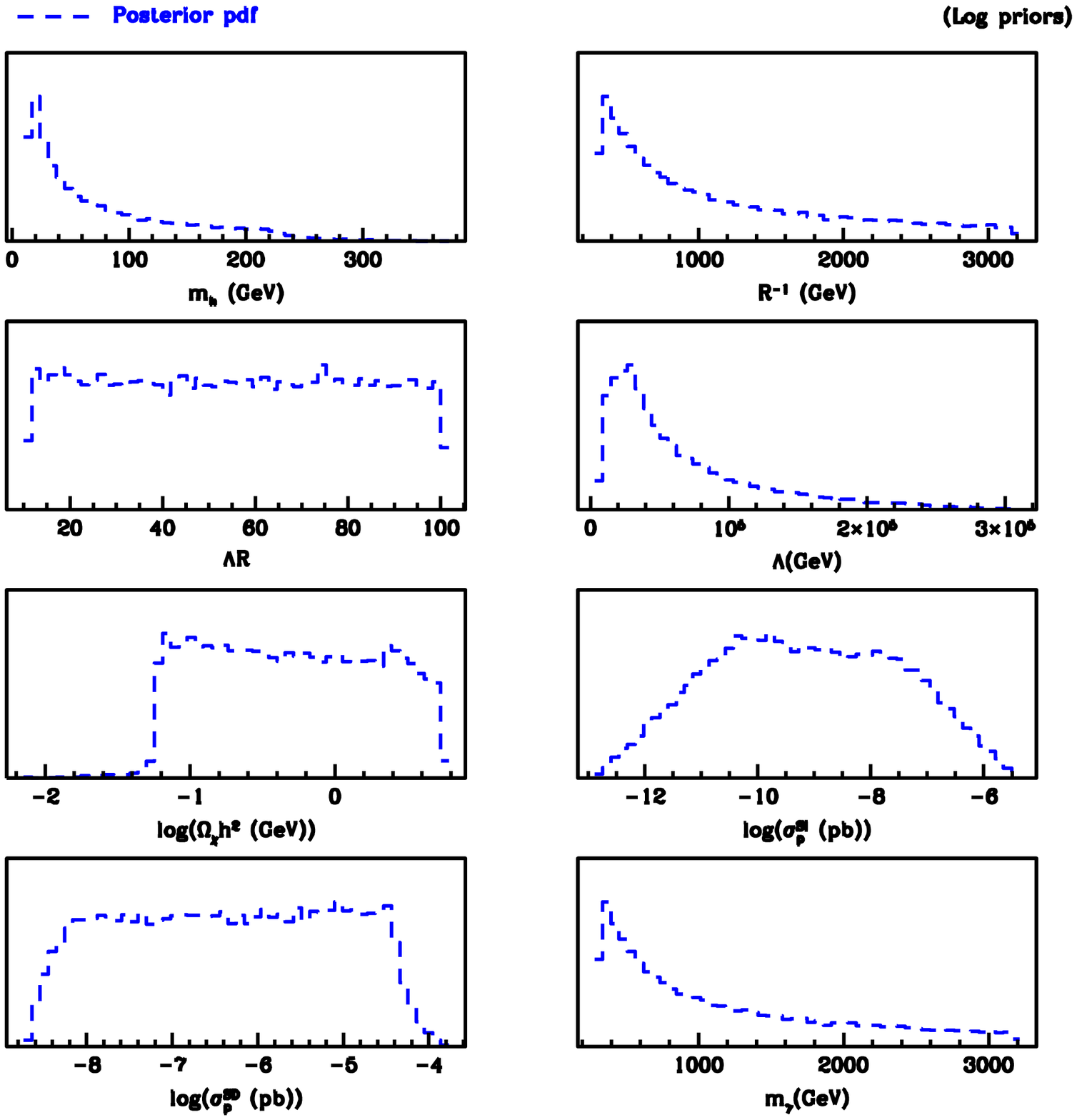} 
\caption{Prior distributions for the input variables and some observables for flat priors on $(m_h, R^{-1})$ (left panel) and log priors (right panel), which are uniform in $(\log(m_h), \log(R^{-1}))$. \label{fig:priors_1D}}
\end{figure*}

\begin{figure*}
\includegraphics[width=.24 \linewidth]{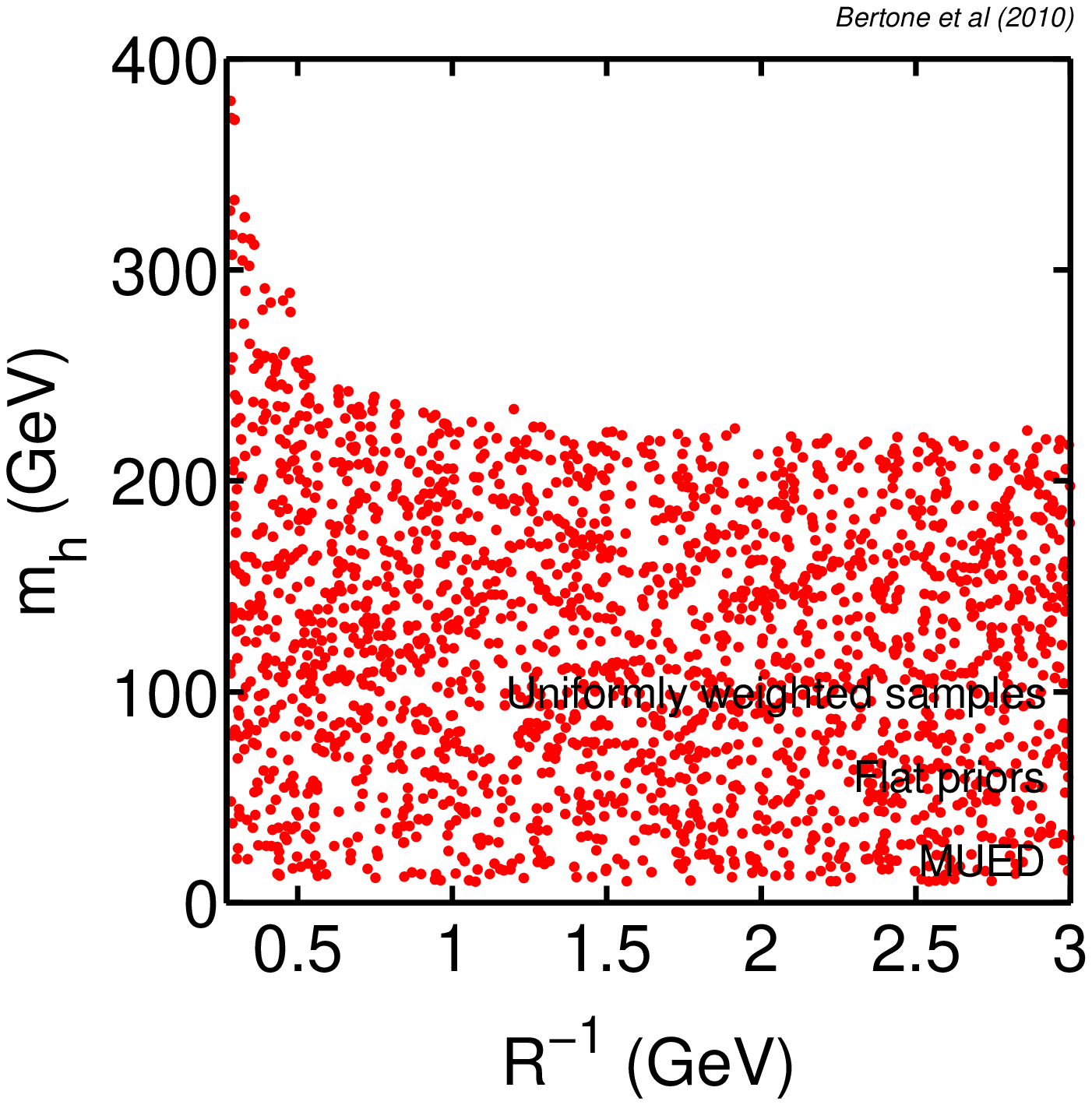} 
\includegraphics[width=.24 \linewidth]{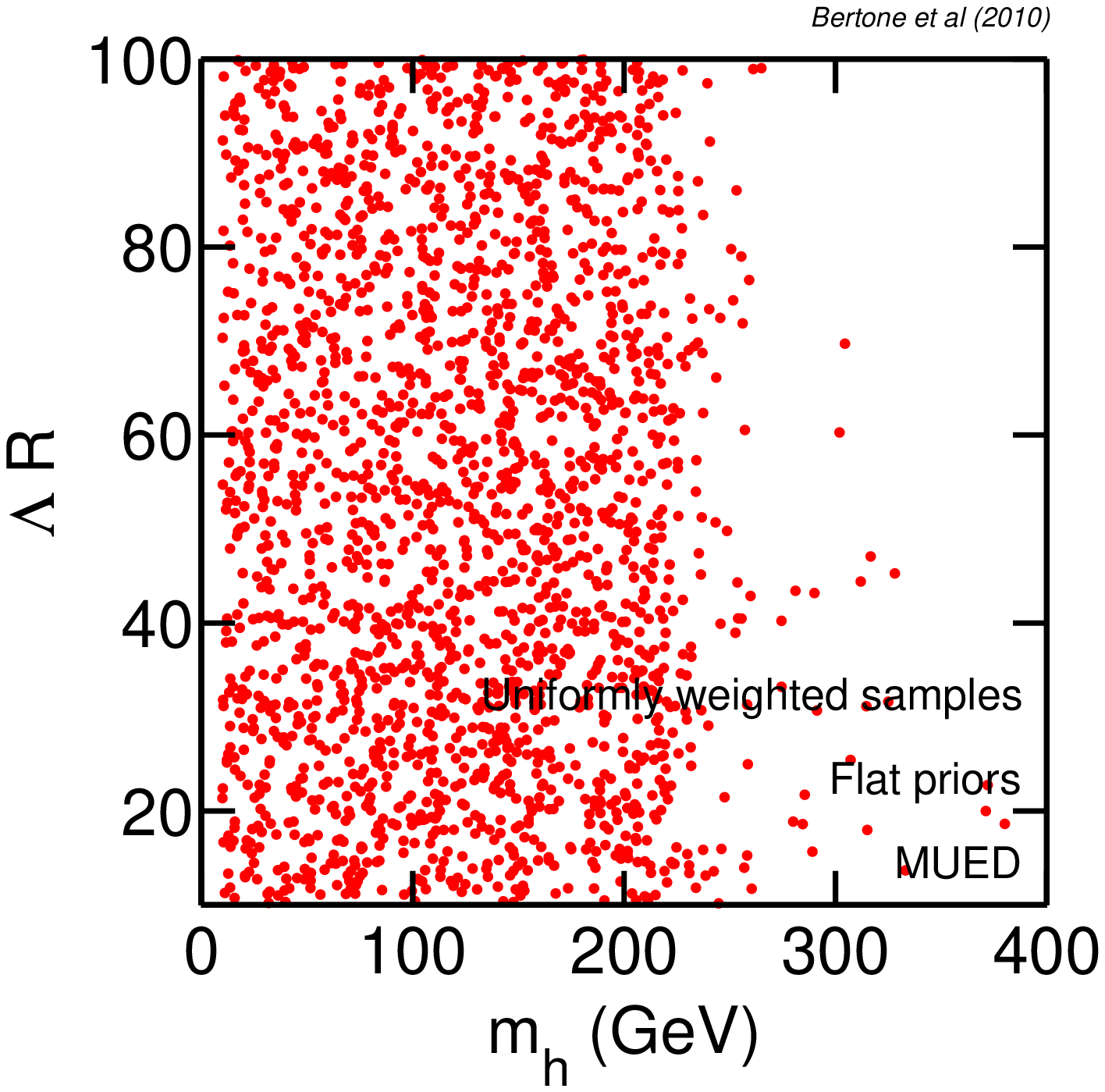} 
\includegraphics[width=.24 \linewidth]{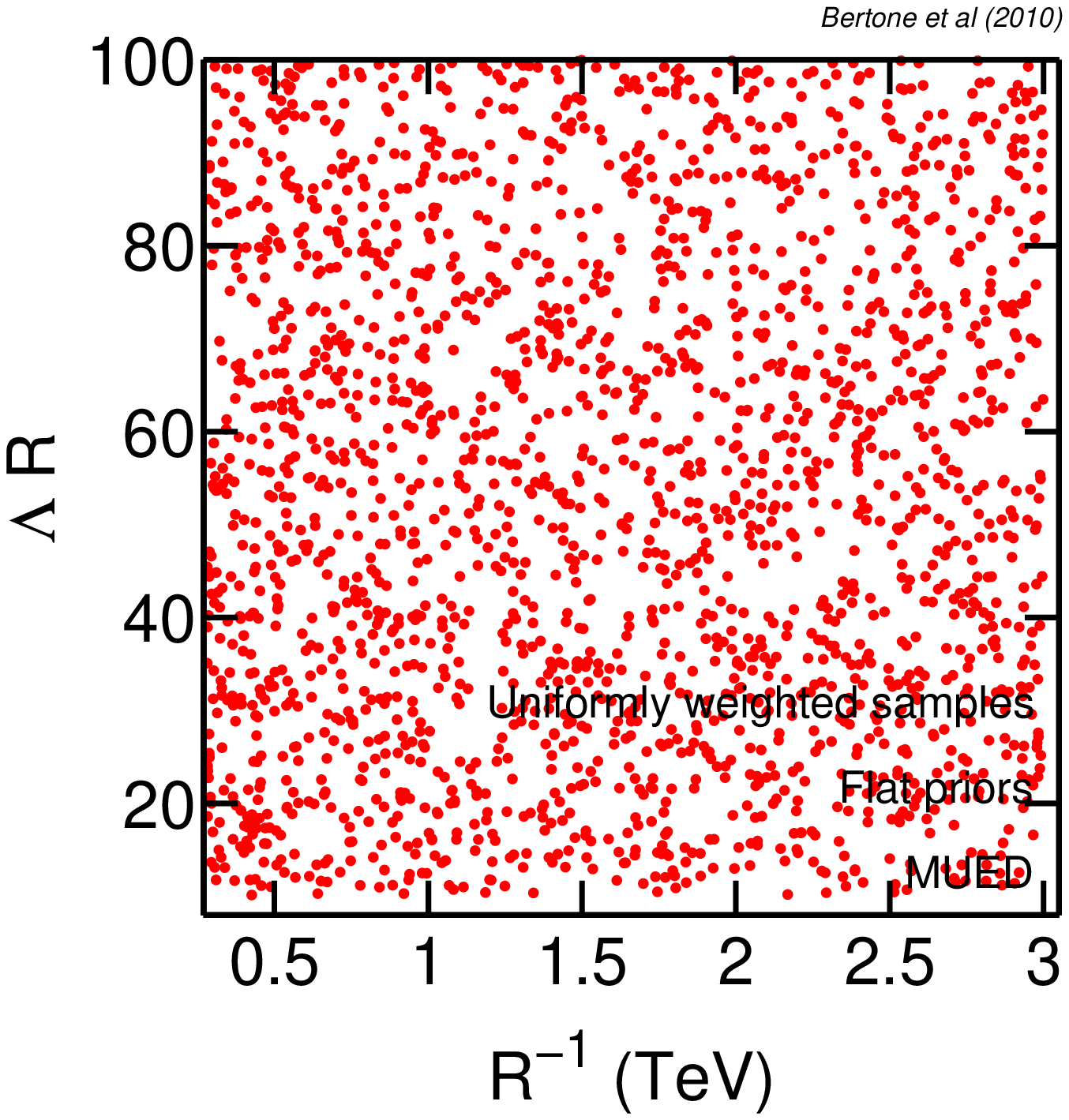}  
\includegraphics[width=.24 \linewidth]{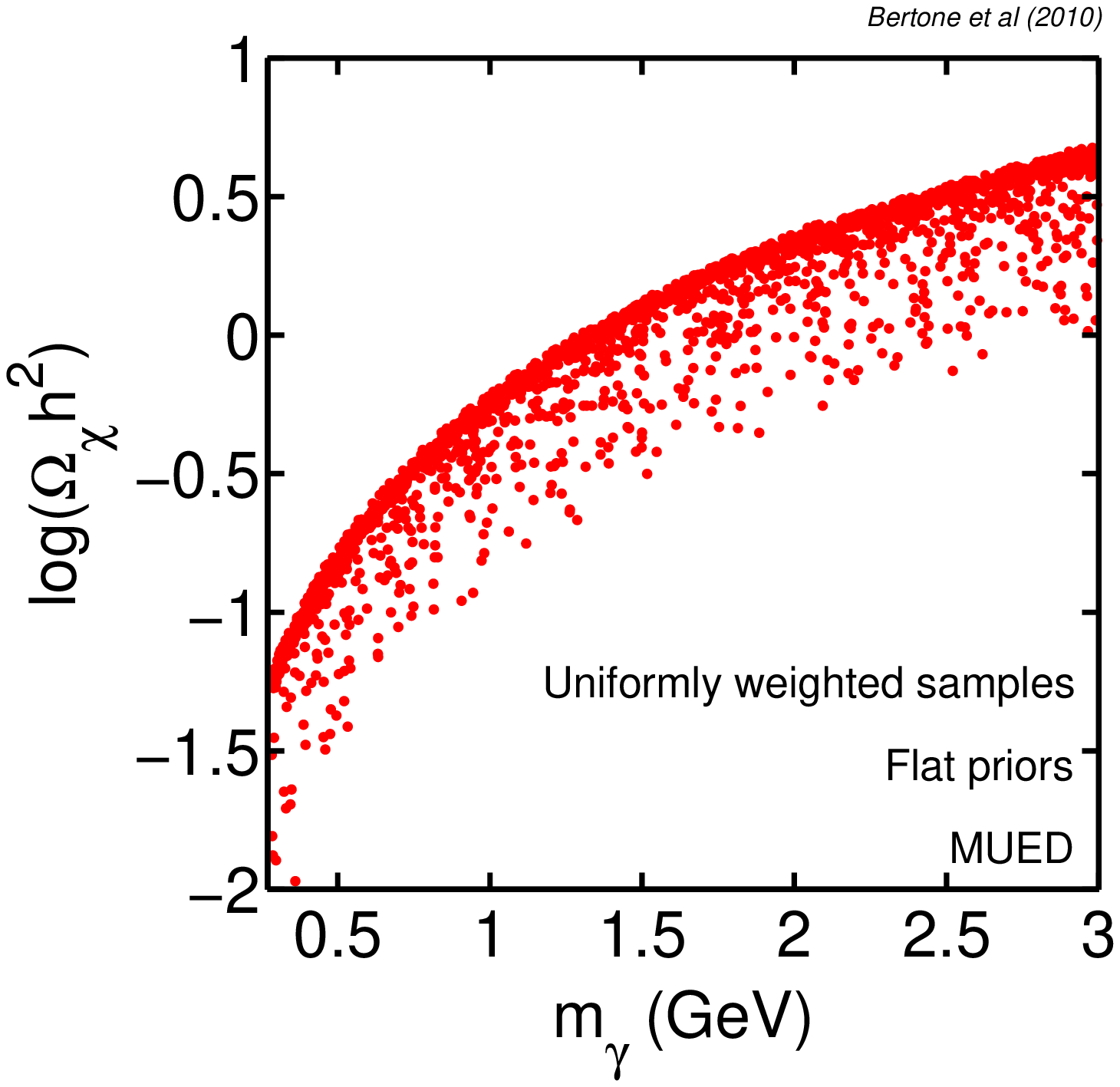} \\ 
\includegraphics[width=.24 \linewidth]{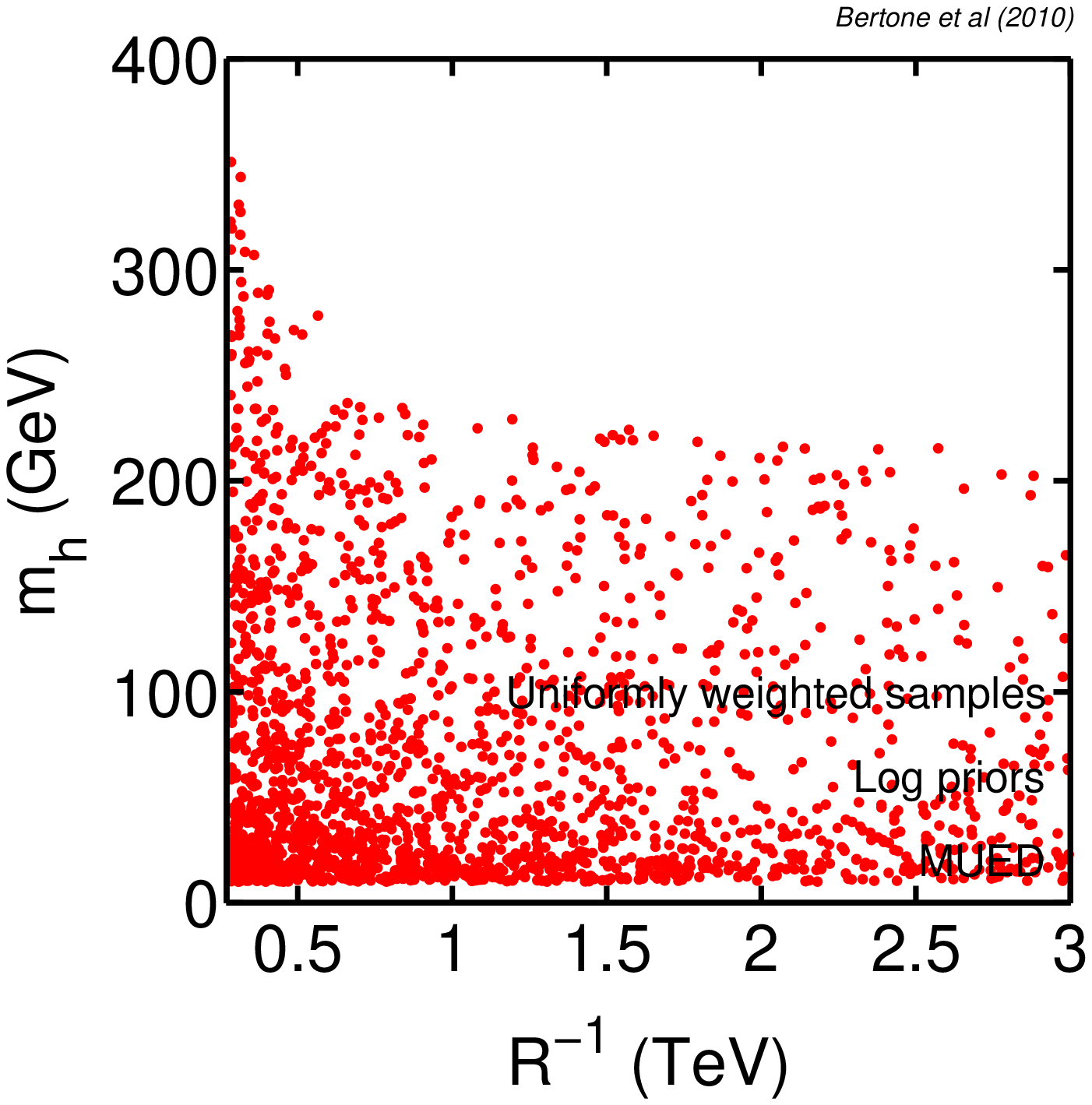} 
\includegraphics[width=.24 \linewidth]{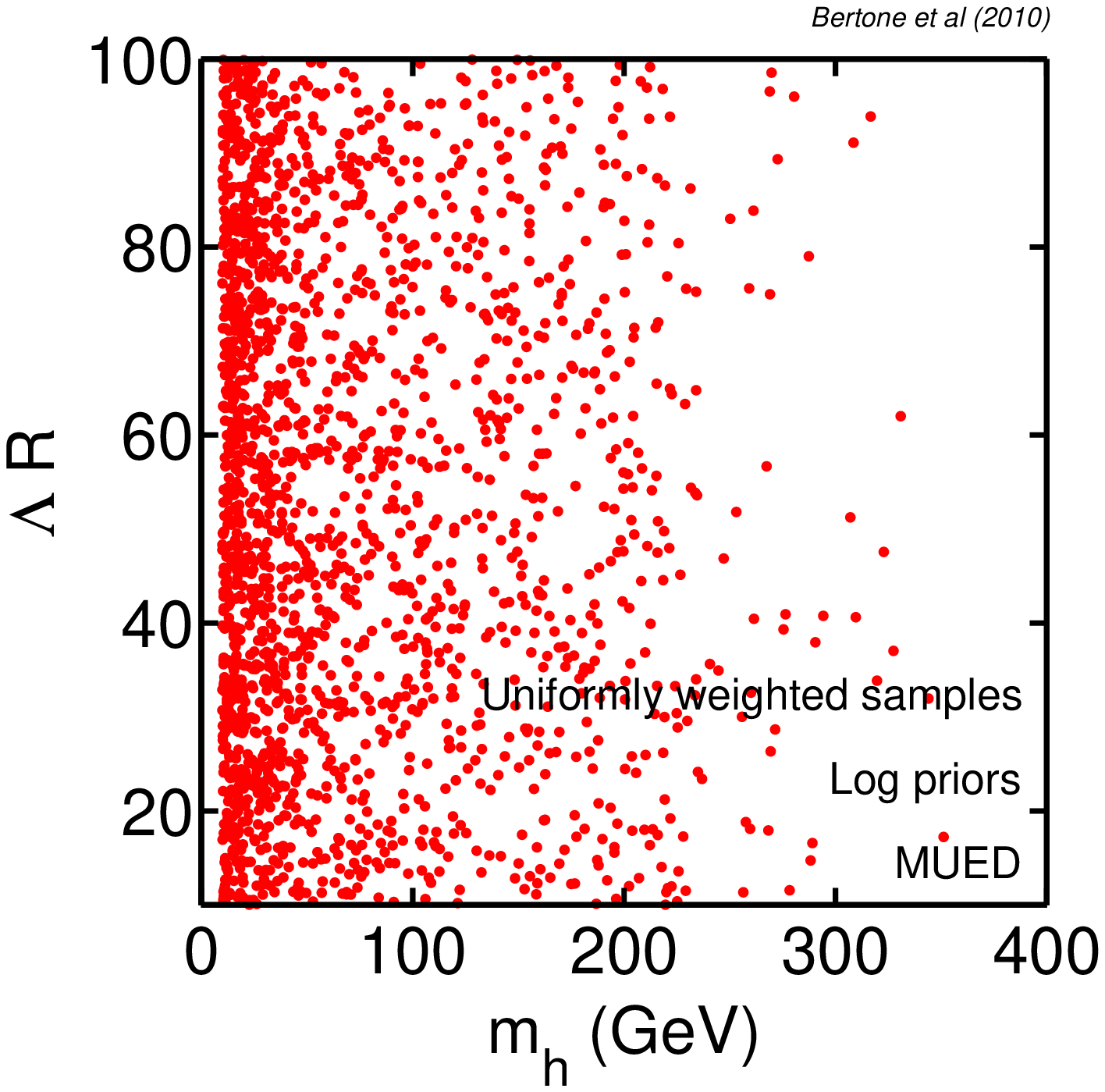} 
\includegraphics[width=.24 \linewidth]{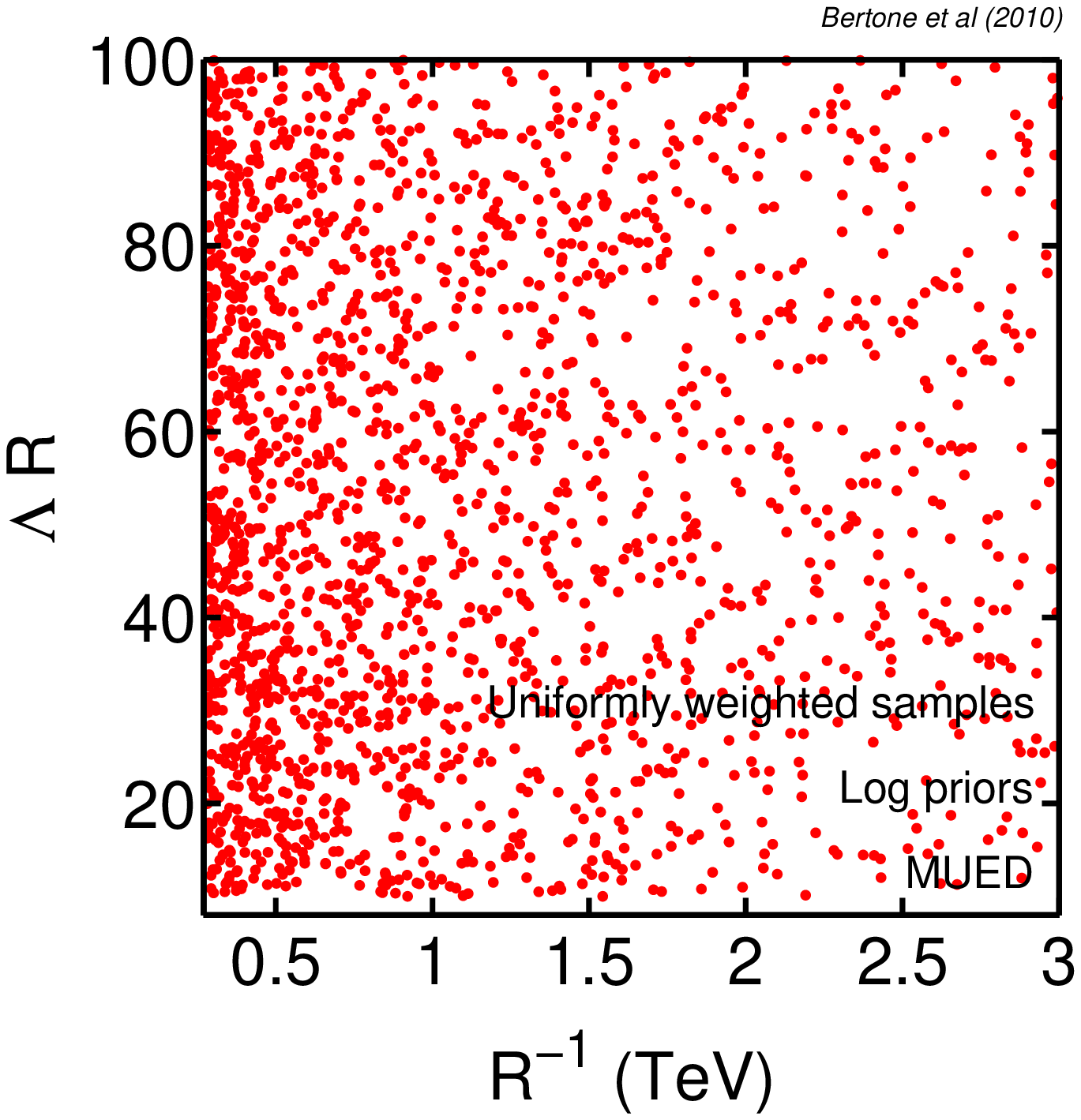}  
\includegraphics[width=.24 \linewidth]{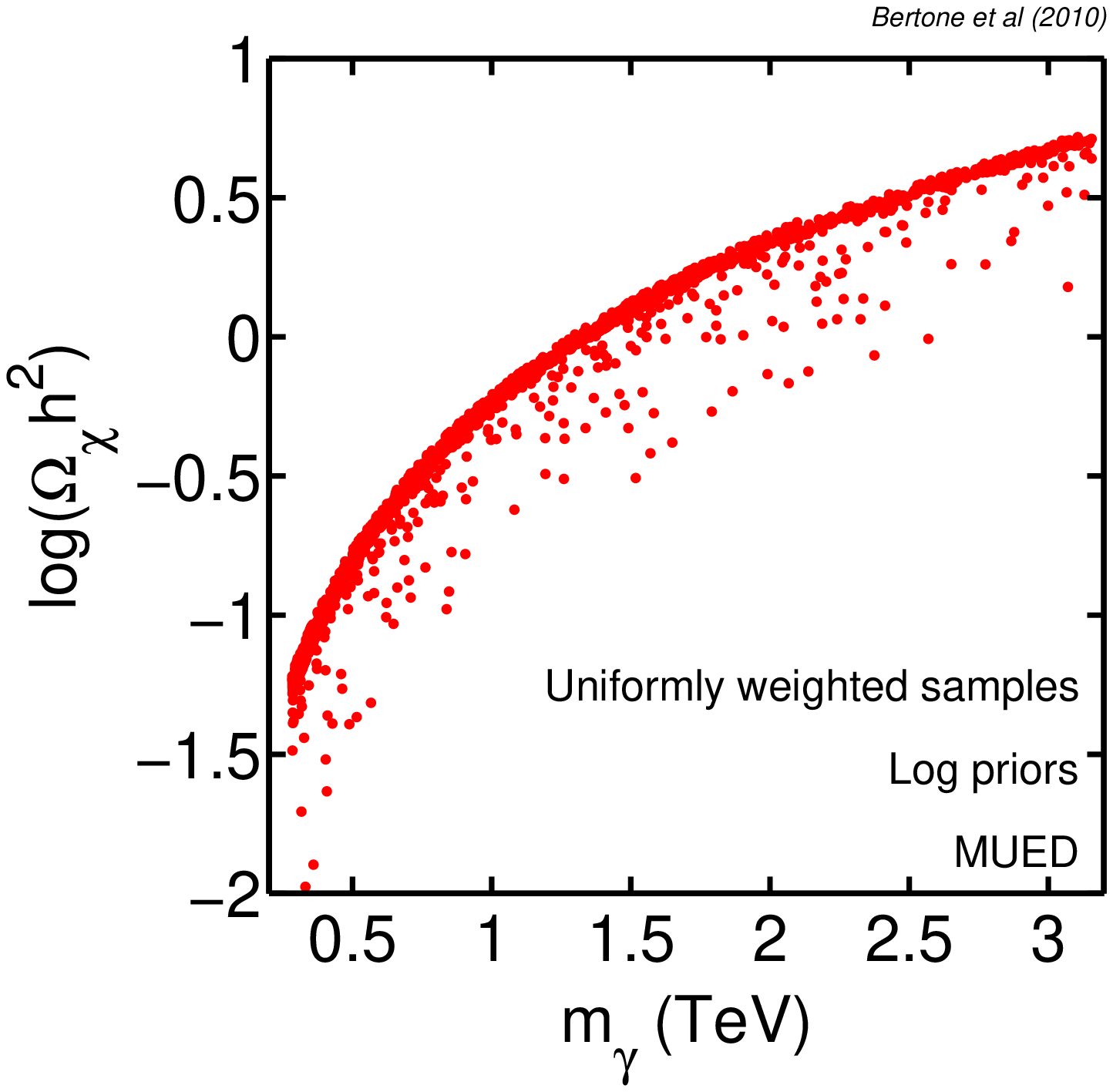} \\
\caption{Equal-weight samples from the prior for flat priors (top panels) and log priors (bottom panel). Density of points reflects prior probability density. \label{fig:priors_2D}}
\end{figure*}

The distribution of probability implied by our choice of priors for the MUED parameters and for some observables is shown in Fig.~\ref{fig:priors_1D} in one dimension, and in 2-dimensional marginal distributions in Fig.~\ref{fig:priors_2D}. We observe the expected uniform distribution in $R^{-1}$ and $\Lambda R$ for the flat prior choice in the left panels of Fig.~\ref{fig:priors_1D}, while the distribution on $m_h$ is flat up to $\sim 250$ GeV and then it falls off sharply. This is a consequence of the fact that the radiative corrections to the KK Higgs are negative and proportional to the 
mass ($m_h$) of the SM Higgs \cite{Cheng:2002iz}. 
The LEP limit on the Higgs mass is around 115 GeV, therefore leaving an allowed mass range 115-250 GeV. 
Therefore, for a given value of $R^{-1}$, there is a value of the SM Higgs mass for which 
the charged KK Higgs becomes lighter than the KK photon \cite{Cembranos:2006gt}, and  
it takes over the role of the KK photon as a DM candidate. 
As clearly charged DM is not allowed from cosmology, we discard points in which this happens. 
This sets an upper limit to the mass of the SM Higgs ($m_h \lsim 400$ GeV, but with a very low probability above$\sim  250$ GeV) \cite{Cembranos:2006gt}, as observed in Fig.~\ref{fig:priors_1D}. 
In the right-hand side panels of Fig.~\ref{fig:priors_1D}, we can observe the impact of the log prior which disfavours large values of $m_h$ and $R^{-1}$.

The distribution of the relic abundance under both priors shows that very small values are not realized, i.e. the prior density goes to 0 for $\Oh \lsim 0.05$. This is explained by Fig.~\ref{fig:priors_2D}, where it is shown how the relic abundance is tightly correlated with $R^{-1}$, which controls the level of degeneracy between the masses in the KK spectrum. At lower values of $R^{-1}$, the spectrum becomes more and more degenerate, hence annihilation is more efficient and the relic abundance is reduced. However, since our prior includes a lower limit $R^{-1}>280$ GeV due to the Tevatron constraints, as explained above, this leads to a lower limit in the distribution of $\Oh$ from the prior. 
While particle physics alone does not provide an upper bound on $R^{-1}$, the thermal relic density of LKPs grows with $R^{-1}$ and LKPs would overclose our universe for $R^{-1} > 1.5$ TeV \cite{Servant:2002aq}.  
Finally, the prior distribution of both the spin-dependent and the spin-independent cross sections is relatively flat and spans several orders of magnitude. This range is to be compared with the much tighter range in the posterior (see below Fig.~\ref{fig:1D_global_constraints}), which means that the posterior distribution for those quantities (to be discussed in detail below) is dominated by the likelihood.

The likelihood function is constructed as follows. For each of the SM parameters, we assume a Gaussian likelihood with mean and standard deviation as given in  
Table~\ref{tab:nuisance}. 
%
%%%%%%%%%%
\begin{table}[t]
\centering    .

\begin{tabular}{|l | l l | l|}
\hline
SM (nuisance)  &   Mean value  & \multicolumn{1}{c|}{Standard deviation} & Ref. \\
 parameter &   $\mu$      & ${\sigma}$  &  \\ \hline
$\mtpole$           &  173.1 GeV    & 1.3 GeV&  \cite{topmass:mar09} \\
$m_b (m_b)^{\overline{MS}}$ &4.20 GeV  & 0.07 GeV &  \cite{pdg07} \\
$\alphas$       &   0.1176   & 0.002 &  \cite{pdg07}\\
$1/\alphaemmz$  & 127.955 & 0.03 &  \cite{Hagiwara:2006jt} \\ \hline
\end{tabular}
\caption{Experimental mean $\mu$ and standard deviation $\sigma$
 adopted for the likelihood function for SM (nuisance) parameters,
 assumed to be described by a Gaussian distribution.
\label{tab:nuisance}}
\end{table}
%%%%%%%%%%%%%
%
To constrain the MUED parameters, we use data from electroweak precision observables (EWPO) 
which can be interpreted as constraints on the parameters given 
by the set \cite{epsilons}
\be
\epsilon = \{ \epsilon_1, \epsilon_2, \epsilon_3\}.
\label{eq:thetavalues}
\ee
The maximum likelihood (ML) value of $\theta$
obtained from LEP1 experiment data is \cite{lep_ewpos}  
$\epsilon_{\text{ML}} = \{5 \times 10^{-3}, -8 \times 10^{-3}, 4.8 \times 10^{-3}\}$. 
The likelihood function from EWPO is then modeled as a multi-dimensional Gaussian centered
at  the observed ML values,
\be \label{eq:likeEWPO}
-2 \ln \like_\text{EWPO} = \chi^2_\text{EWPO} = (\epsilon - \epsilon_{\text{ML}})^t C^{-1} (\epsilon_{\text{ML}}),
\ee
where the covariance matrix $C$ is given in Table~\ref{tab:covmat}.

%%%%%%%%%%%
\begin{table}[t]
\centering    

\begin{tabular}{| l | l l l |}
\hline
&  $\epsilon_1$ & $\epsilon_2$ &$\epsilon_3$ \\ \hline 
$\epsilon_1 $ & $5.78 \times 10^6$  & $-1.71 \times 10^6$ &  $-4.65 \times 10^6$ \\
$\epsilon_2$ &      &    $1.39 \times 10^6$ &  $8.93 \times 10^5$ \\ 
$\epsilon_3$ &      &        &  $5.01 \times 10^6$  \\ \hline
\end{tabular}
\caption{EWPO covariance matrix employed in the analysis~\cite{Barbieri:2004qk}.
\label{tab:covmat}}
\end{table}
%%%%%%%%%%%%%

We also include constraints from the WMAP 5-years measurement of the cosmological DM relic density \cite{Komatsu:2008hk}, which give for $\Oh$ a mean value $\muW= 0.1099$ and a standard deviation $\siW =  0.0062$ (notice that using the updated WMAP 7-years values would not change our result considerably).
When assuming that the LKP makes up the whole of the DM, we impose a Gaussian likelihood with the above mean and standard deviation, to which we add a 10\% theoretical error in quadrature. We shall be interested in relaxing the requirement that all of the DM is made of LKPs, and in this case we use the WMAP measurement only as an upper bound. We show in the Appendix that in this case the correct effective likelihood is given by the expression 
\be \label{eq:upperbound}
\like_\text{WMAP}(\OhKK) = \like_0 \int_{\OhKK/\siW}^\infty 
%\exp\left(-\frac{1}{2}(x-r_\star) \right) x^{-1}{\rm d}x 
e^{-\frac{1}{2}(x-r_\star)} x^{-1}{\rm d}x ,
\ee
where $ \like_0$ is an irrelevant normalization constant, $r_\star \equiv \muW/\siW$ and $\OhKK$ is the predicted relic density of the LKP as a function of the MUED and SM parameters being considered. Notice that this is slightly different from what is usually adopted in the literature, namely either a sharp upper bound say 2$\sigma$ above the WMAP mean, or a one-sided Gaussian which starts to drop at the WMAP mean and is flat below (see Fig.~\ref{fig:bound}).

The total log-likelihood is thus given by the sum of the log-likelihoods defined above, i.e.
\be
-2 \ln \like_\text{tot} = \chi^2_\text{tot} = \chi^2_\text{EWPO} + \chi^2_\text{SM} + \chi^2_\text{WMAP}.
\ee

The posterior distribution $P(\params|\data)$ is determined numerically by drawing samples from it. Markov Chain Monte Carlo techniques can be used to this aim, but in this paper we employ the MultiNest code, which implements the nested sampling algorithm (For a detailed description of the algorithm, see \cite{Feroz:2007kg,Feroz:2008xx,Trotta:2008bp}). To perform our statistical analysis, we use a modified version of the \superbayes{} code~\cite{rtr,Trotta:2008bp}\footnote{See www.superbayes.org} which includes the MultiNest algorithm. Compared to standard MCMC methods, MultiNest provides a higher efficiency, guarantees a better exploration of degeneracies and multimodal posteriors and computes the Bayesian evidence as well (which is difficult to extract from MCMC methods). In our MultiNest scans, we use 20,000 live points and a tolerance factor 0.5. We collect a total of about 220,000 samples from the posterior, which guarantees an adequate exploration of the parameter space.

%%%%%%%%%%%%%%%%%%%%%%%%%%%%%%%%%%%%%%%%%%%%%%%%%%%%%%%%%%%%%
\section{Results} \label{sec:results}
%%%%%%%%%%%%%%%%%%%%%%%%%%%%%%%%%%%%%%%%%%%%%%%%%%%%%%%%%%%%

%%%%%%%%%%%%%%%%%%%%%%%%%%%%%%%%%%%%%%%%%%%%%%%%%%%%%%%%%%%%%%%%%%%%%
\subsection{MUED parameter constraints} 
%%%%%%%%%%%%%%%%%%%%%%%%%%%%%%%%%%%%%%%%%%%%%%%%%%%%%%%%%%%%%%%%%%%%%

\begin{table}
\begin{center}
\begin{tabular}{| l |l |l |l |l|}
\hline
Parameter & Mean  & Best fit & 68\% range & 95\% range \\
\hline
\multicolumn{5}{|c|}{LKP: the sole constituent of DM}\\
\hline
$m_h$ (GeV) & 198.4 & 215  & [173 : 222.3] & [ 135.3 : 233.8] \\
$R^{-1}$ (GeV) & 640.9 & 641.6  & [574.1 : 707.5] & [536.5 : 843.5] \\
$\Lambda R$ & 55 & 38  & [23 : 86] & [12 : 98] \\\hline
$m_\gamma$ (GeV) & 641 & 642  & [574.7 : 707.8] & [537.3 : 843.4] \\
$\OhKK$ &  0.115 & 0.111  & [0.1 : 0.128] & [0.091 : 0.145] \\
$\log(\siSI \text{ (pb)})$ & -11.1 & -11.2  & [-11.4 : -10.8]  & [-11.7 : -10.5] \\
$\log(\siSD \text{ (pb)})$ & -5.7 & -5.7  & [-6 : -5.5] & [-6.3 : -5.2] \\
\hline
\multicolumn{5}{|c|}{LKP: subdominant constituent of DM}\\
\hline
$m_h$ (GeV) & 224  & 226.7 & [202.4 : 245.4] & [163.6 : 265.4] \\
$R^{-1}$ (GeV) & 602.9 & 607.4 & [528.9 : 677.2] & [477.1 : 795.5] \\
$\Lambda R$ & 55 & 66 & [25 : 86] & [12 : 98] \\ \hline
$m_\gamma$ (GeV) & 603.5 & 607.9 & [529.7 : 677.5] & [478 : 795.4] \\
$\OhKK$ & 0.08 & 0.08 & [0.057 : 0.108] & [0.035 : 0.127]\\
$\log(\siSI \xi \text{ (pb)})$ & -11.3 & -11.4 & [-11.6 : -11] & [-11.8 : -10.6]\\
$\log(\siSD \xi \text{ (pb)})$ & -5.8 & -5.9  & [-6.1 : -5.5] & [-6.3 : -5.2] \\
\hline
\end{tabular}
\end{center}
\caption{Posterior mean and best fit values for the input MUED parameters and some relevant observables, both for the case where the LKP is the sole constituent of DM (top section) and where it is allowed to be a subdominant component (bottom section). 
We also give the 68\% and 95\% Bayesian equal-tails 
credibility intervals. While these figures are for the flat prior choice, the log prior choice gives very similar results and is therefore not shown. } 
\label{tab:global_constraints}
\end{table}

\begin{figure*}
\includegraphics[width=.48 \linewidth]{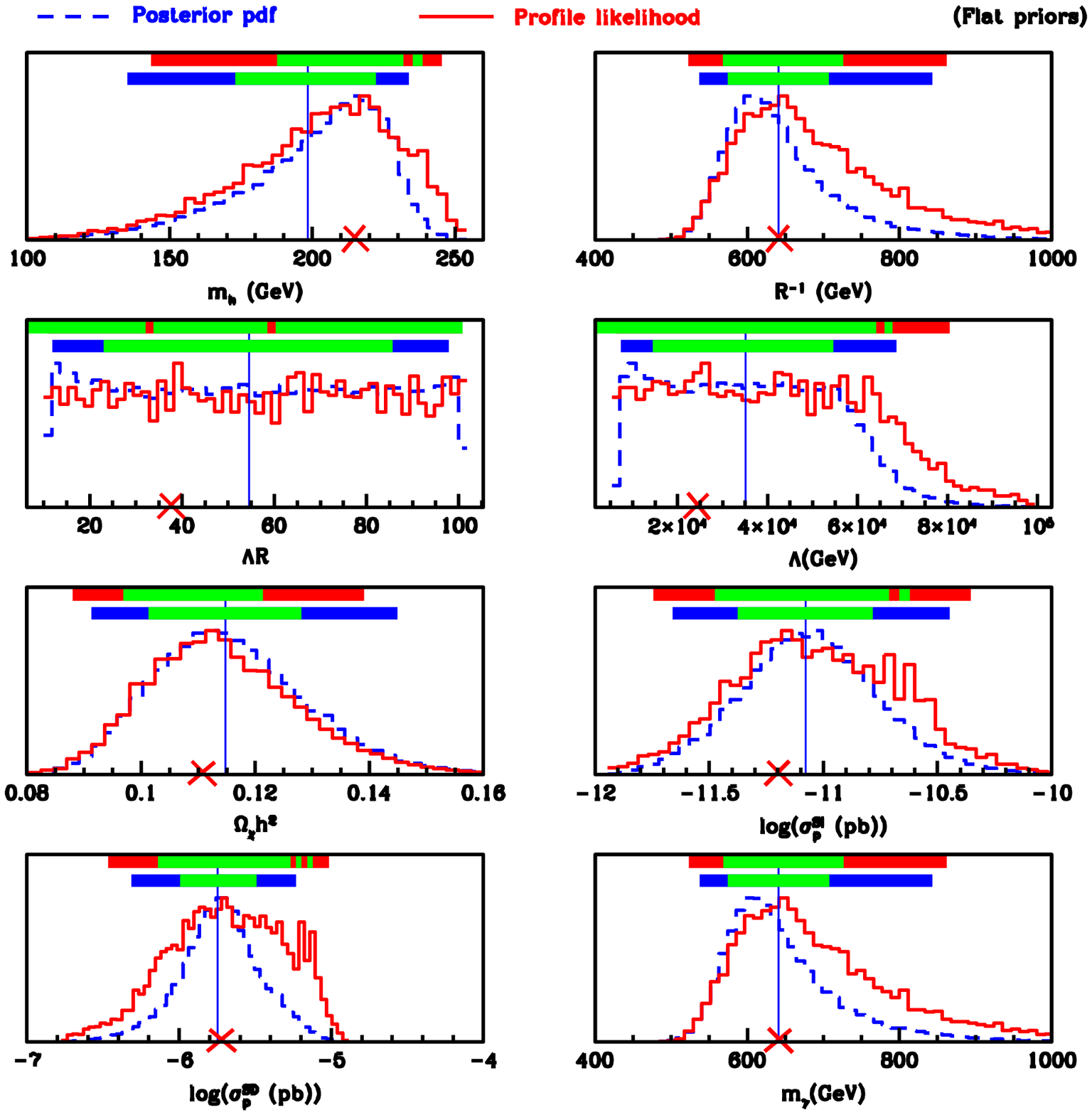} \hfill
\includegraphics[width=.48 \linewidth]{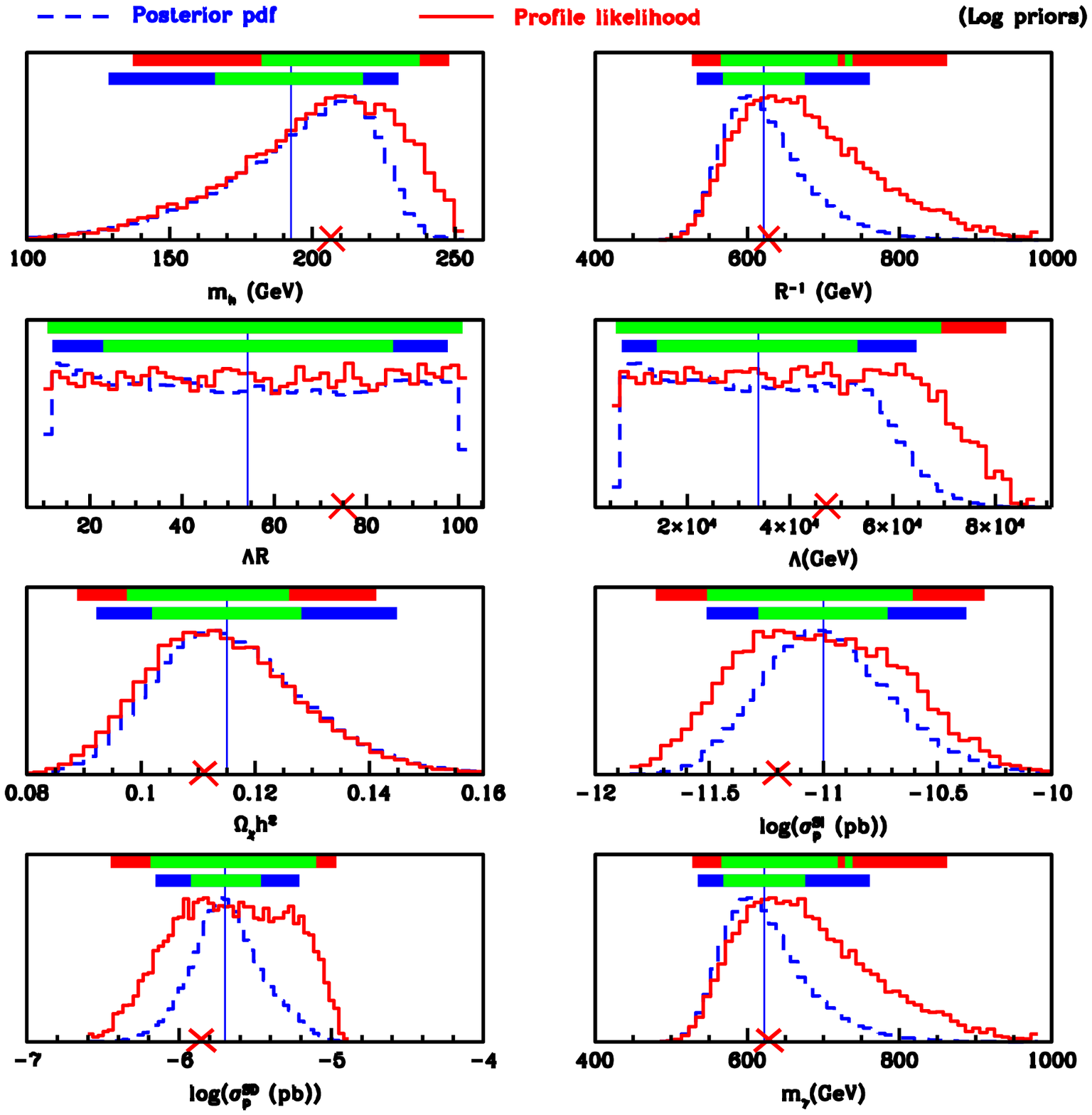}
\caption{Global constraints on the MUED parameters for two different choices of priors, assuming the LKP is the sole constituent of DM. The red cross gives the best fit, the vertical line the posterior mean. The horizontal blue/green bands give the 68\%, 95\% marginalized Bayesian posterior intervals; the red/green bands represent the 68\%, 95\% confidence intervals from the profile likelihood. There is only a very mild dependence in the constraints on the prior used or the choice of statistics. \label{fig:1D_global_constraints}}.
\end{figure*}

We begin by  showing in Fig.~\ref{fig:1D_global_constraints} the constraints on the MUED parameters and on some of the observable quantities. We plot the posterior distribution (blue), obtained by marginalizing over the posterior in the dimensions not show, and the profile likelihood (red), which maximizes the likelihood function over the variables not shown. The left panel is for the flat prior choice, while the right panel is for the log prior choice. It is clear that there is very little prior dependency and that the input parameters are well constrained by the data. Therefore, from now on we will only show results from the flat prior choice. The profile likelihood is also in good agreement with the posterior, which signals that one expects little prior dependency. Hence our results can be deemed to be robust with respect to changes in the choice of priors and statistical approach. 

\begin{figure*}
\includegraphics[width=.32 \linewidth]{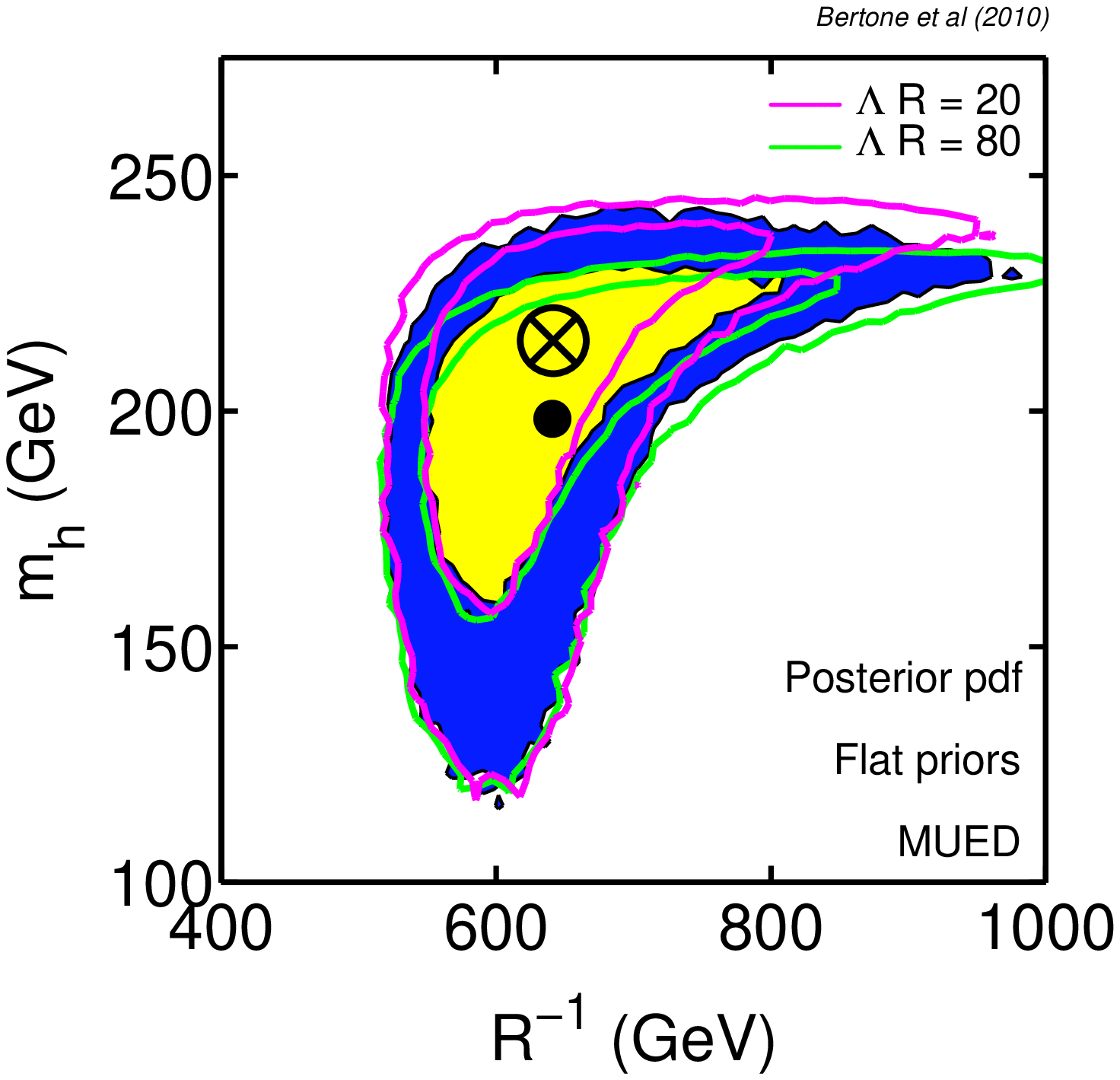} 
\includegraphics[width=.32 \linewidth]{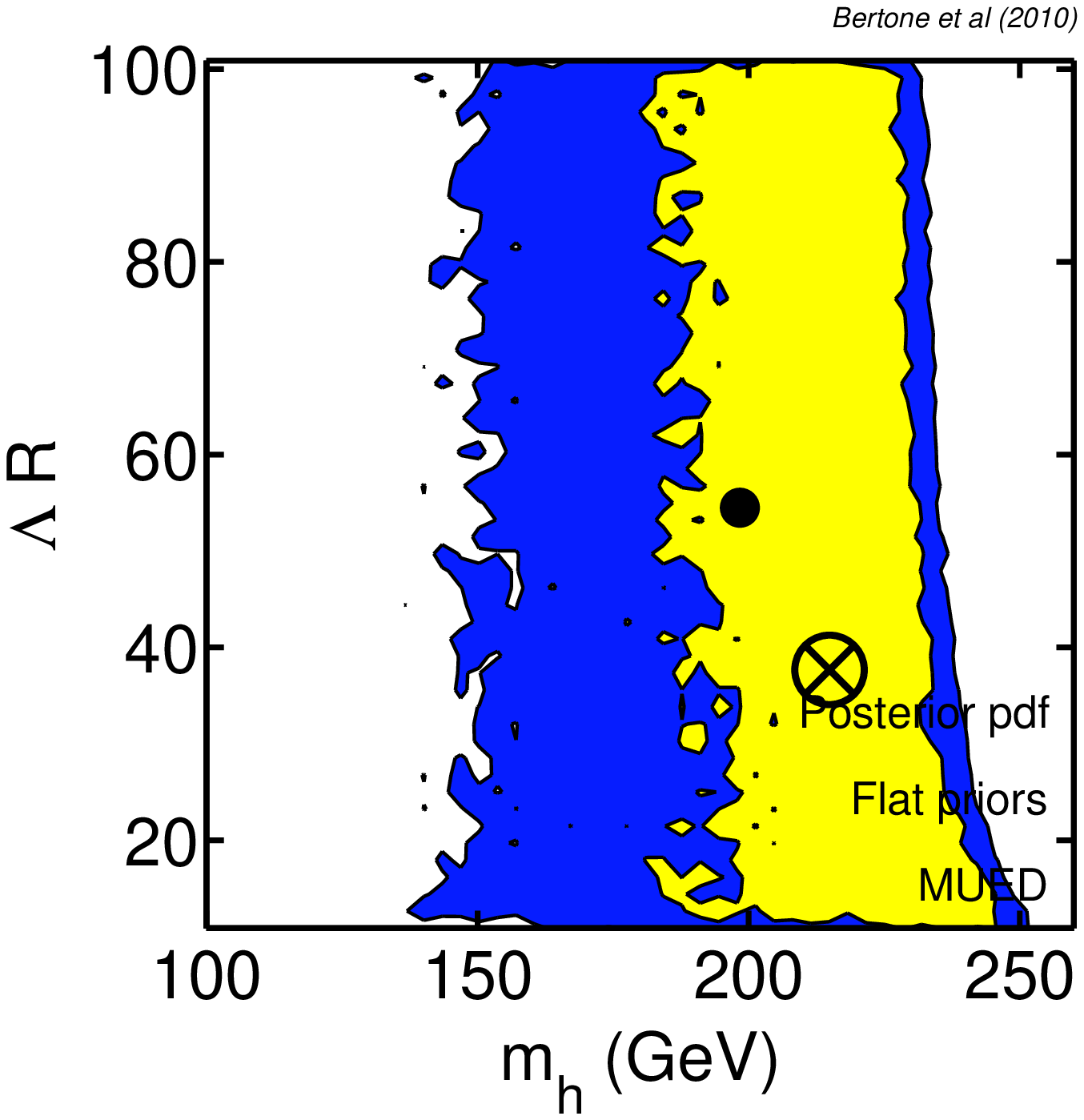} 
\includegraphics[width=.32 \linewidth]{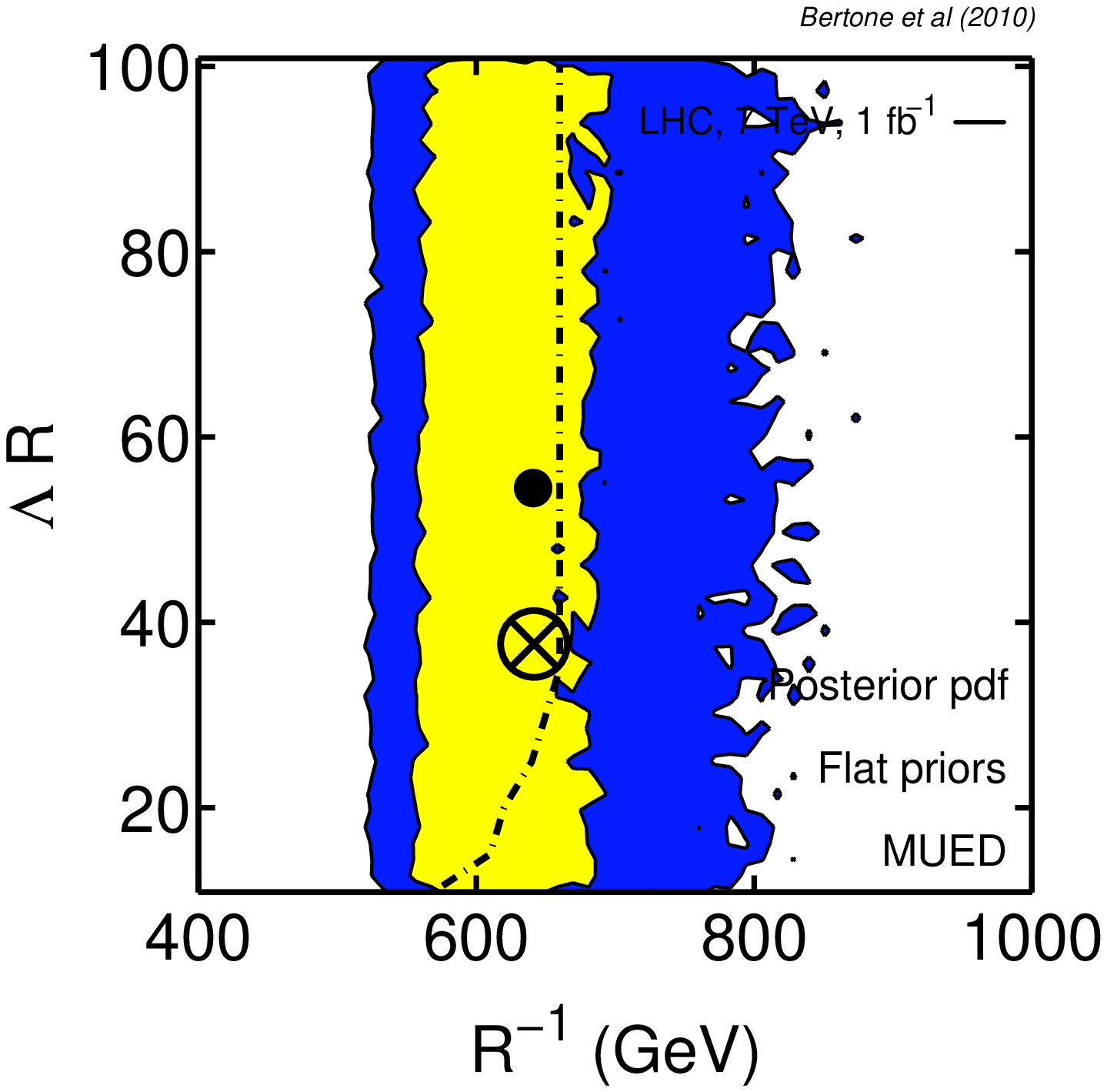}  
\includegraphics[width=.32 \linewidth]{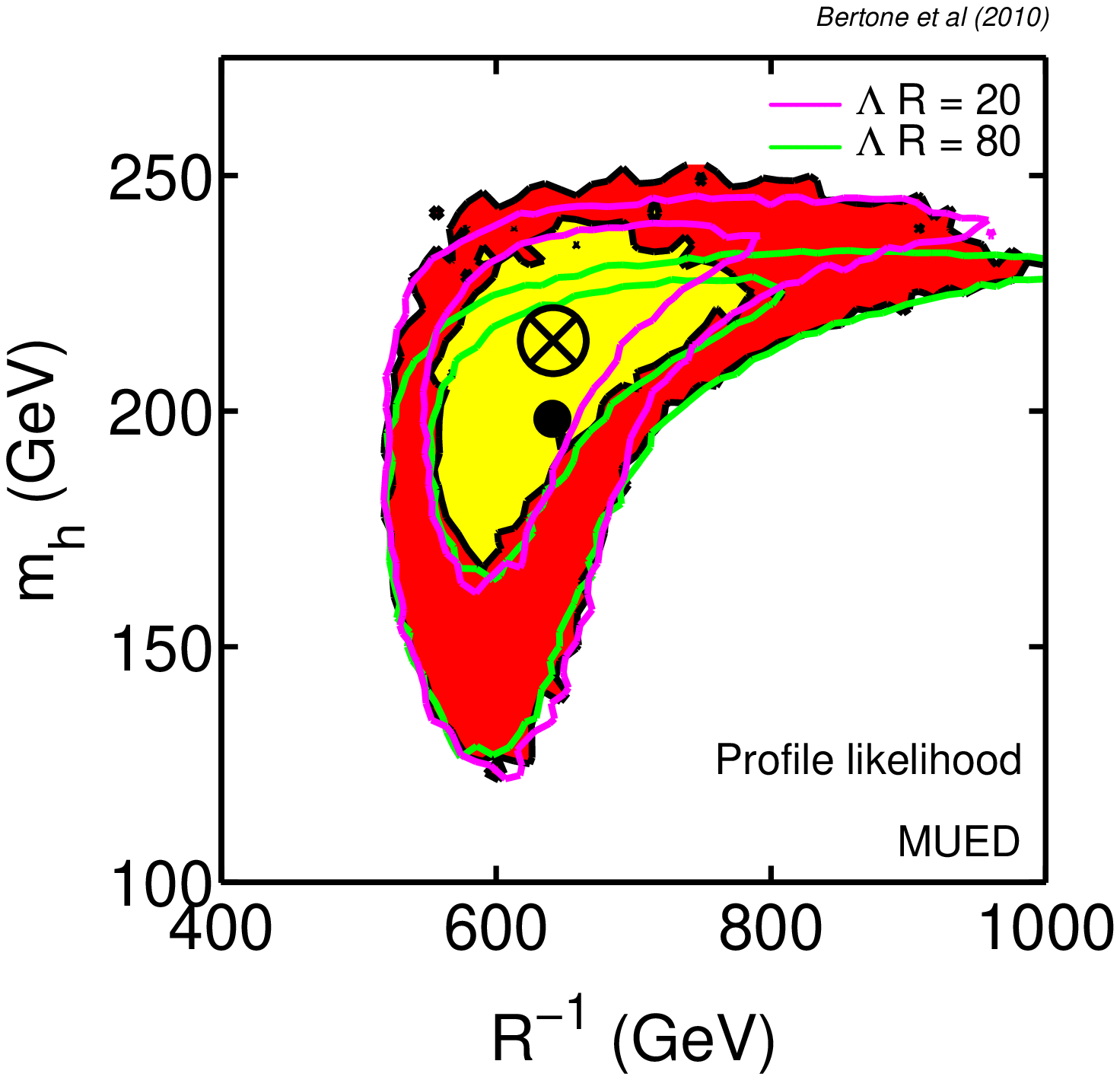} 
\includegraphics[width=.32 \linewidth]{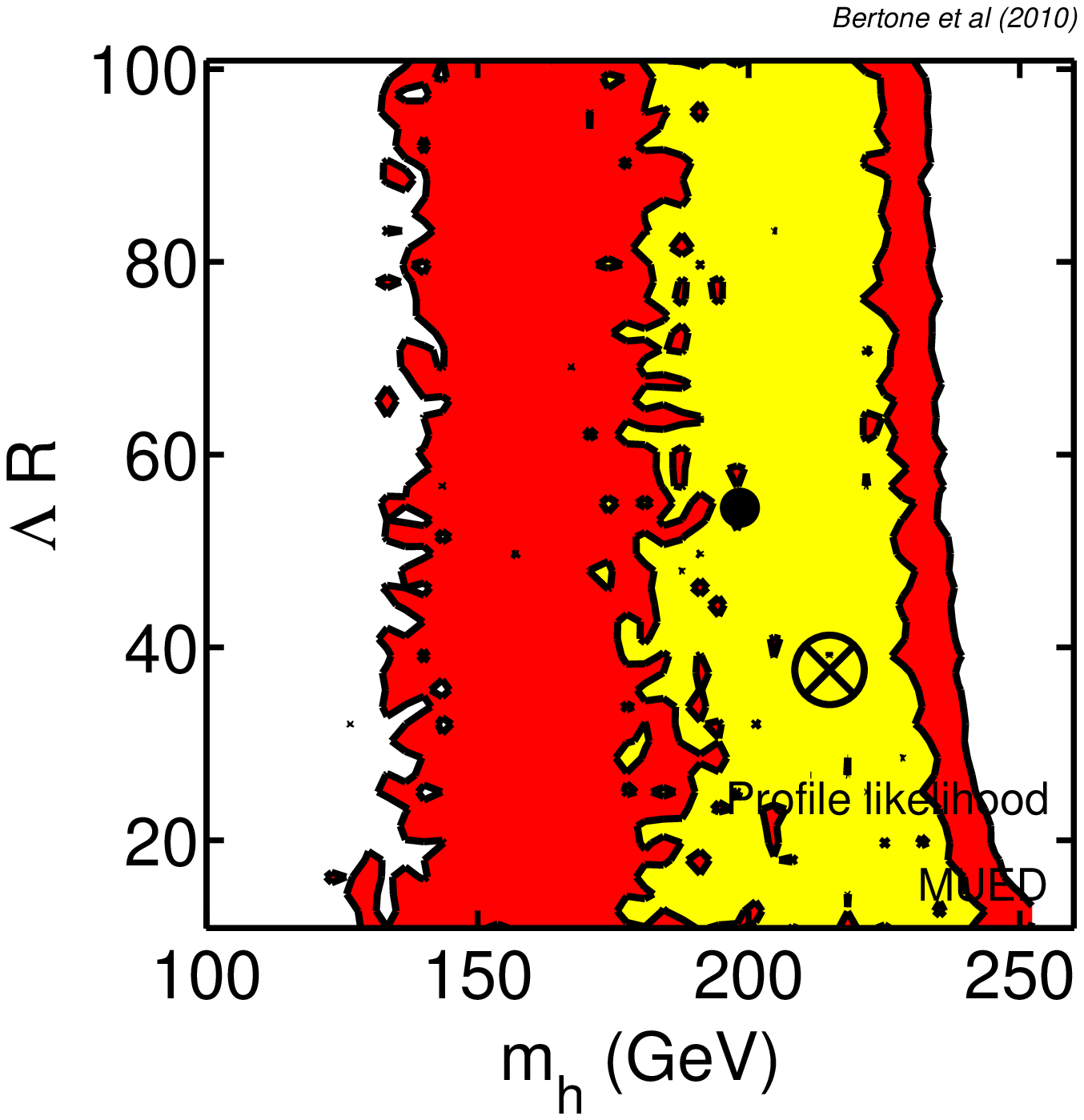} 
\includegraphics[width=.32 \linewidth]{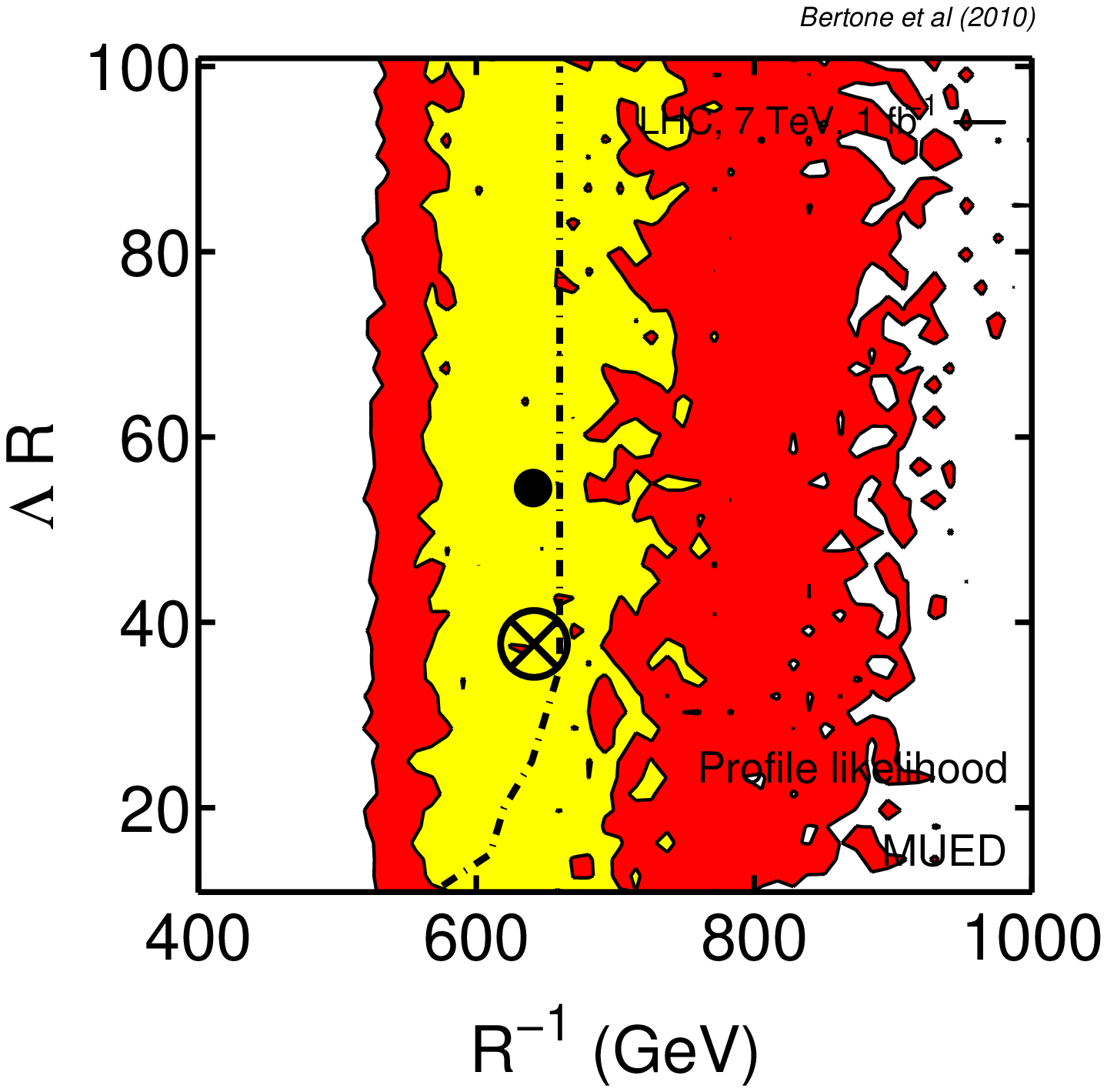}  
\caption{2D global constraints on the MUED parameters for the case where the LKP is the sole constituent of DM. The top row shows 68\% (yellow) and 95\% (blue) regions from the posterior pdf (assuming flat priors), while the bottom row gives confidence regions from the profile likelihood, with 68\% confidence level region in yellow and 95\% in red. We notice that the two statistics agree very well. The encircled cross gives the location of the best fit, the filled black dot of the posterior mean. In the $m_h$ vs $R^{-1}$ plane we plot in magenta/green the 68\% (inner contours) and 95\% (outer contour) regions for the case where $\Lambda R$ is fixed to 20 and 80, respectively. In the $\Lambda R$ vs $R^{-1}$ figure we show the LHC reach with $7$ TeV and 1 fb$^{-1}$ of integrated luminosity in the trilepton channel. 
 \label{fig:2D_global_constraints} }
\end{figure*}

The posterior distribution for $R^{-1}$ peaks near the the best fit value $R^{-1}=641.6$ GeV. The current Tevatron limit on $R^{-1}$, as we have seen, is 280 GeV but by the end of 2011, Tevatron is expected to have 100 times more data, pushing up the limit closer to our best fit. By that time, the LHC should have collected 1 fb$^{-1}$ of data, and 
it should therefore be able to discover MUED or at least rule out the best fit (see the reach of the LHC in the right panels of Fig.~\ref{fig:2D_global_constraints}). As for the mass of the Higgs, the posterior peaks near the best fit value $m_h=215$ GeV, 
for which the Higgs dominantly decays into $W^+W^-$ and $ZZ$.
This mass range of the Higgs is challenging for the 7 TeV LHC with 1 fb$^{-1}$.

We do not find any constraints on the value of $\Lambda R$ (see also Fig.~\ref{fig:2D_global_constraints}).
This can be understood as follows.
In general, a change in $\Lambda R$ modifies the mass spectrum, but the dependence is only logarithmic and it affects masses of 
strongly interacting particles only at the order of $\sim 10\%$ or less.
For electroweak particles, the dependence is almost flat in variation of $\Lambda R$.
For instance, the KK lepton mass changes by 2\% or so from $\Lambda R=10$ to $\Lambda R=40$ and 
the KK photon is barely affected.
In the computation of the relic density, the dominant contribution arises from self annihilation of the KK photon 
and coannihilation with $SU(2)_W$-single KK leptons, while coannihilation with the KK quark is only subdominant 
(one reason is the coupling strength (hypercharge) and the other reason is the heaviness of the KK particles).
Therefore the effect of a variation in $\Lambda R$ is expected to be small in the relic density calculation.
Similarly, the DD and ID are expected to be rather insensitive to variation in $\Lambda R$.
Therefore, as far as the constraints considered here are concerned, only the values of $R^{-1}$ and $m_h$ are important.
However the collider phenomenology may be affected by the value of $\Lambda R$, and this will require a further, dedicated investigation. We notice that our results (and errorbars on the parameters) fully account for the lack of constraints on the value of $\Lambda R$.

\begin{figure*}
\includegraphics[width=.32 \linewidth]{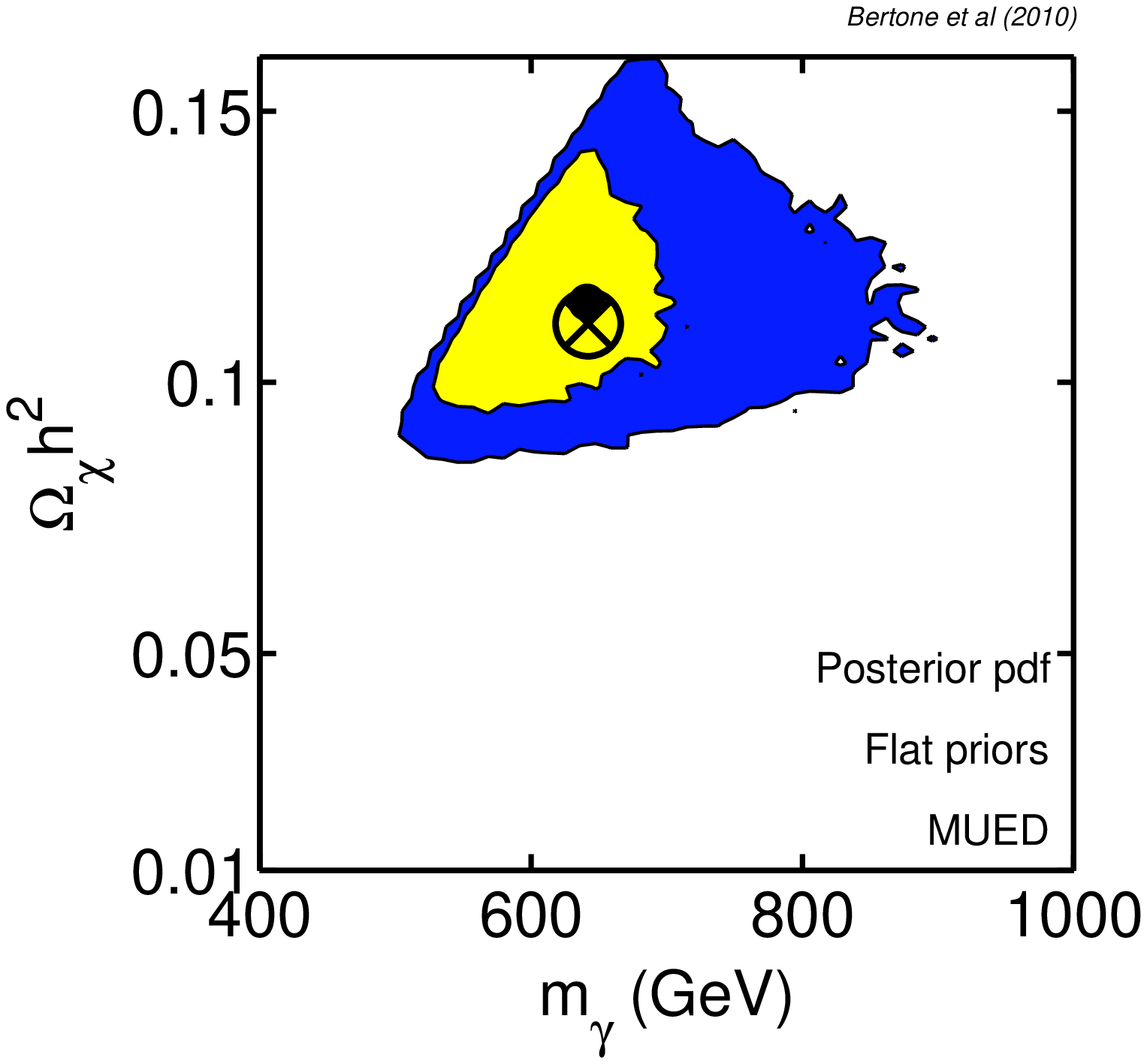} \hfill 
\includegraphics[width=.32 \linewidth]{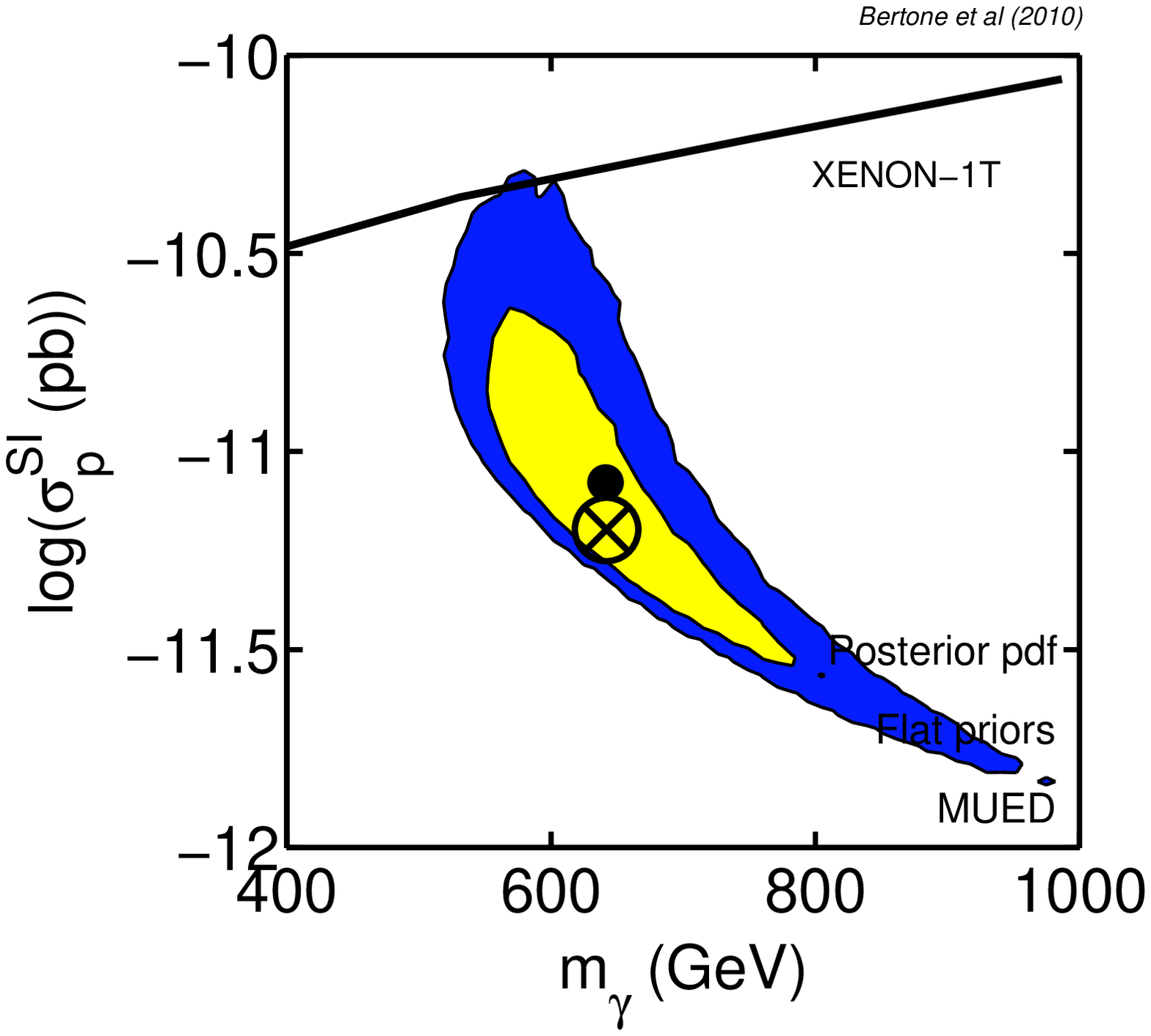} \hfill
\includegraphics[width=.32 \linewidth]{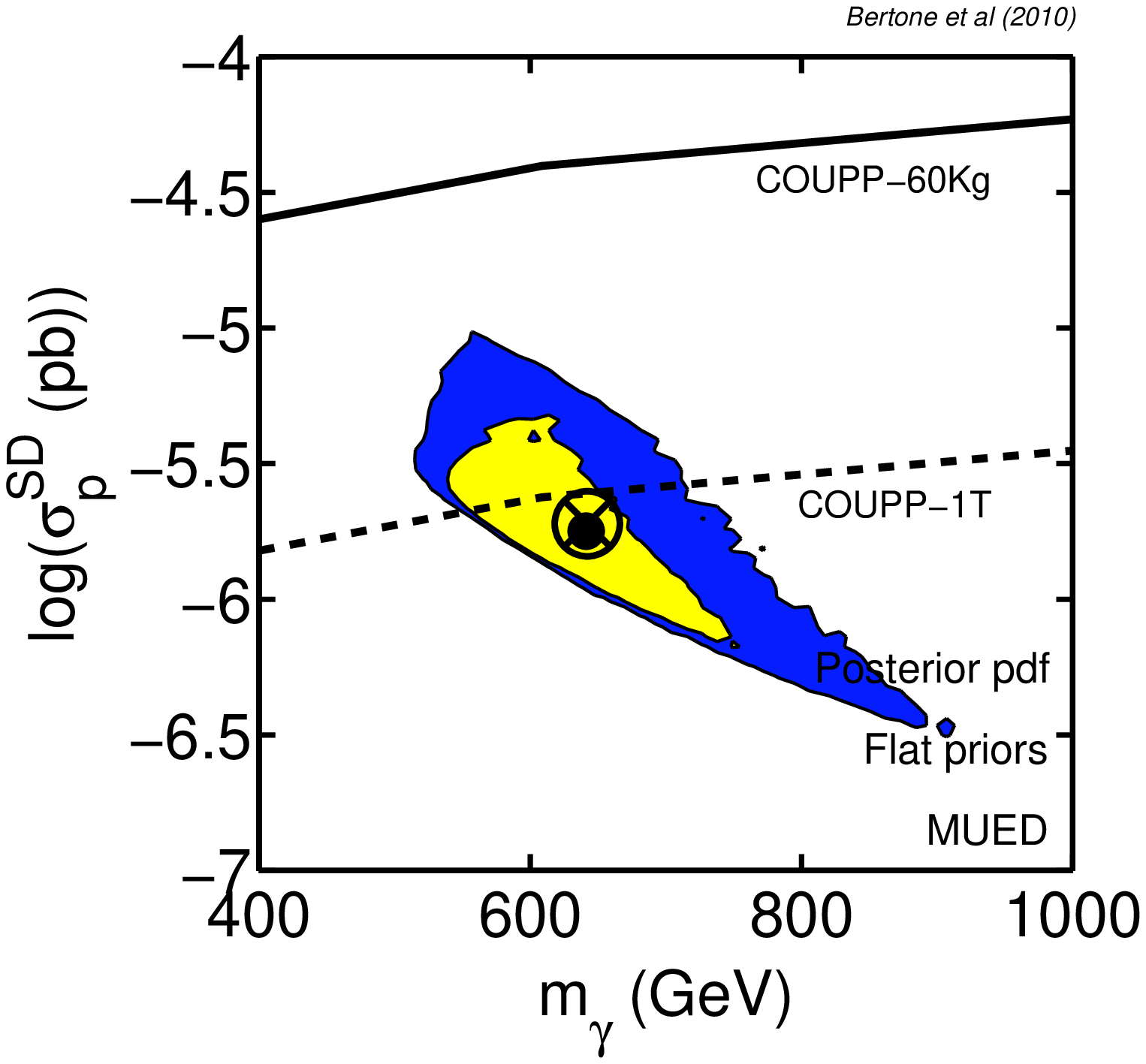}  
\includegraphics[width=.32 \linewidth]{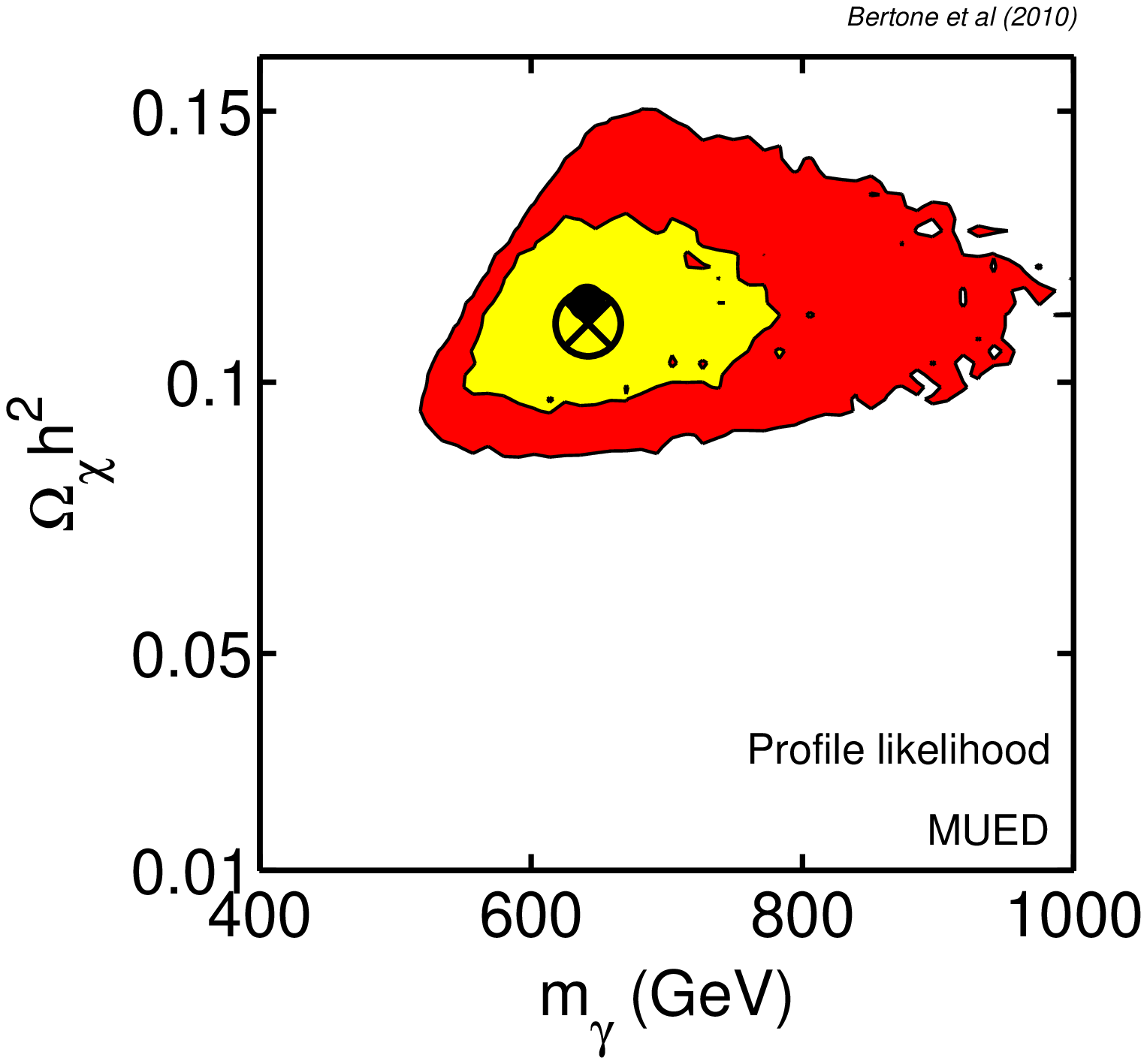} \hfill 
\includegraphics[width=.32 \linewidth]{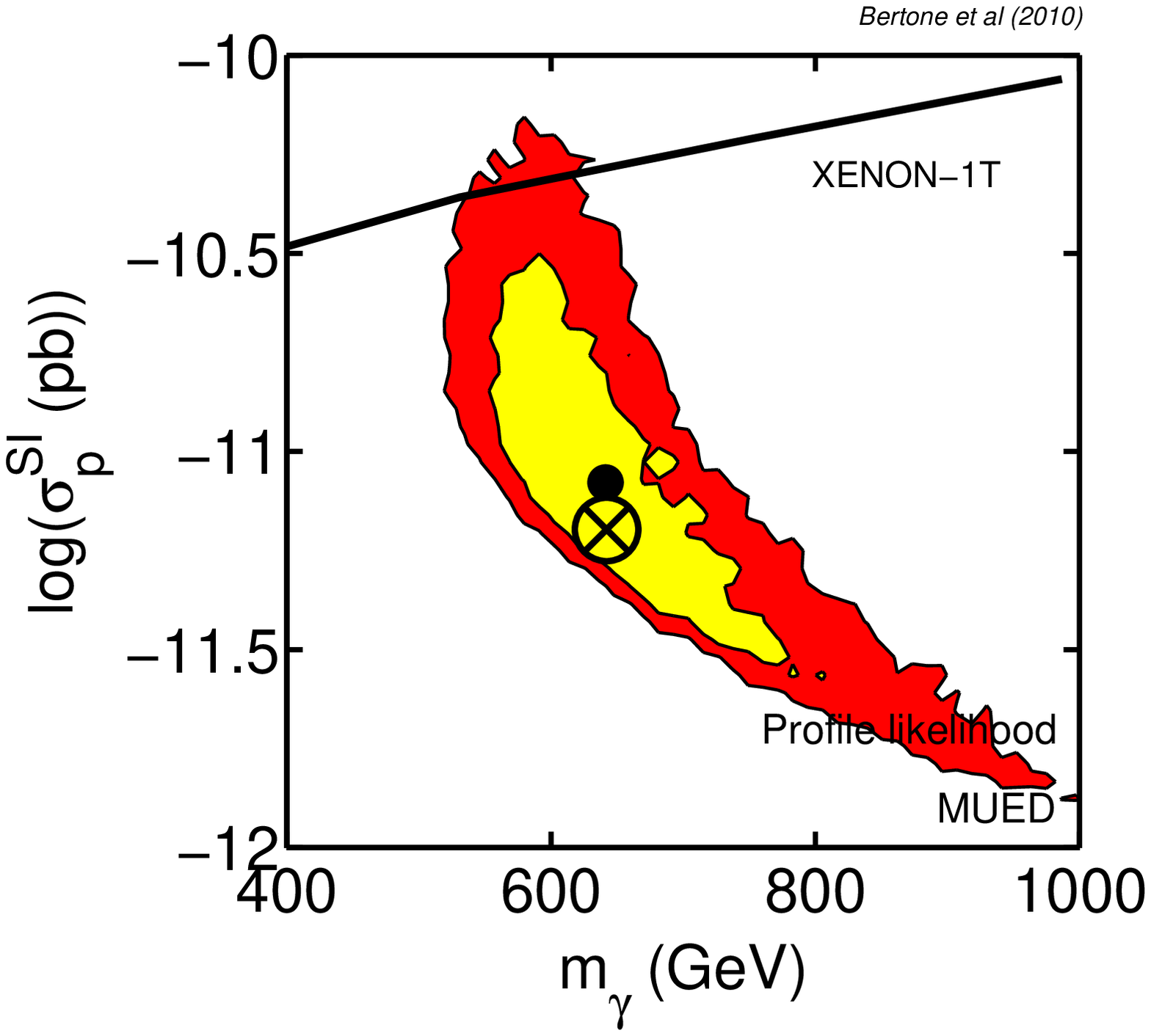} \hfill
\includegraphics[width=.32 \linewidth]{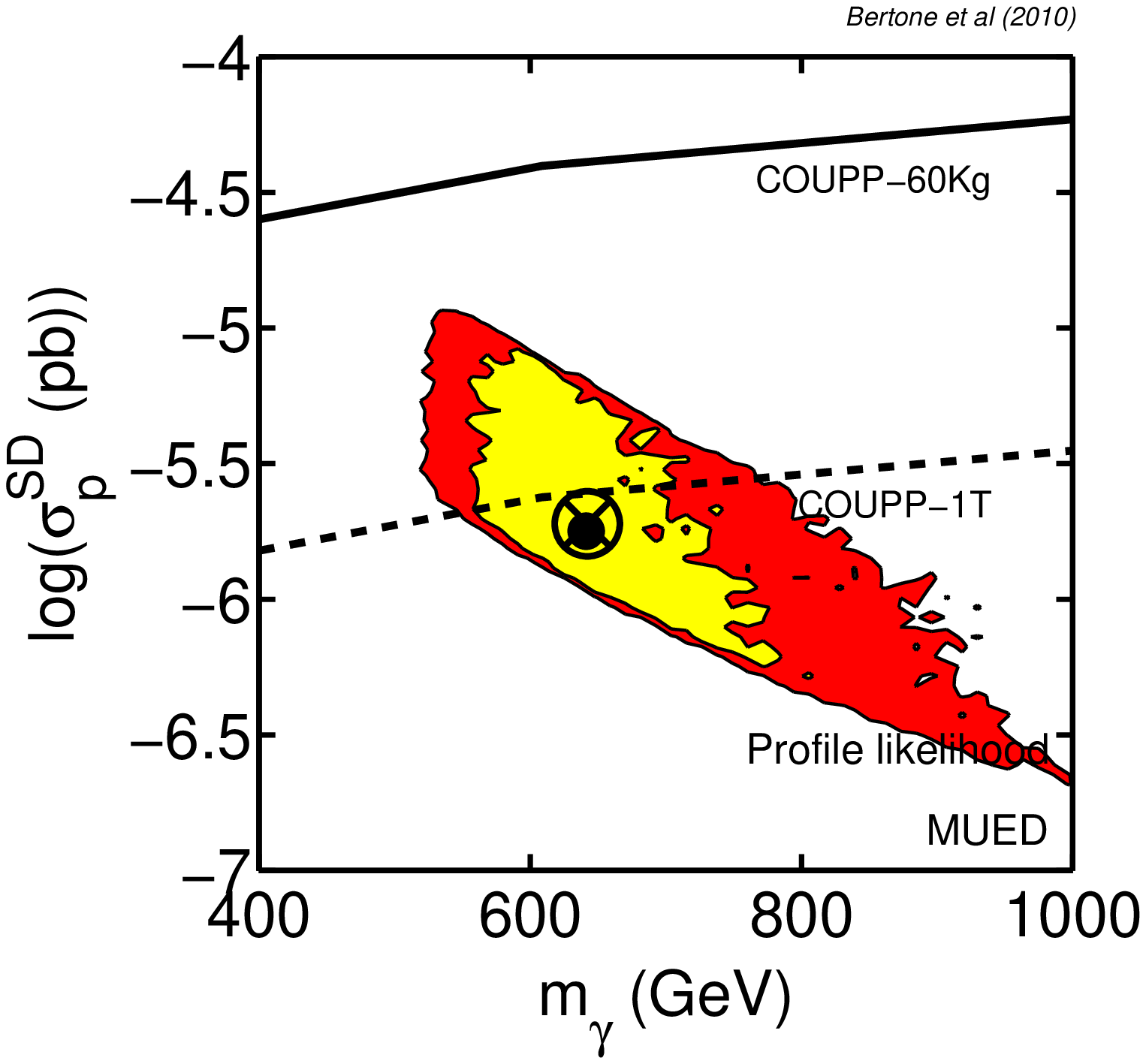}  
\caption{2D correlations among some relevant observables and MUED parameters, for the scenario where the \bone{} particle is the sole constituent of the DM, with colour coding as in Fig.~\ref{fig:2D_global_constraints}. We show the Bayesian posterior (top row) and the profile likelihood (bottom row). In the central and right-hand panels we display the reach of future direct detection experiments.  \label{fig:2D_derived_constraints} }
\end{figure*}

Fig.~\ref{fig:2D_global_constraints} shows 2D correlation plots for the MUED parameters, both for the Bayesian posterior (top row, for the flat prior choice) and the profile likelihood (bottom row). Contour delimit regions of 68\% and 95\% probability. We see also in this figure the relatively good constraints on $R^{-1}$ and $m_h$, and the lack of constraints on  $\Lambda R$. We stress once more the reassuring agreement between the posterior and the profile likelihood, which implies little dependence of the priors. 
In Fig.~\ref{fig:2D_derived_constraints} we plot the ensuing favoured regions for some of the observables.

\begin{figure}[t]
\includegraphics[width=0.45 \textwidth]{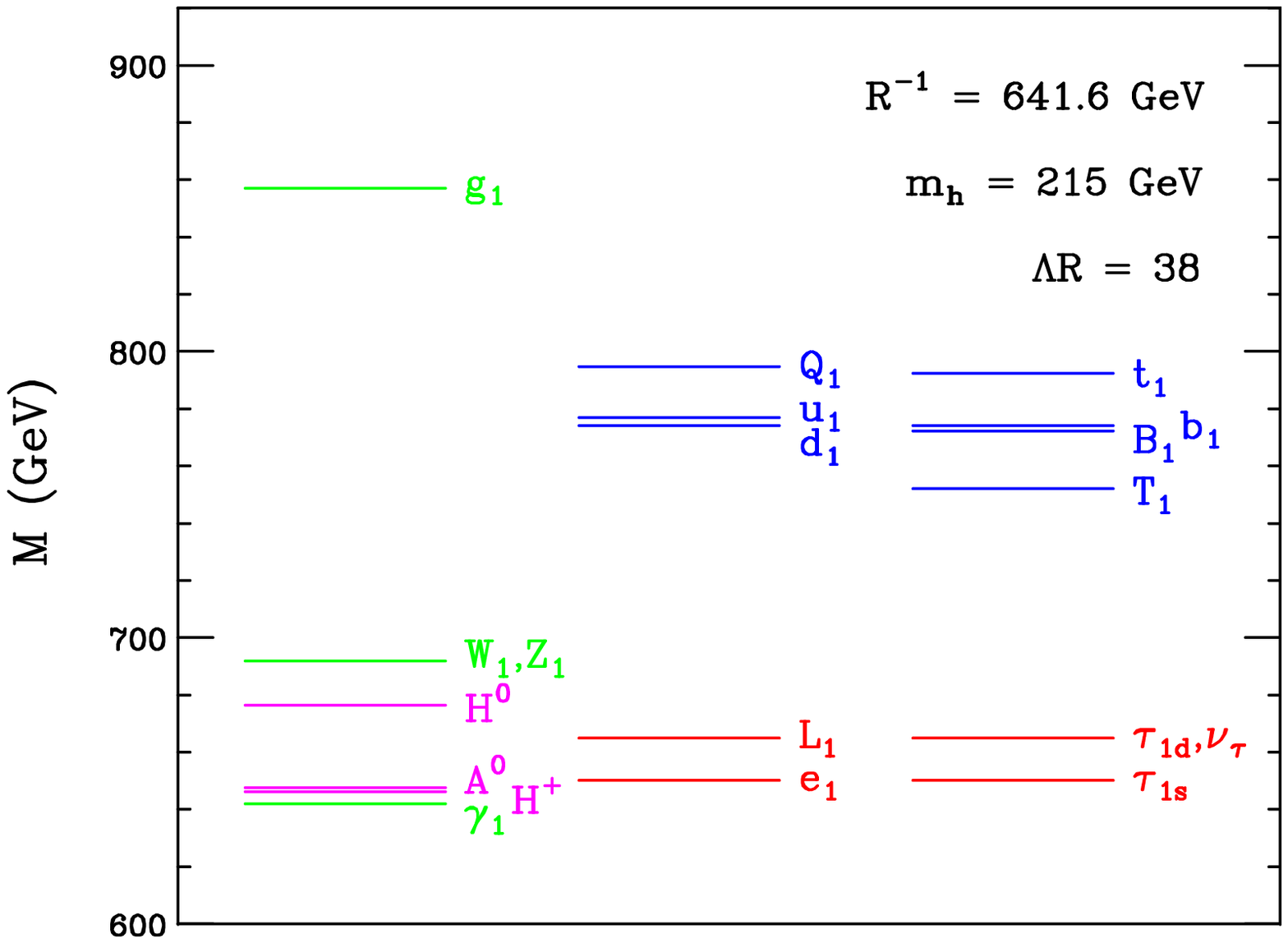}
\caption{Best fit mass spectrum of the first KK level from our global, assuming the LKP is the sole constituent of DM. \label{fig:spectrum} }
\end{figure}

In Fig. \ref{fig:spectrum} we show the mass spectrum of the first KK level for the best fit of the flat prior scan (the log prior case is very similar), under the assumption that the LKP makes up the whole of the DM.
The KK bosons (gauge (in green) and Higgs bosons (in magenta)) are shown in the left column, 
while the first two generations of quarks (in blue) and leptons (in red) are shown in the middle and 
the third generation in the right column. 
KK particles denoted by lower (upper) case are singlets (doublets) under $SU(2)_W$.
The mass spectrum and decay patterns from our best fit agree well with those shown in literature \cite{Cheng:2002iz,Cheng:2002ab,Cembranos:2006gt}
but here the corresponding scales are $R^{-1} = 642$ GeV and $m_h=215$ GeV. 

A natural question to ask is whether EWPO or relic density constraints influence more significantly the posterior. To understand this, we have performed two additional analyses, keeping EWPO constraints and discarding the relic density bound in the first case, and viceversa in the second. As expected, removing  the bound on EWPO opens up the parameter space at small values of $R^{-1}$, which results in a fairly flat pdf across the whole range of allowed Higgs mass values, and in pronounced volume effects on $R^{-1}$, for which the posterior distribution disagrees with the profile likelihood. Removing the constraint on the relic density {\it completely} (as opposed to setting an upper bound on this quantity, as discussed below), and keeping only EWPO constraints, allows larger values of $R^{-1}$, which has the effect of pushing down the predictions for direct detection cross sections.

%A natural question to ask is whether EWPO or relic density constraints influence more significantly the posterior. To understand this, we have perform two additional analyses, keeping EWPO constraints and discarding relic density bound in the first, and viceversa in the second. As expected, relaxing the bound on EWPO opens up the parameter space at small values of $R^{-1}$, which in this case peaks around $400$ GeV. The posterior for the SM Higgs mass is instead similar to the case where EWPO are also included, with the bulk of the posterior in the 200--250 GeV range. {\bf Gf: KC, can you comment on why the bound on the relic density disfavors low $m_h$ masses, and therefore selects this interval?} . Relaxing the constraint on the relic density {\it completely} (as opposed to setting an upper bound on this quantity, as discussed below), and keeping into account only EWPO, allows larger values of $R^{-1}$, which has the effect of pushing down the predictions for direct detection cross sections.

\begin{figure*}
\includegraphics[width=.48 \linewidth]{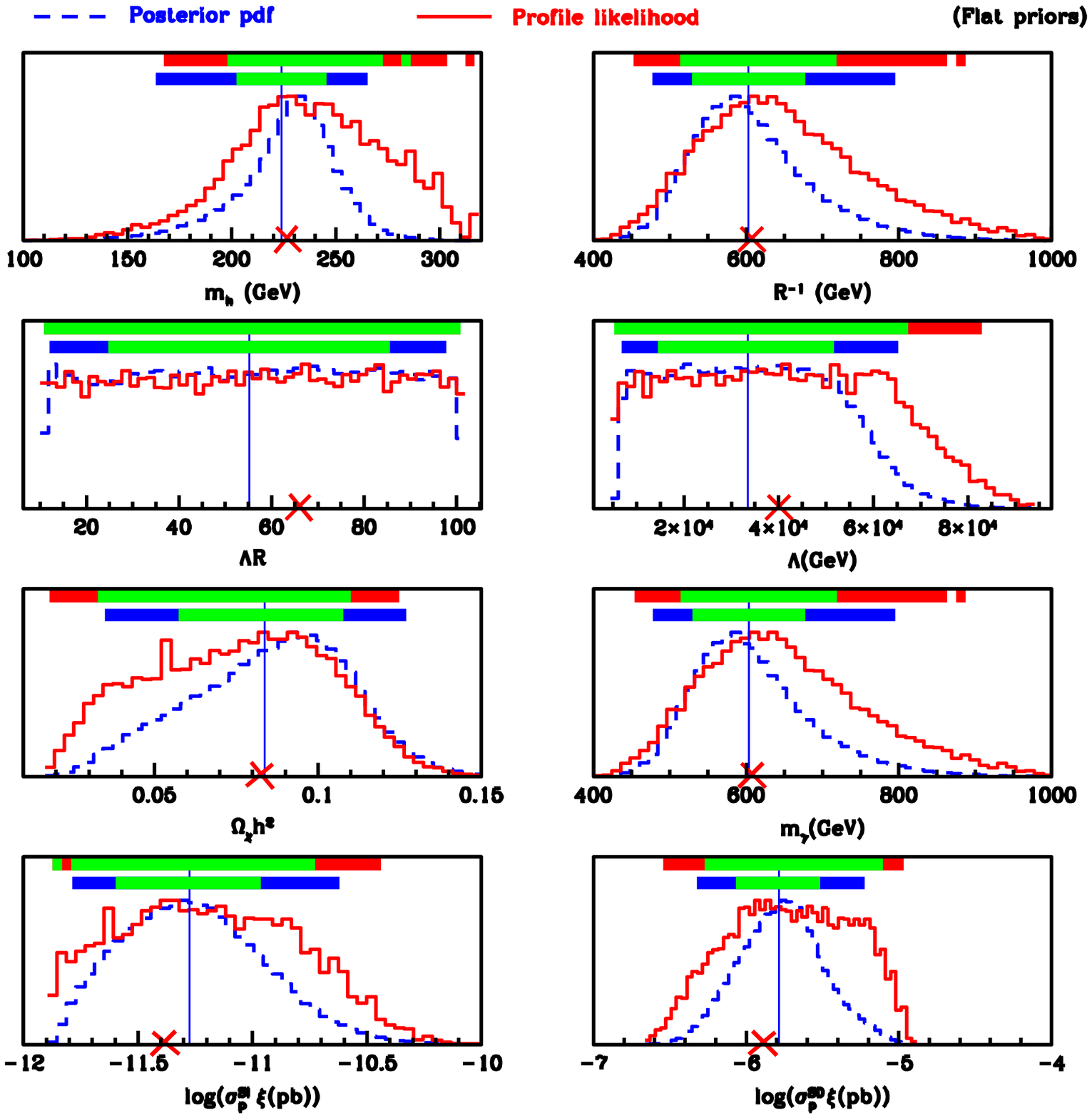} \hfill
\caption{Global constraints on the MUED parameters for the flat prior, dropping the assumption that the \bone{} particle is the sole constituent of DM. The results with the log prior are very similar and are therefore not shown. \label{fig:1D_constraints_relaxed}}
\end{figure*}

\begin{figure*}
\includegraphics[width=.32 \linewidth]{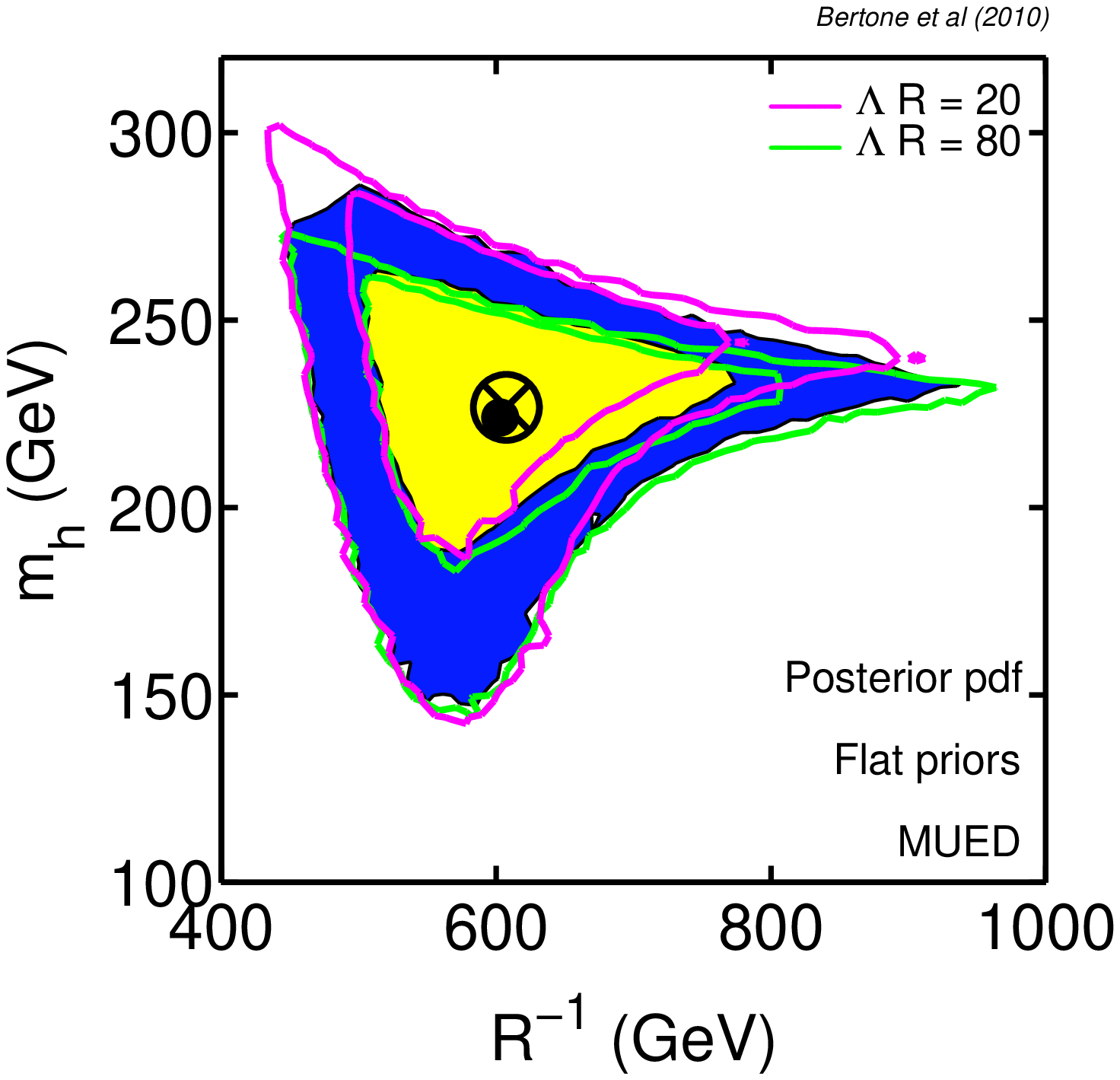} \hfill 
\includegraphics[width=.32 \linewidth]{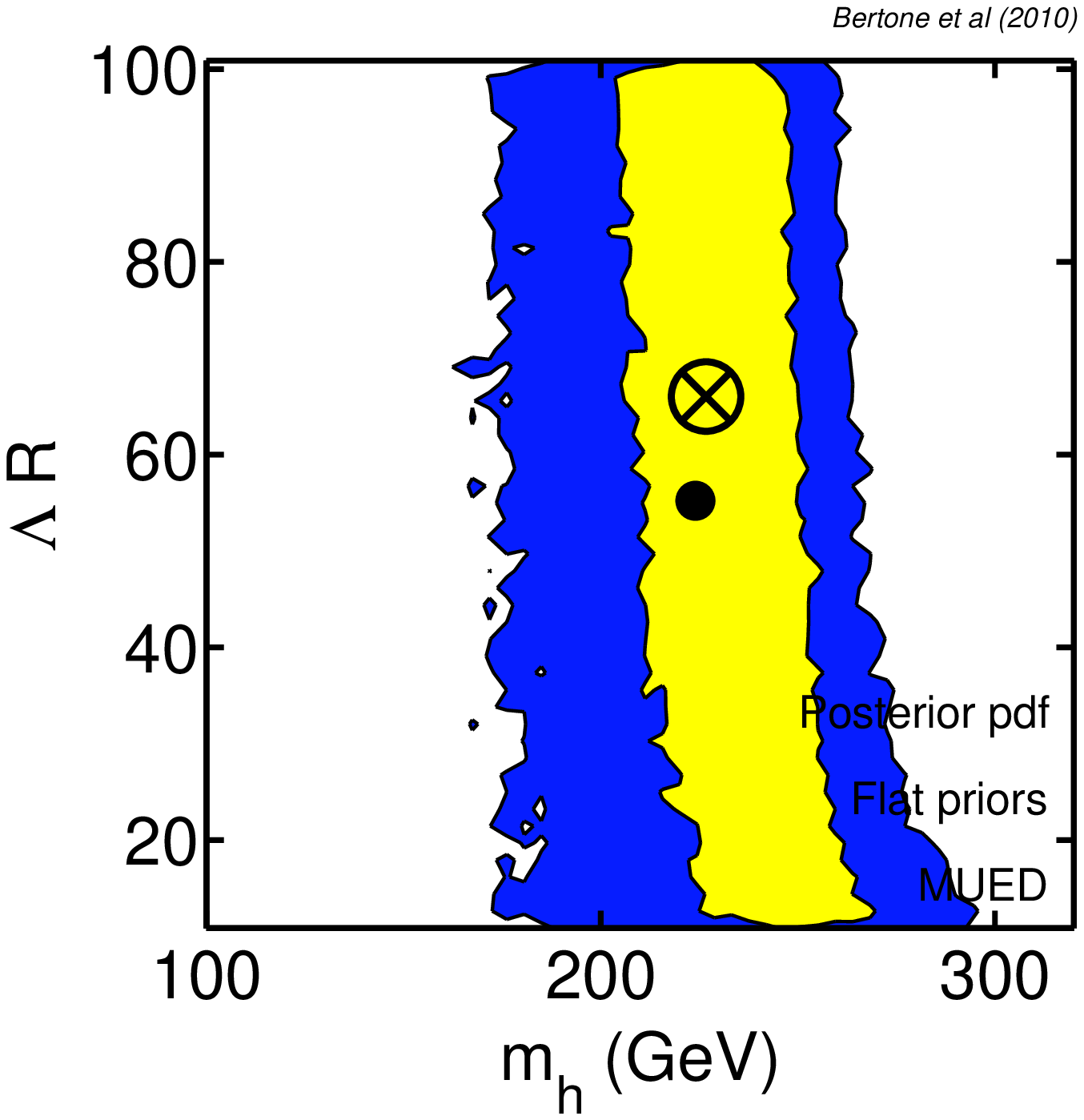} \hfill
\includegraphics[width=.32 \linewidth]{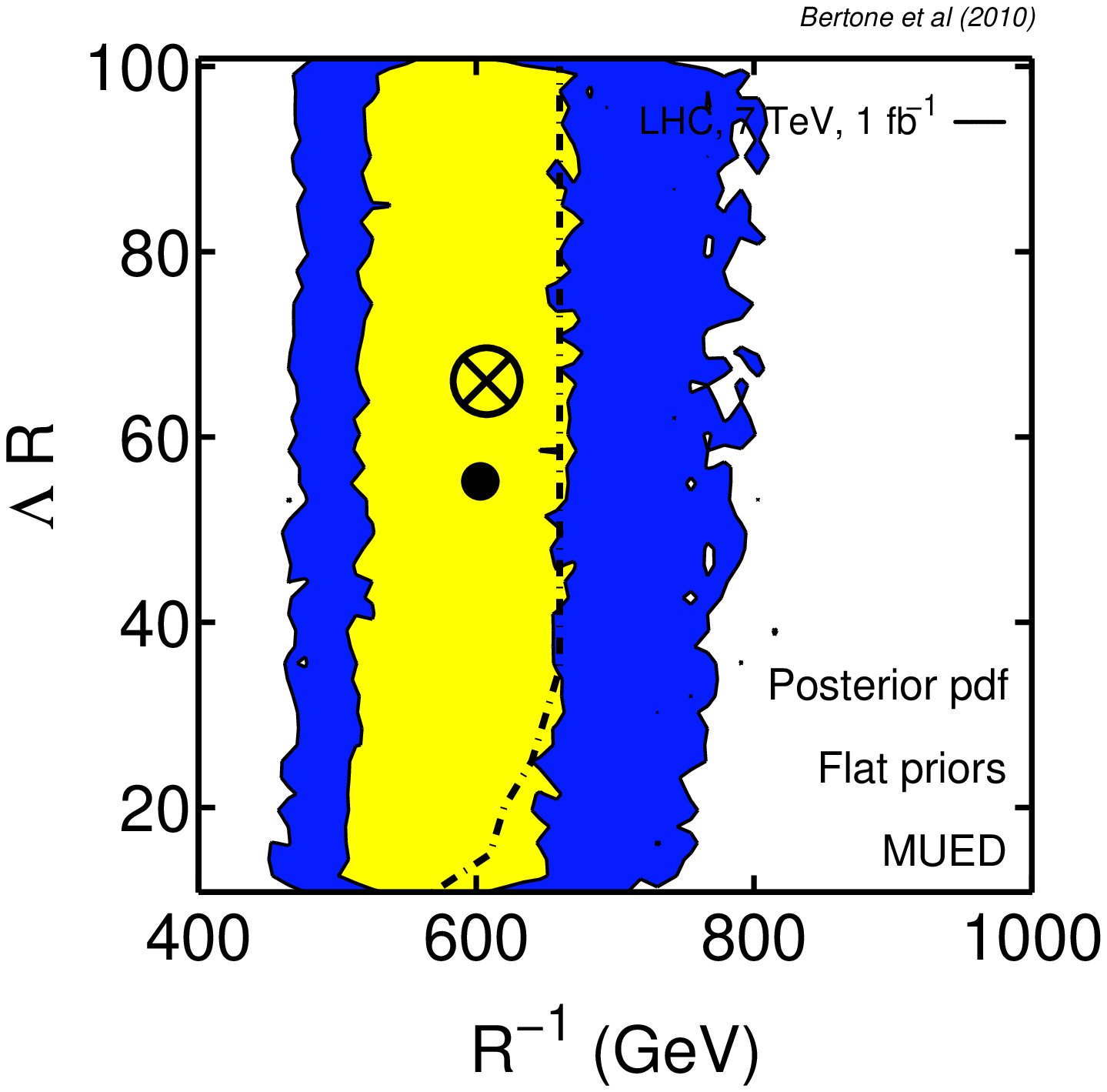}  \\ 
\includegraphics[width=.32 \linewidth]{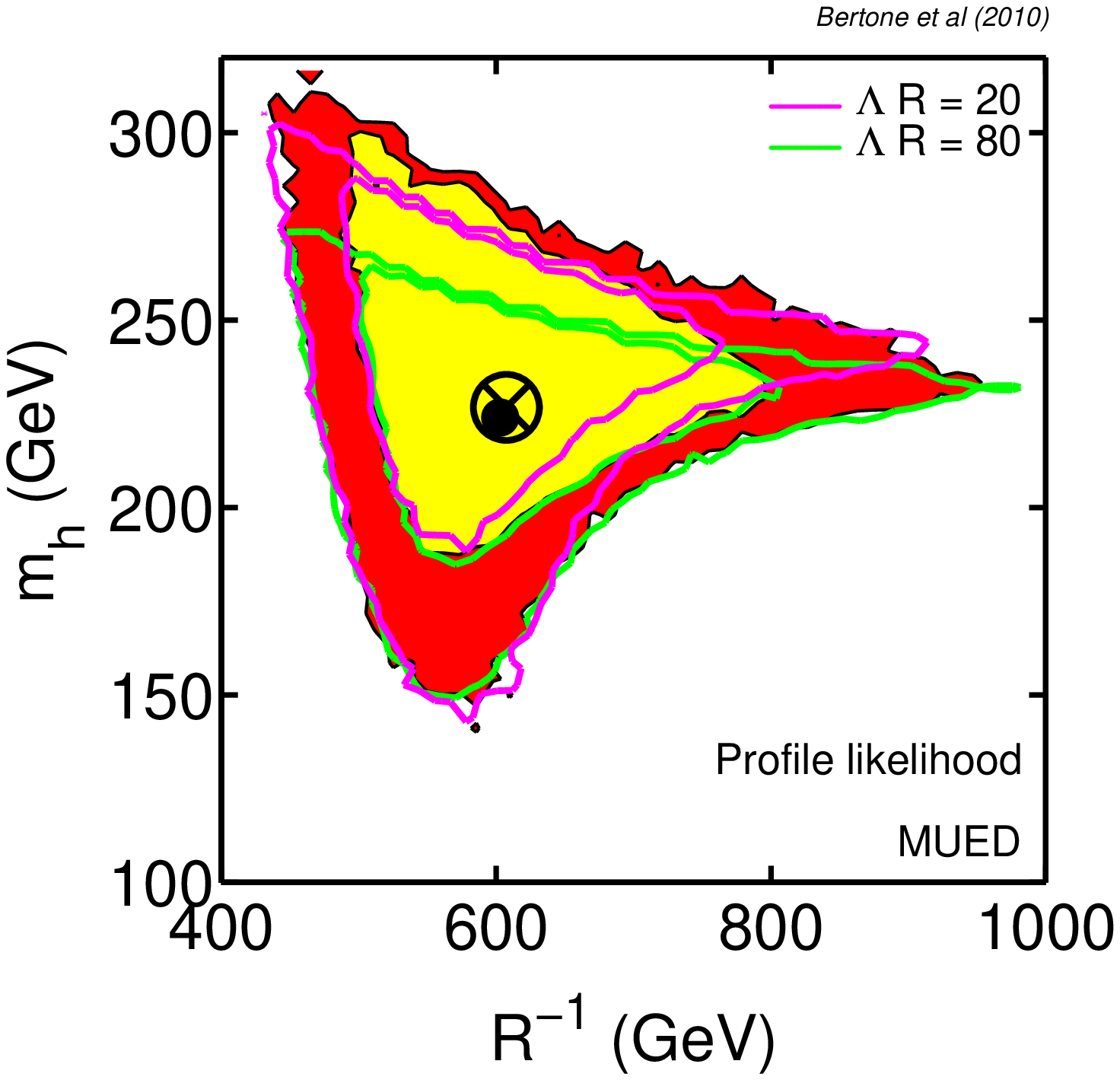} \hfill 
\includegraphics[width=.32 \linewidth]{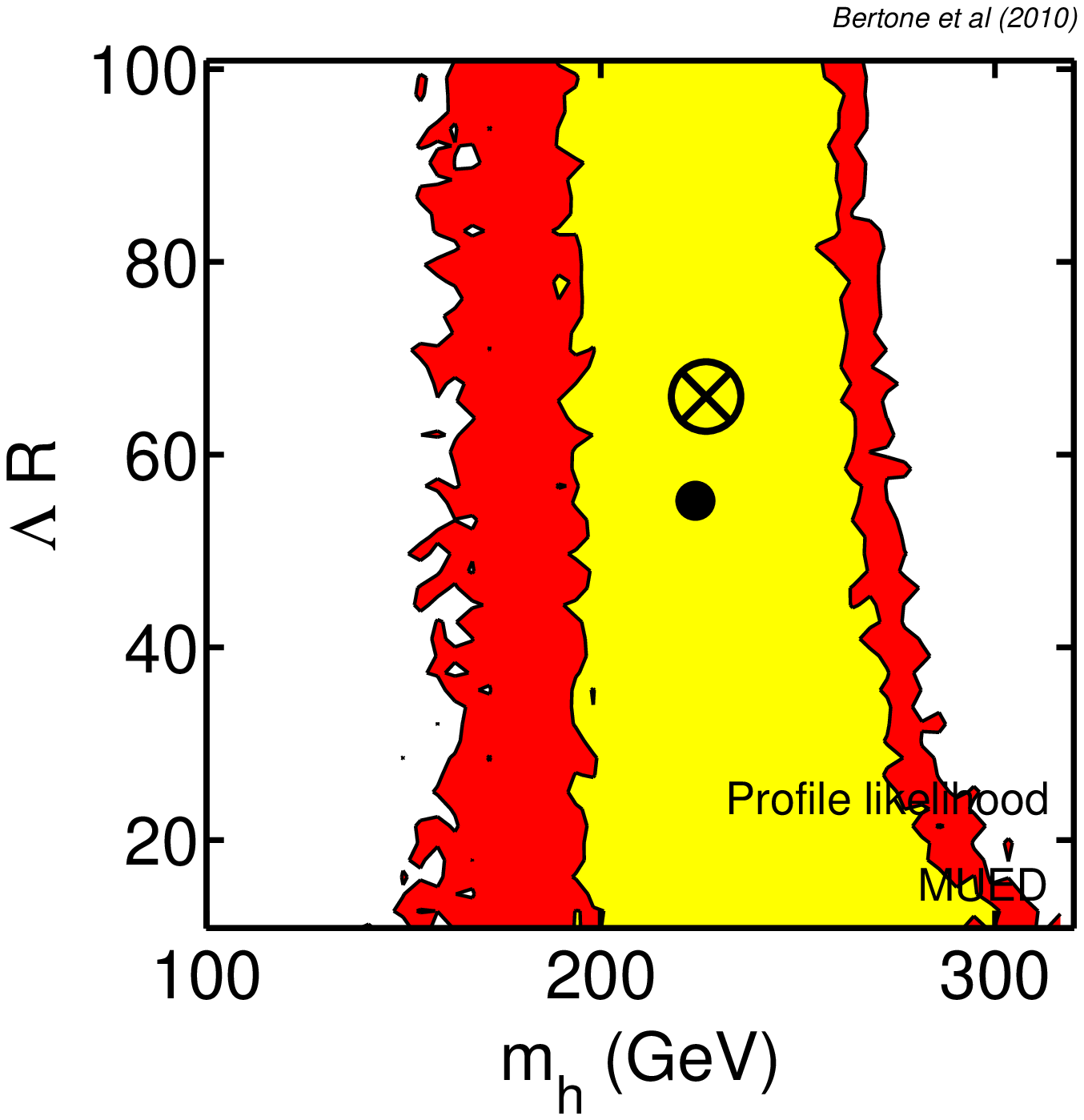} \hfill
\includegraphics[width=.32 \linewidth]{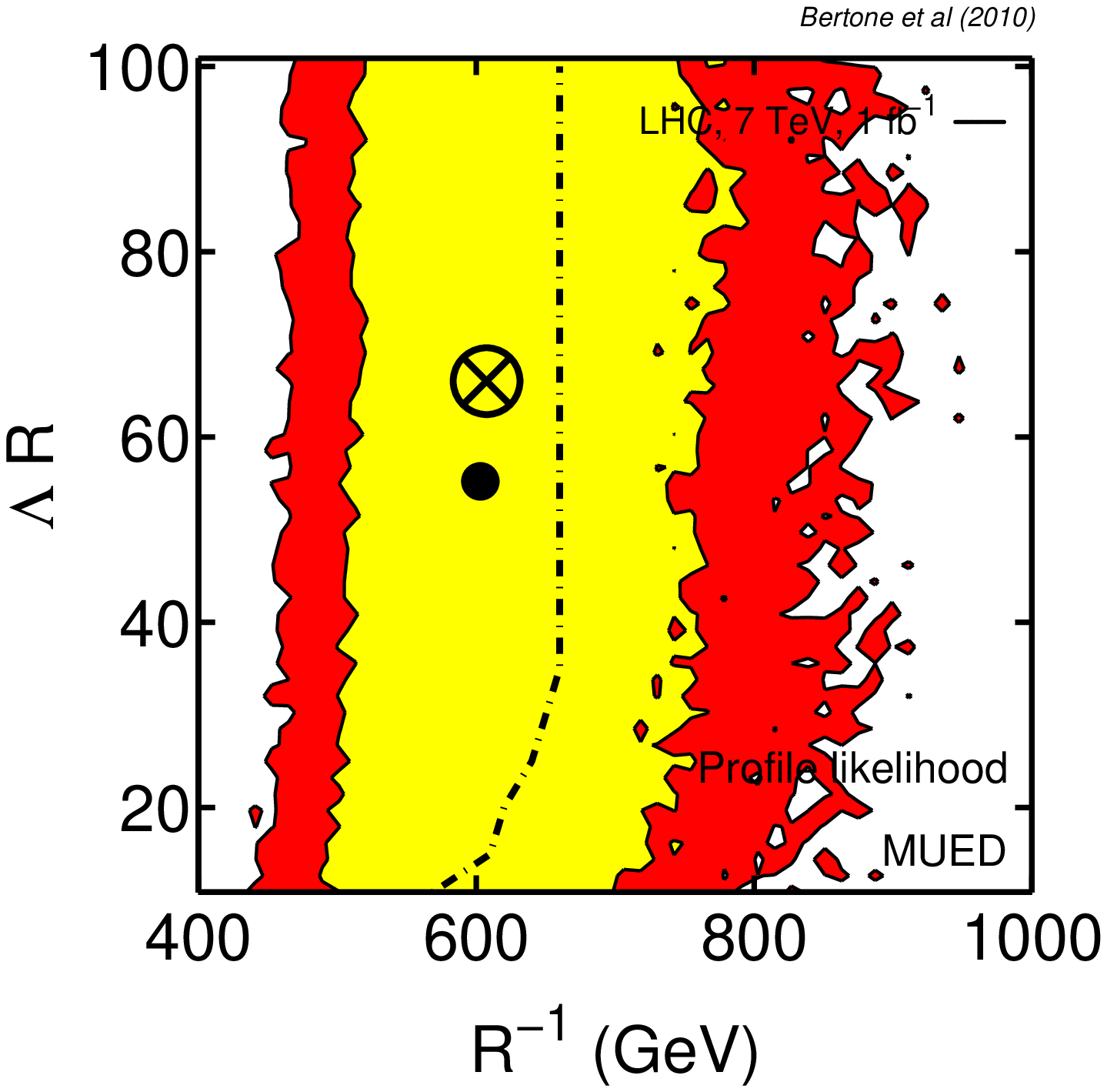} 
\caption{As in Fig.~\ref{fig:2D_global_constraints} but dropping the assumption that the \bone{} particle is the sole constituent of DM, and imposing only an upper bound instead.\label{fig:2D_constraints_relaxed} }
\end{figure*}

To relax the strong assumption that the \bone{} particle makes up the whole of the relic DM, we adopt the upper limit on the relic abundance, represented by the likelihood in Eq.~\eqref{eq:upperbound}. This modifies the posterior in the $\Omega_\gamma$ vs $m_\gamma$ plane in an obvious fashion, since we are allowing the case $\Omega_\gamma < \Omega_{DM}$. The resulting constraints and corresponding favoured regions for the observables are shown in Figs.~\ref{fig:1D_constraints_relaxed} and \ref{fig:2D_constraints_relaxed}. In both those figures the EWPO constraints have been applied. There are small quantitative differences with respect to the case where the \bone{} makes all of the DM, and the implications for the reach of the LHC and the mass of the Higgs are qualitatively unchanged. 

In terms of the best-fit $\chi^2$, we find that the relic density constraints contributes about 0.01 units to the total $\chi^2$ for the best fit point, which means that the WMAP value can be reproduced very well by the model (both when it is taken as a Gaussian constraint and as an upper bound). The EWPO constraints contribute a $\chi^2 \sim 1.6$ at the best-fit point. So our best-fit $\chi^2$ is approximately 1.6, for 1 nominal degree of freedom (4 data points for 3 free parameters; we count 4 data points as the EWPO covariance matrix has 3 independent eigenvalues, plus the relic density constraint). However, as we have mentioned above, the parameter $\Lambda R$ is effectively unconstrained by the data, so it is not clear whether it should count in the computation of the number of degrees of freedom. In summary, our best-fit $\chi^2 = 1.6$ is statistically acceptable both when counting 1 degree of freedom in the fit, or (even more so) when discounting $\Lambda R$  and therefore assuming 2 degrees of freedom.

\begin{figure*}
\includegraphics[width=.32 \linewidth]{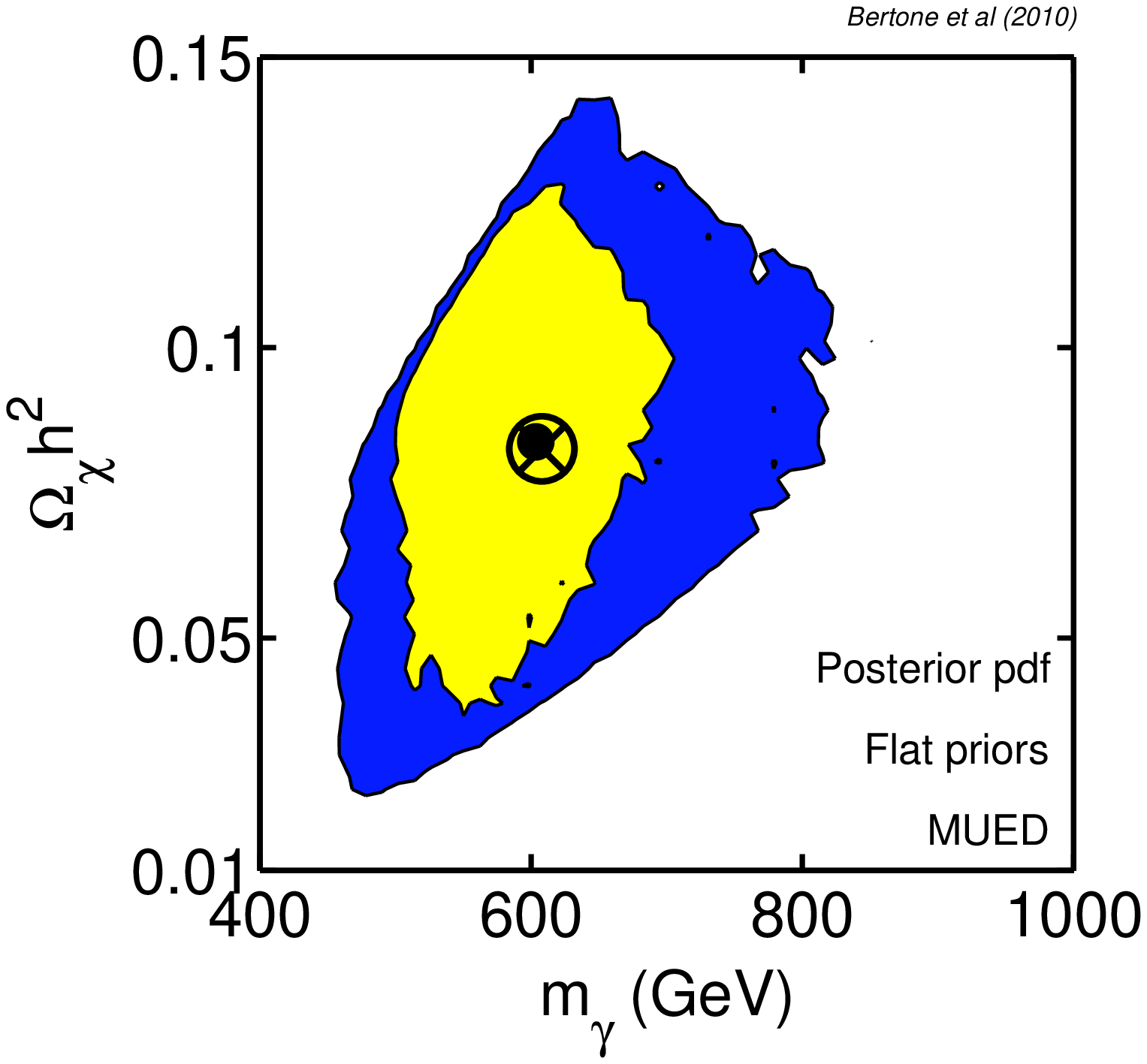} \hfill 
\includegraphics[width=.32 \linewidth]{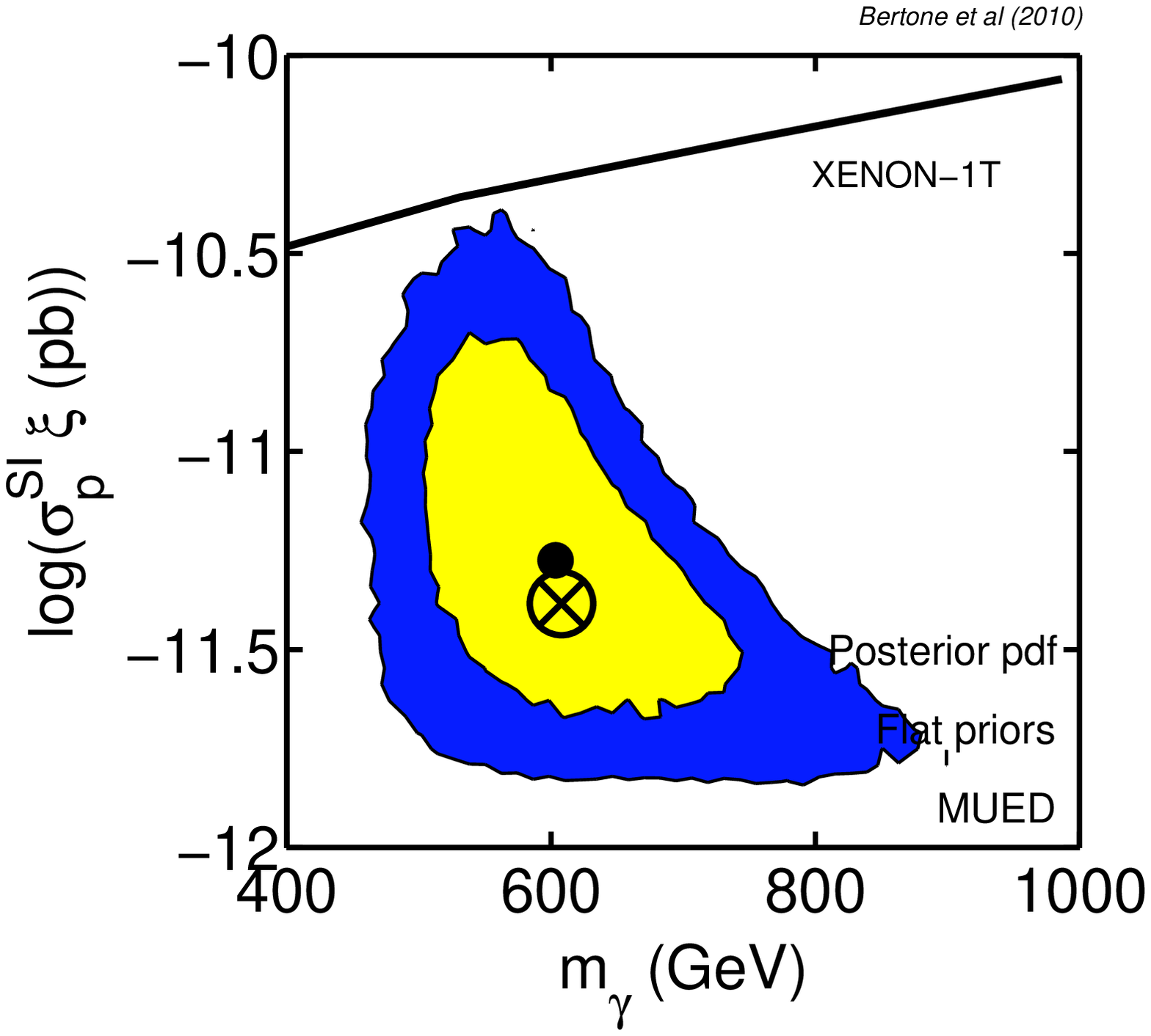} \hfill
\includegraphics[width=.32 \linewidth]{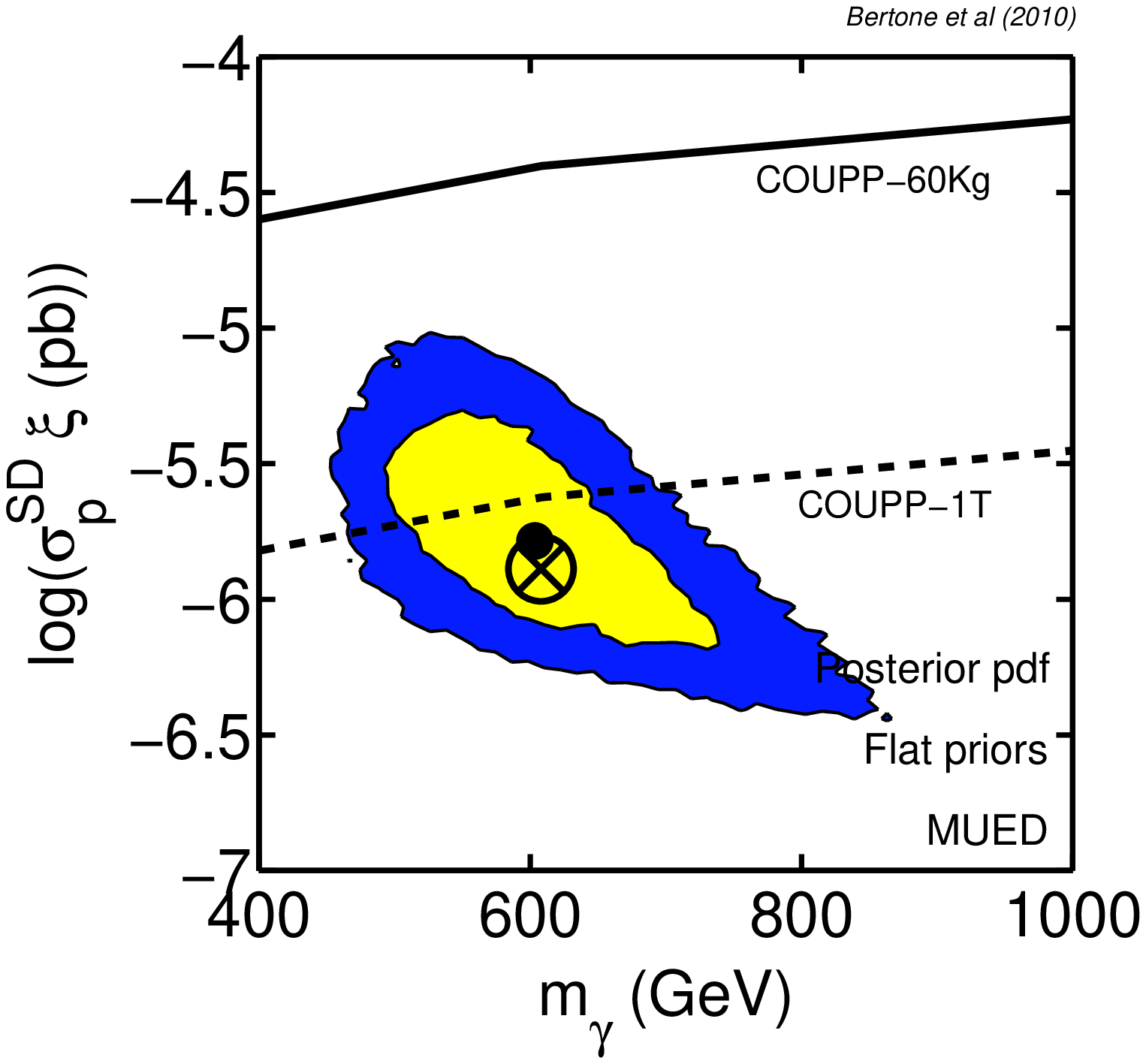}
\includegraphics[width=.32 \linewidth]{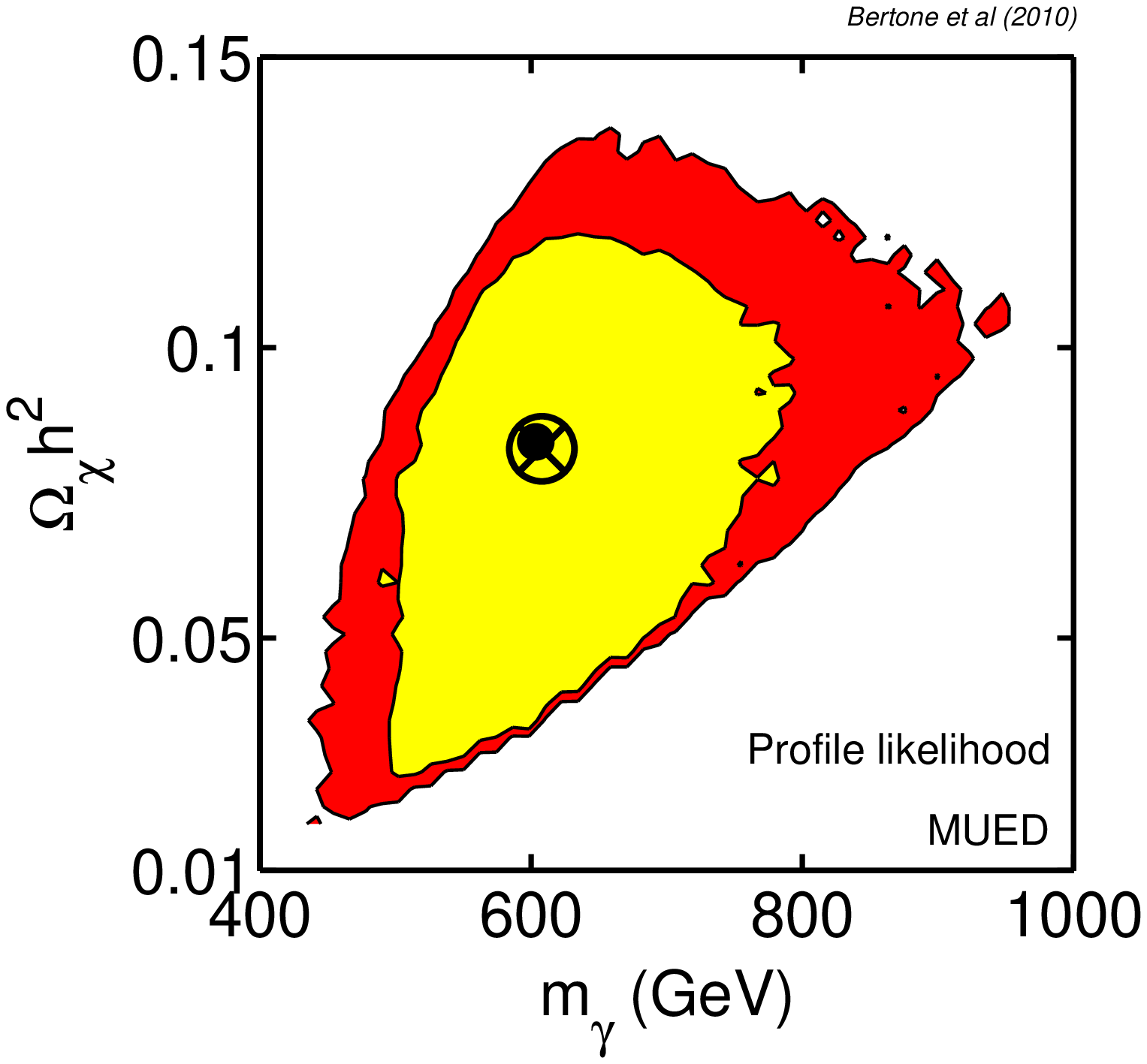} \hfill 
\includegraphics[width=.32 \linewidth]{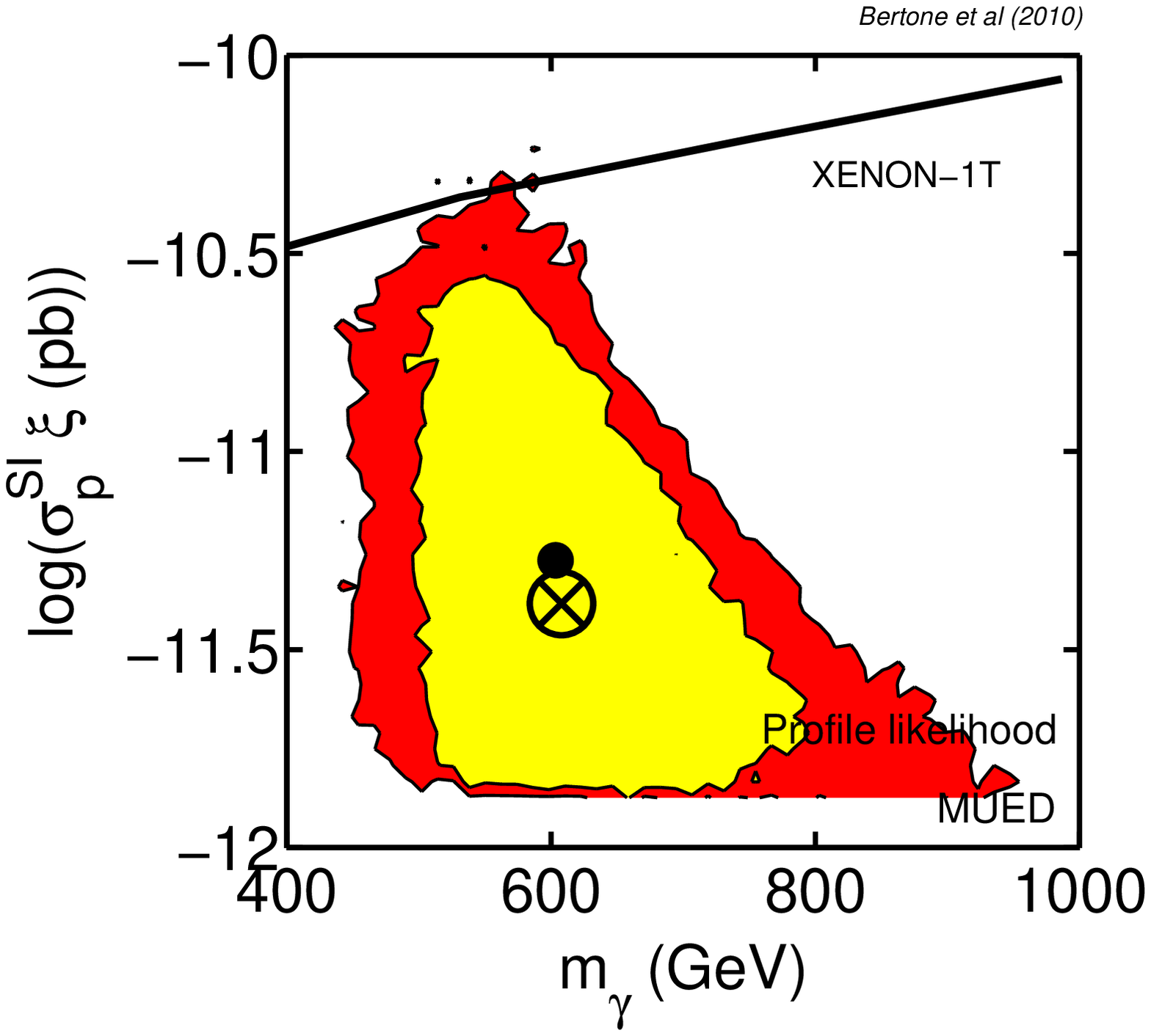} \hfill
\includegraphics[width=.32 \linewidth]{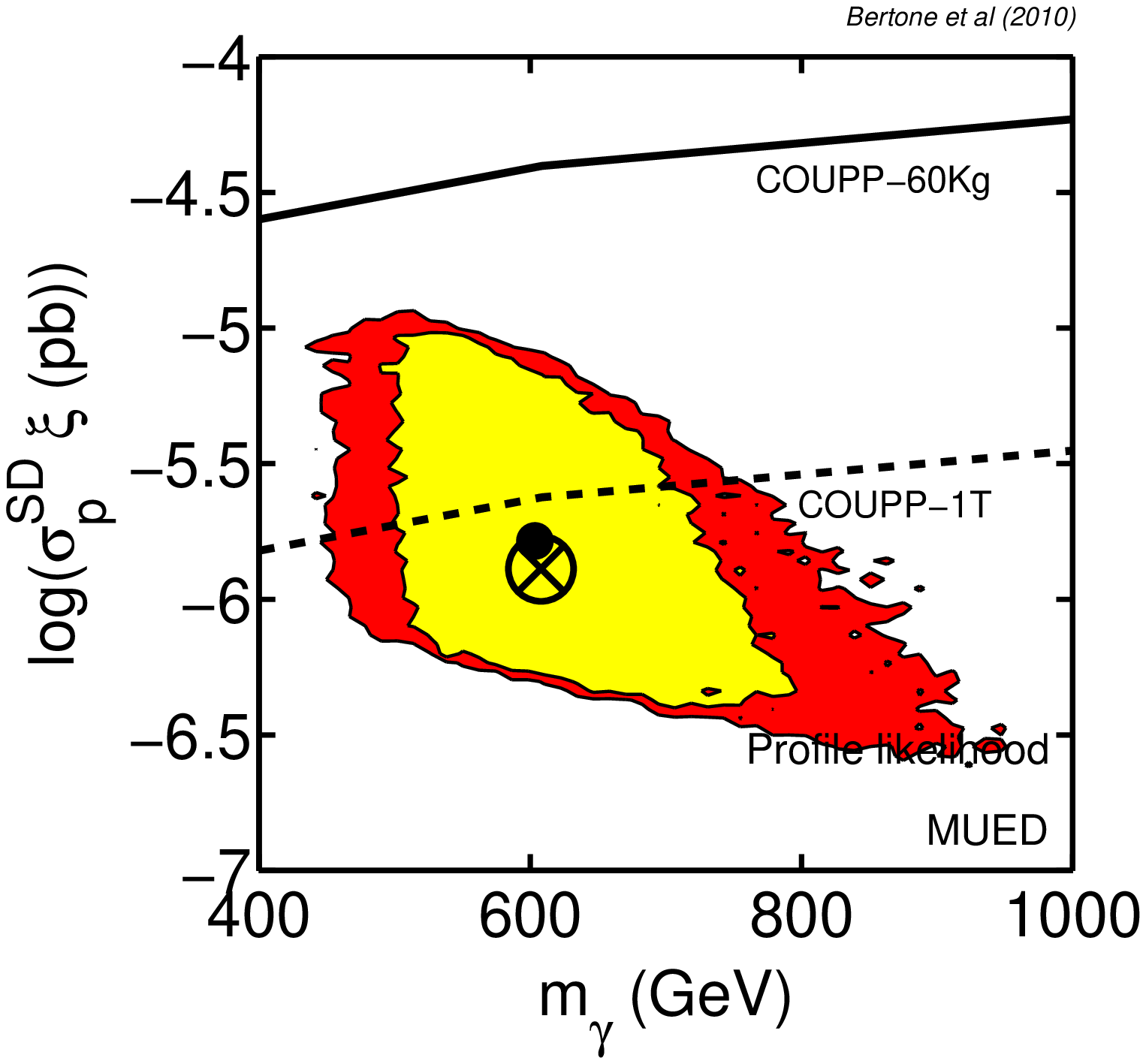}
\caption{As in Fig.~\ref{fig:2D_derived_constraints} but dropping the assumption that the \bone{} particle is the sole constituent of DM, and imposing only an upper bound instead. \label{fig:2D_derived_constraints_relaxed}}
\end{figure*}

%%%%%%%%%%%%%%%%%%%%%%%%%%%%%%%%%%%%%%%%%
\subsection{Prospects for MUED discovery }

We now move on to discuss the implications of our results for prospects of various experimental approaches to discover MUED.

The rate of events in a direct detection experiment is obviously proportional to the product of the spin-independent cross-section, $\sigma^{\rm SI}_{\rm p}$, times the local density of the LKP,  $\rho_\gamma$. This quantity can be smaller than $\rho_\text{DM}$, in the case where the LKP is not the only constituent of DM. 
In order to assess the prospects of detection, we therefore multiply $\sigma^{\rm SI}_{\rm p}$ by $\xi$, i.e. the ratio between the local KK density and the local DM density, which, following Ref. \cite{Bertone:2010rv}, we assume to be equal to the ratio of the cosmic abundances of the two species, $\xi \equiv \rho_\gamma / \rho_{\rm DM}= \Omega_\gamma / \Omega_{\rm DM}$. For $\Omega_{\rm DM}$ we adopt the central value of the WMAP determination, while for $\rho_{\rm DM}$ we adopt, following Ref. \cite{Catena:2009mf}, the value $ \rho_\chi=0.385 \mbox{ GeV} \mbox{ cm}^{-3} $ (see also \cite{Strigari:2009zb,Salucci:2010qr,Weber:2009pt,Pato:2010yq}). We note that the actual DM local density is probably larger, due to the larger density of DM in the stellar disk \cite{Pato:2010yq}, but we do not take this into account in order to be conservative.
 As one can see from Figs.~\ref{fig:2D_derived_constraints} and \ref{fig:2D_derived_constraints_relaxed}, the 2-$\sigma$ contours of the posterior lie below the sensitivity even of future experiments such as Xenon1T~\cite{Aprile:2009yh}. An experiment attempting to probe KK DM in MUED should therefore be much bigger. The fact that the posterior is concentrated over one order of magnitude in  $\siSI$, and that it lies right below the sensitivity of 1 ton experiments, suggests that an eventual generation of 10-ton experiments, would be able to probe most of the favouted parameter space.

The 1D posterior of  $\siSD$, which is the key-quantity for indirect DM searches with neutrino telescopes, shown in Figs.~\ref{fig:1D_global_constraints} and ~\ref{fig:1D_constraints_relaxed}, allows us to make a robust prediction on the prospects for detecting KK DM with the IceCube telescope, currently under construction at the South Pole, and already taking data. In fact, recent analyses of the sensitivity of IceCube to DM particles (see e.g. \cite{Halzen:2009vu}) estimate the minimum cross-section that can be probed by this experiment to be $\approx 5 \times 10^{-5}$pb or larger, in the relevant range of masses and after 5 years of data taking. However, the posterior pdf for $\siSD$ peaks one order of magnitude below this value, and it rapidly decreases for larger masses. Therefore we conclue that MUED searches at IceCube are unlikely to be successful.

Turning now to the prospects at colliders, by the end of Run II, 
Tevatron is expected to deliver more than 10 fb$^{-1}$ of data and  
will greatly improve the current bound (280 GeV at 95\% C.L.), making it closer to our best fit point. 
From the LHC side, the reach for level 1 KK particles in MUED has been calculated in \cite{Cheng:2002ab}, 
where the gold-plated $4\ell \met$ signature is considered. 
The 4 leptons are obtained from the decay of KK Z, which is produced by the decay of KK quarks.
This is quite similar to the production of the second lightest neutralino in supersymmetry.
In MUED, however, the branching fraction of KK Z into 2 leptons is large (1/6 for each generation) and 
the production cross sections of KK gluon and KK quarks are 5-10 times larger than those in SUSY 
\cite{Datta:2005zs,Datta:2005vx}.
The 14 TeV LHC can probe MUED up to $R^{-1} \sim $ 1.5 TeV (1 TeV) with 100 fb$^{-1}$ (1 fb$^{-1}$). 
A compactification scale of $R^{-1} \sim 600$ GeV (close to our best fit point) could be discovered or ruled out 
by the 14 TeV Run with 100 pb $^{-1}$ \cite{Cheng:2002ab}. 
The prospect for discovery of level 2 KK particles is discussed in \cite{Datta:2005zs,Battaglia:2005ma} in terms of 
dilepton resonance. The reach is worse than the level 1 case due to the heaviness of level 2 particles.
Very recently, the reach at 7 TeV LHC has been studied in \cite{Bhattacherjee:2010vm}.
It turns out that the opposite sign dilepton channel is the most promising 
discovery mode with 1 fb$^{-1}$ of data. It is shown that MUED can be discovered 
if $R^{-1}$ is less than 700 GeV, so this kind of search would be able to probe our $1\sigma$ region for $R^{-1}$.

%%%%%%%%%%%%%%%%%%%%%%%%%%%%%%%%%%%%%%%%%%%%%%%%%%%%%%%%%%%%%%%%%%%%
\subsection{Distinguishing the MUED scenario from the CMSSM with direct detection}
%%%%%%%%%%%%%%%%%%%%%%%%%%%%%%%%%%%%%%%%%%%%%%%%%%%%%%%%%%%%%%%%%%%%%

We now turn to the question of how to distinguish a MUED scenario from a supersymmetric one, for which we will take the paradigmatic case of the constrained minimal supersymmetric standard model (CMSSM) for simplicity.  We briefly summarize in the following the approach taken here to constraining the parameters of the CMSSM (for details, see Ref.~\cite{Cabrera:2008tj}). 

Apart from the scalar mass $m_0$, the gaugino mass $m_{1/2}$ and the trilinear coupling $A_0$ assumed to be universal 
at $M_{\rm GUT}$, the CMSSM can be parameterized in terms of 
the bilinear scalar coupling $B$, the usual Higgs mass term in the 
superpotential and the SM-like parameters $s$. The latter include the 
$SU(3)\times SU(2)\times U(1)_Y$ gauge couplings, 
$g_3, g, g'$, and the Yukawa couplings, which in turn determine the fermion
masses and mixing angles. 

In Ref. \cite{Cabrera:2008tj} was shown that considering $M_Z^{\rm exp}$ 
as experimental data in the likelihood one can integrate out $\mu$ via 
marginalization.
This procedure automatically accounts for the fine-tunning in the sense
that the posterior distribution is penalized in regions of the parameter space 
with large fine-tunning. 
Similarly the Yukawa couplings are easily integrated out when they 
are profitably traded by the physical fermion masses. 
Besides, it is highly advantageous to trade the initial $B-$parameter by 
the derived $\tan \beta$ parameter which is defined as the relative value of 
the two expectation values of the two Higsses.

The resulting posterior in function of the usual variables 
$\{m_0, m_{1/2}, A_0, \tan\beta)\}$ introduces a global Jacobian factor 
in the posterior distribution which carries the penalization of fine-tuned 
regions. Let us stress that the Jacobian is not ``subjective" at all.
Thus
\begin{eqnarray}
\label{eff_prior}
P(g_i, m_f, m_0, m_{1/2}, A_0, \tan\beta| \ \data)\  &=&  \nonumber \\
    J|_{\mu=\mu_Z}\  P(g_i, y_f,  m_0, m_{1/2}, A_0, B, \mu=\mu_Z)\, ,
\end{eqnarray}
where $J$ is the Jacobian of the transformation 
$\{\mu, y_f, B\}\ \rightarrow\  \{M_Z, m_f, \tan\beta\}$ 
(see Ref. \cite{Cabrera:2008tj} for an explicit expression for $J$),  
$\mu_Z$ is the value of $\mu$ that reproduces the experimental value
of $M_Z$ for the given values of $\{s, m_0, m_{1/2}, A_0, B\}$ and 
$P(s, m, M, A, B, \mu)$ is the prior in the initial parameters
(For a detailed discussion on the chosen priors see 
Ref.~\cite{Cabrera:2009dm}).
%
%\bea
%\label{approx_eff_prior}
%J \ \propto \   \left[\frac{E}{R_\mu^2}\right]\ 
%\frac{y}{y_{\rm low}} \frac{\tan\beta^2-1}{\tan\beta(1+\tan\beta^2)} 
%\frac{B_{\rm low}}{\mu_Z}, \
%\eea
%

One of the main consequences of this approach is that the results 
exhibit a remarkable robustness under changes of the priors 
(see Ref. \cite{Cabrera:2009dm}), showing an absence of dependences on 
the initial chosen ranges for the CMSSM parameters.
Moreover the results are compatible with likelihood 
based analyses \cite{Buchmueller:2009fn}.

\begin{figure*}[t]
\includegraphics[width=0.45 \linewidth]{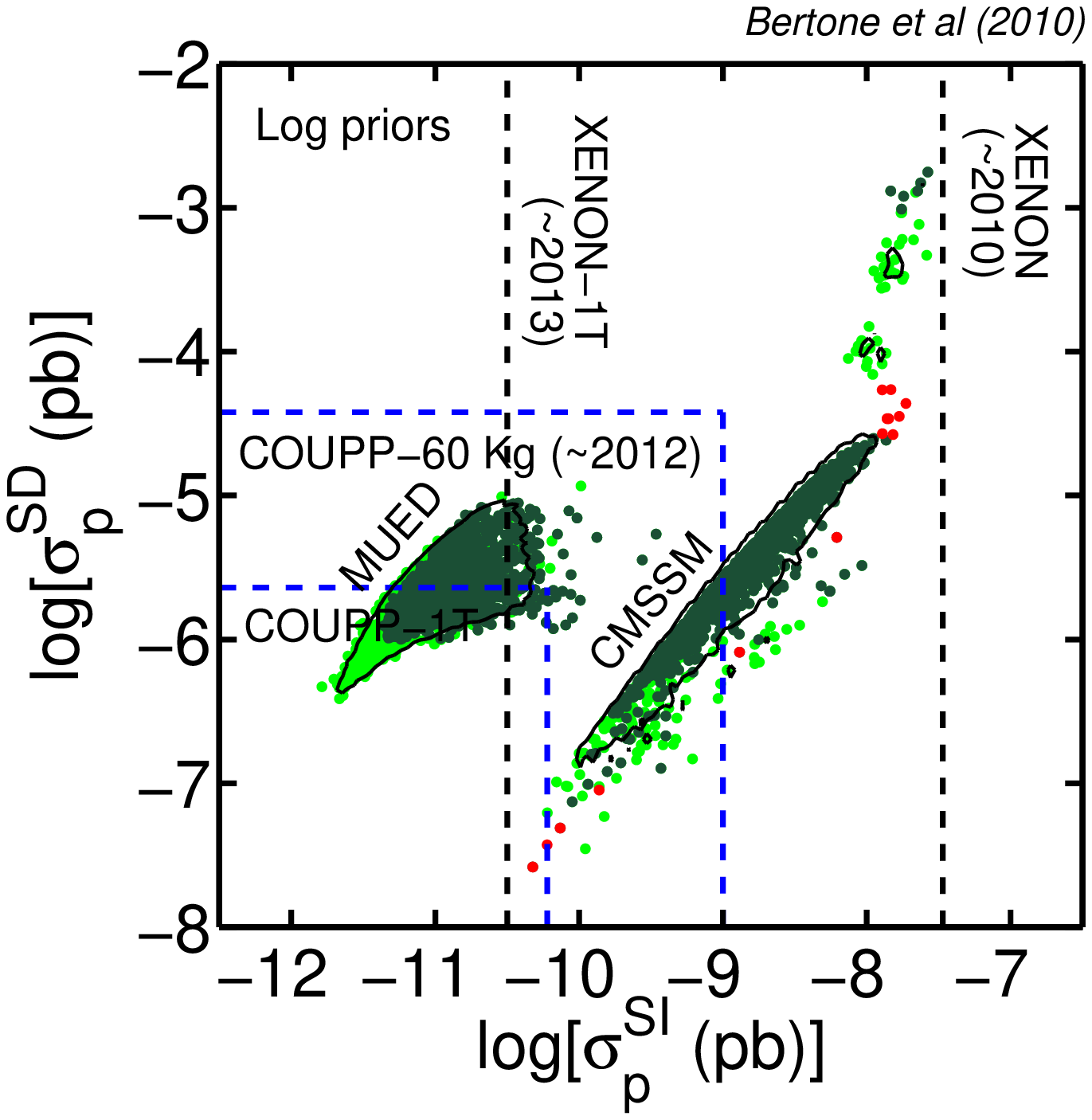}
\includegraphics[width=0.45 \linewidth]{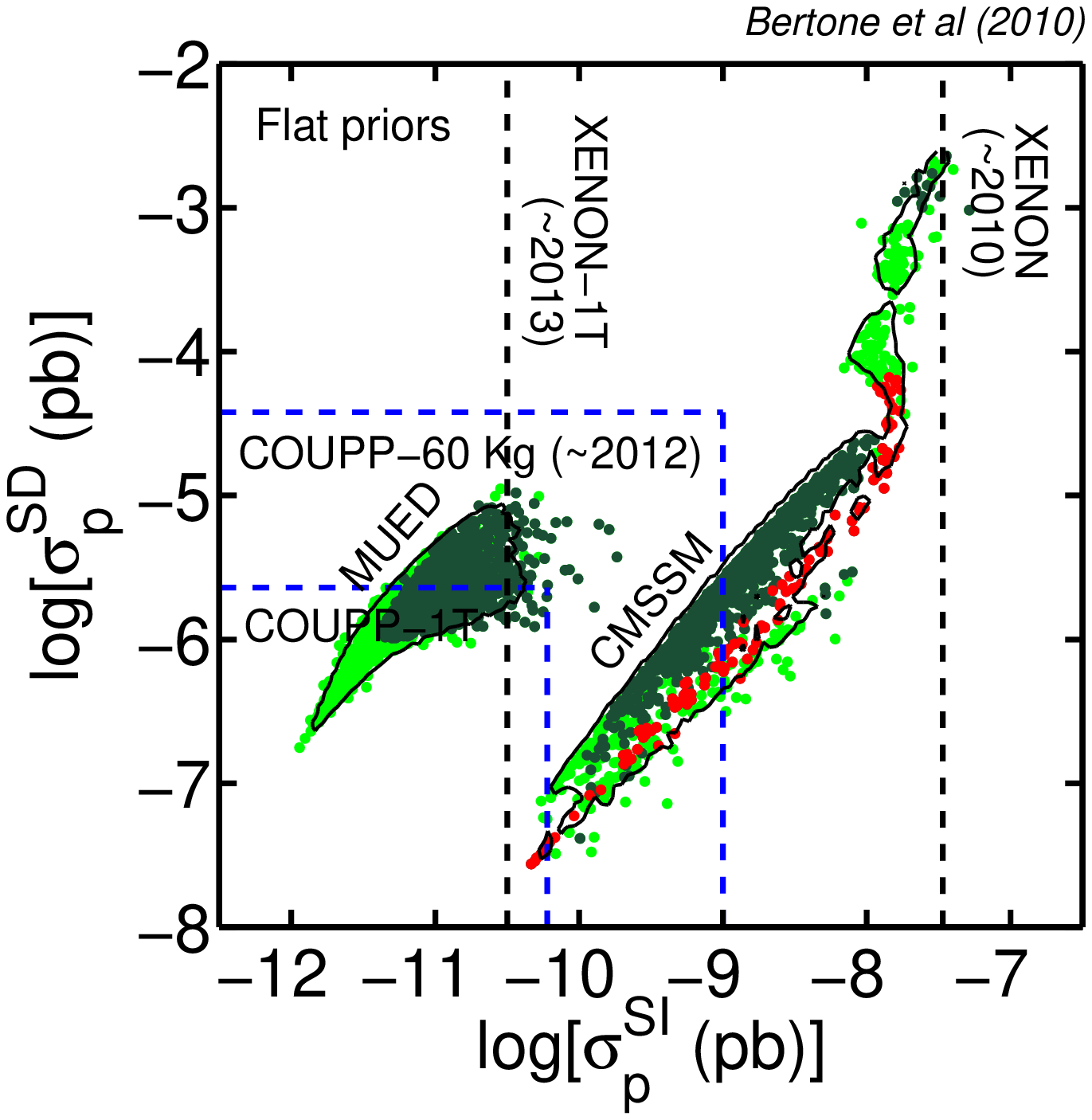}
\caption{Favoured regions for the spin-independent and spin-dependent scattering cross section for the MUED scenario (bottom left cloud) and for the CMSSM. Points are equally-weighted posterior samples for each model, accounting for all relevant present-day constraints. Dark green (light green) regions are within the reach of the LHC with 7 TeV and 1 fb$^{-1}$ (with 14 TeV and 100 fb$^{-1}$) integrated luminosity~\cite{Baer:2009dn,Baer:2010tk}, while red points are outside the reach of the LHC. Closed black contours delimit the 95\% region for each model. Dashed lines give the approximate reach of future direct detection probes.\label{fig:sigma_SI_SP} } 
\end{figure*}

Fig.~\ref{fig:sigma_SI_SP} compares the favoured regions for the spin-dependent and spin-independent scattering cross section for the MUED and the CMSSM (see also Ref. \cite{Barger:2008qd}, where a similar analysis is performed). The experimental data used in constraining the latter are given in Table 2 of \cite{Cabrera:2009dm}. Regions in light green (dark green) are within the reach of the LHC with 7 TeV and 1 fb$^{-1}$ (with 14 TeV and 100 fb$^{-1}$, respectively) for both models, while red regions are outside the reach of the LHC. Thus we can see that with 14 TeV and 100 fb$^{-1}$ the LHC is going to probe the whole of the favoured region for the MUED scenario. Also shown in Fig.~\ref{fig:sigma_SI_SP} are the sensitivities of various existing and upcoming direct detection experiments. One sees from this plot that the detection of DM off spin-independent targets would point towards SUSY, rather than KK, DM, which is consistent with the findings of Ref.~\cite{Bertone:2007xj}. The detection of KK DM in fact appears very problematic in astroparticle experiments. As we have seen in the previous section the spin-dependent coupling are such that the neutrino flux from DM annihilations in the Sun fall below the sensitivity of IceCube, even after 5 years of data taking. 
%Give here explanation of how reach was defined (ref: hep-ph/0205314, fig. 4, {\bf the reach is explained in the earlier section.}).

The only viable search strategy appears to be the detection in an experiment sensitive to spin-dependent cross-section with a large exposure. The COUPP collaboration, for instance, has been operating an ultraclean, room-temperature bubble chamber containing 1.5 kilograms of superheated CF$_3$I, that produced interesting limits on the spin-dependent coupling \cite{Behnke:2008zza}. The plans for the future include the operation of a 60 kg chamber at Snolab, that could allow a substantial improvement in sensitivity.  In Fig.~\ref{fig:sigma_SI_SP} we show for reference the reach of the 60 kg version of COUPP, as well as the case of a 1 ton chamber. If existing techniques for spin-dependent detection turn out to be scalable to such large volumes, and if one is not limited by some form of background, then KK DM could be within their reach.

%%%%%%%%%%%%%%%%%%%%%%%%%%%%%%%%%%%%%%%%%%%%%%%%%%%%%%%%%%%%%%%%%%%%%
\section{Discussion and conclusions}
\label{sec:conc}
%%%%%%%%%%%%%%%%%%%%%%%%%%%%%%%%%%%%%%%%%%%%%%%%%%%%%%%%%%%%%%%%%%%

We have discussed the prospects for detecting KK DM at accelerators and with DM experiments with a a Bayesian analysis of the minimal UED scenario. We have derived in particular the most probable range of mass and scattering cross sections off nucleons, keeping into account cosmological and electroweak precision constraints. The value of the three free parameters of the model at our best fit point are $R^{-1} =641.6$ GeV, $\Lambda R = 38$ and $m_h=215$ GeV if the KK DM explains all of the DM in the universe and $R^{-1} =607.4$ GeV, $\Lambda R = 66$ and $m_h=226.7$ GeV if KK DM is assumed to be a subdominant constituent. As we have seen, the current Tevatron limit on $R^{-1}$ is $\sim 280 GeV$, but by the end of 2011, Tevatron is expected to have 100 times more data, pushing up the limit closer to our best fit. By the time, the LHC should have collected 1 fb$^{-1}$ of data, and 
it should therefore be able to discover MUED or at least rule out the best fit.
For the two DM scenarios our best fit points for the Higgs mass are 
$\sim 215-227$ GeV, for which the Higgs dominantly decays into $W^+W^-$ 
and $ZZ$.
This mass range of the Higgs is challenging for the 7 TeV LHC with 1 fb$^{-1}$.

Our analysis has dramatic consequences for the detectability of the MUED scenario with astrophysical DM experiments. Figs.~\ref{fig:2D_derived_constraints} and \ref{fig:2D_derived_constraints_relaxed} clearly show that the 2-sigma contours in the $ \siSI$ vs. mass plane fall below $10^{-10}$ pb, i.e. even beyond the reach of future ton-scale experiments. This implies that if new particles are actually found with direct detection experiments, they are unlikely to be associated with KK DM. 
Direct detection is however not hopeless, provided that current experiments with to spin-dependent targets rapidly improve their sensitivity. We have seen that experiments such as COUPP might probe the relevant portion of the parameter space if they can go beyond the upcoming scale of 60 kg. The analysis presented here therefore provides an additional motivation to build such detectors, in case the MUED scenario is discovered at accelerators, in which case one could perform a combined analysis of accelerator and direct detection data, following the approach suggested in Ref.\cite{Bertone:2010rv}.

Indirect detection prospects are not very promising, with the most probable flux from of neutrinos from KK DM annihilations at the center of the Sun below the sensitivity of IceCube, even after 5 years of observation. We haven't discussed explicitly the possibility of detecting gamma-rays, anti-matter or synchrotron emission from KK DM annihilations in the halo, because the prospects for detection depend strongly on the assumptions made on astrophysical parameters, and when conservative choices are made for these parameters, the predicted fluxes are below the astrophysical backgrounds. This is easy to understand, since the annihilation fluxes typically scale like $\phi \sim \sigma v/m_\gamma^{2}$, and the annihilation cross section is $ \sigma v \sim m_\gamma^{-2}$, it follows that $\phi \sim m_\gamma^{-4}$. Therefore, given that the most probable range of mass is centered around the relatively large value of 600 GeV, all annihilation fluxes are quite suppressed \cite{Bertone:2004pz,Bertone:2010fn}.

Although we discussed a minimal version of the UED scenario, various extensions beyond MUED have been suggested and 
their rich phenomenology of Kaluza-Klein dark matter has been investigated in Refs. \cite{Arrenberg:2008wy,Flacke:2008ne,Matsumoto:2007dp,Shah:2006gs,Dobrescu:2007ec,Cembranos:2006gt,Kong:2010qd}. An analysis of these non-minimal scenarios will be the subject of a dedicated forthcoming paper.

%%%%%%%%%%%%%%%%%%%%%%%%%%%%%%%%%%%%%%%%%%%%%%%%%%%%%%%%%%%%%%%%

%\begin{figure*}[t]
%\includegraphics[width=.5 \linewidth]{figs/mn_flat_nuis_all_lbr10_1D.ps} 
%\includegraphics[width=.48 \linewidth]{figs/mn_flat_nuis_all_lbr40_1D.ps} 
%\caption{Marginalized constraints on the UED parameters for flat priors and $\Lambda R = 10$ (left) and $\Lambda R = 40$ (right). The red cross gives the best fit, the vertical thing line the posterior mean. \label{fig:1D_constraints} }
%\end{figure*}

{\em Acknowledgements.} 
The authors would like to thank Tim Tait for useful comments. The work of R. RdA has been supported in part by MEC (Spain)
under grant FPA2007-60323, by Generalitat Valenciana under grant
PROMETEO/2008/069 and by the Spanish Consolider Ingenio-2010 program
PAU (CSD2007-00060).  R. RdA would like to thank the support of the Spanish MICINN's
  Consolider-Ingenio 2010 Programme under the grant MULTIDARK
  CSD2209-00064.
SLAC is supported in part by the DOE under contract DE-AC02-76SF00515. RT would like to thank SLAC for hospitality during the completion of this work and the EU FP6 Marie Curie
Research and Training Network ``UniverseNet'' (MRTN-CT-2006-035863)
for partial support. The use of Imperial College High Performance Computing Service is gratefully acknowledged.

\appendix

\section{Derivation of the relic density upper bound likelihood}
\newcommand{\omKK}{\omega_\text{KK}}
\newcommand{\omDM}{\omega_\text{DM}}

In this appendix, we derive the likelihood given in Eq. ~\eqref{eq:upperbound}. There exist in the literature various expressions for the likelihood function in the case where the WMAP result is taken to be only an upper bound to the DM density. Often, the likelihood is taken to be flat up to an arbitrary cutoff value (e.g., the 95\% upper range of the WMAP likelihood) and zero above it. Ref.~\cite{FerozAllanach08}  advocated using a likelihood function which is flat below the WMAP central value, and falls off as a half-Gaussian above it. Here we derive the correct expression, which is functionally slightly different from what has been previously used. 

\begin{figure}[t]
\includegraphics[width=0.45 \textwidth]{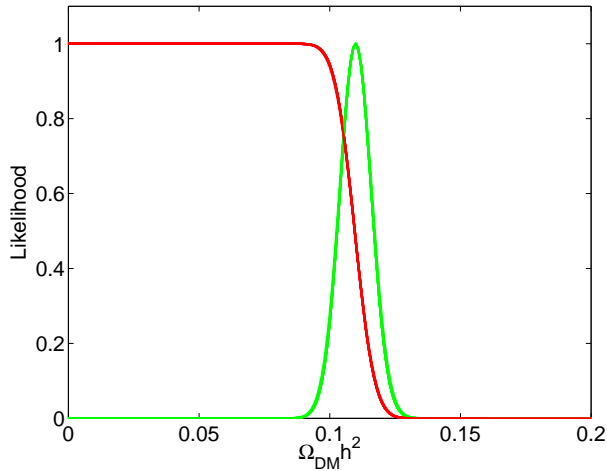}
\caption{Comparison between the WMAP likelihood when the LKP is the sole constituent of dark matter (green, Gaussian shape) and when the LKP is a subdominant component (red, upper bound). Both likelihoods are normalized to their peak value. \label{fig:bound}}
\end{figure}

We define the following shortcut notation: $\omKK \equiv \OhKK$ is the relic density of KK particles, while $\omDM \equiv \Oh$ is the relic density of all dark matter, which might comprise a secondary component beside the LKP, i.e. $\omKK \leq \omDM$. The WMAP measured mean value is given by $\muW$, and its uncertainty is $\siW$. We thus want to determine the effective likelihood 
\begin{widetext}
\be \label{eq:effective_like}
\like_\text{WMAP}(\omKK) \equiv p(\muW | \omKK) = \int p(\muW | \omDM)p(\omDM|\omKK) {\rm d} \omDM, 
\ee
\end{widetext}
where $p(\muW | \omDM)$ is a Gaussian in $\omDM$ with mean $\muW$ and standard deviation $\siW$, i.e. $\omDM \sim {\mathcal N}(\muW, \siW^2)$. In order to determine $p(\omDM|\omKK)$, we use Bayes Theorem to obtain
\be \label{eq:conditional}
p(\omDM|\omKK) = p(\omKK|\omDM) \frac{p(\omDM)}{p(\omKK)}.
\ee
On the RHS of Eq.~\eqref{eq:conditional} the first term is the conditional probability for the LKP relic density given a specified total DM density. Since we are considering the case $\omKK \leq \omDM$ and nothing else is known about the relative densities between the LKP and a secondary dark matter component, we set 
\be
 p(\omKK|\omDM) = 
 \left\{ \begin{array}{rl}
 \omDM^{-1} &\mbox{ if $\omKK \leq \omDM$,} \\
  0 &\mbox{ otherwise.}
       \end{array} 
       \right.
 \ee 
To specify the priors $p(\omDM), p(\omKK)$ in Eq.~\eqref{eq:conditional}, we appeal to the principle of indifference. Lacking any other information about the relative densities of the LKP particle and the total DM density,  we should take the two priors to be equal for both components. The ratio therefore must be 1 everywhere except in the unphysical region of negative energy density, where we set it to 0 to enforce positivity of the energy densities. Thus we have
\be \label{eq:priorratio}
 \frac{p(\omDM)}{p(\omKK)} = 
 \left\{ \begin{array}{rl}
 1 &\mbox{ if $0 \leq \omKK, \omDM \leq \Omega$,} \\
  0 &\mbox{ otherwise,}
       \end{array} 
       \right.
 \ee 
 where $\Omega$ is some large cut-off value whose precise value is irrelevant for the end result, as it will be shown below. Using Eqs.~(\ref{eq:conditional}--\ref{eq:priorratio}) into Eq.~\eqref{eq:effective_like} we obtain, taking the limit $\Omega \rightarrow \infty$:
\begin{widetext}
\begin{align} \label{eq:effective_like2}
\like_\text{WMAP}(\omKK) & = \lim_{\Omega \rightarrow \infty} \frac{1}{\sqrt{2\pi} \siW} \int_0^\Omega \exp \left(-\frac{1}{2}\frac{(\omDM-\muW)^2}{\siW^2} \right) \omDM^{-1}\Theta(\omKK-\omDM){\rm d} \omDM, \\
 & = \frac{1}{\sqrt{2\pi} \siW^2} \int_{\omKK/\siW}^\infty \exp\left(-\frac{1}{2}(x-r_\star)^2 \right) x^{-1}{\rm d} x,
\end{align}
\end{widetext}
 which is Eq. ~\eqref{eq:upperbound} (notice that although this expression is not normalized this is immaterial as we only need the likelihood up to an overall normalization constant). This effective likelihood is plotted in Fig.~\ref{fig:bound}, where it is compared to the Gaussian likelihood for the case when the DM is made entirely of LKP. The effective likelihood is flat for $\omKK \ll \muW$, then falls off exponentially for $\omKK \gg \muW$, as one would expect. Notice that the likelihood at  $\omKK = \muW$ is precisely half is asymptotic value for $\omKK \ll \muW$, which reflects the fact that for the WMAP central value we are agnostic as to which fraction of DM is made of KK particles. 

%%%%%%%%%%%%%%%%%%%%%%%%%%%%%%%%%%%%%%%%%%%%%%%%%%%%%%%%%%%%%%%%%%%%%%%%%%%%%%%%%%%%%%%%

\end{document}